\documentclass[twoside,12pt]{Classes/CUEDthesisPSnPDF}

\usepackage{wrapfig}
\usepackage{multirow}
\usepackage{dsfont}
\usepackage{slashed}
\usepackage[small,bf]{caption}
\usepackage[nottoc]{tocbibind}

\ifpdf
    \pdfcatalog { /PageMode (/UseOutlines)
                  /OpenAction (fitbh)  }
\fi

\title{QCD thermodynamics with dynamical fermions}

\ifpdf
  \author{\href{mailto:endrodi@general.elte.hu}{Gergely Endr\H{o}di}}
  \university{\href{http://www.elte.hu}{E\"otv\"os Lor\'and University\\Faculty of Natural Sciences}}
  \crest{\includegraphics[width=30mm]{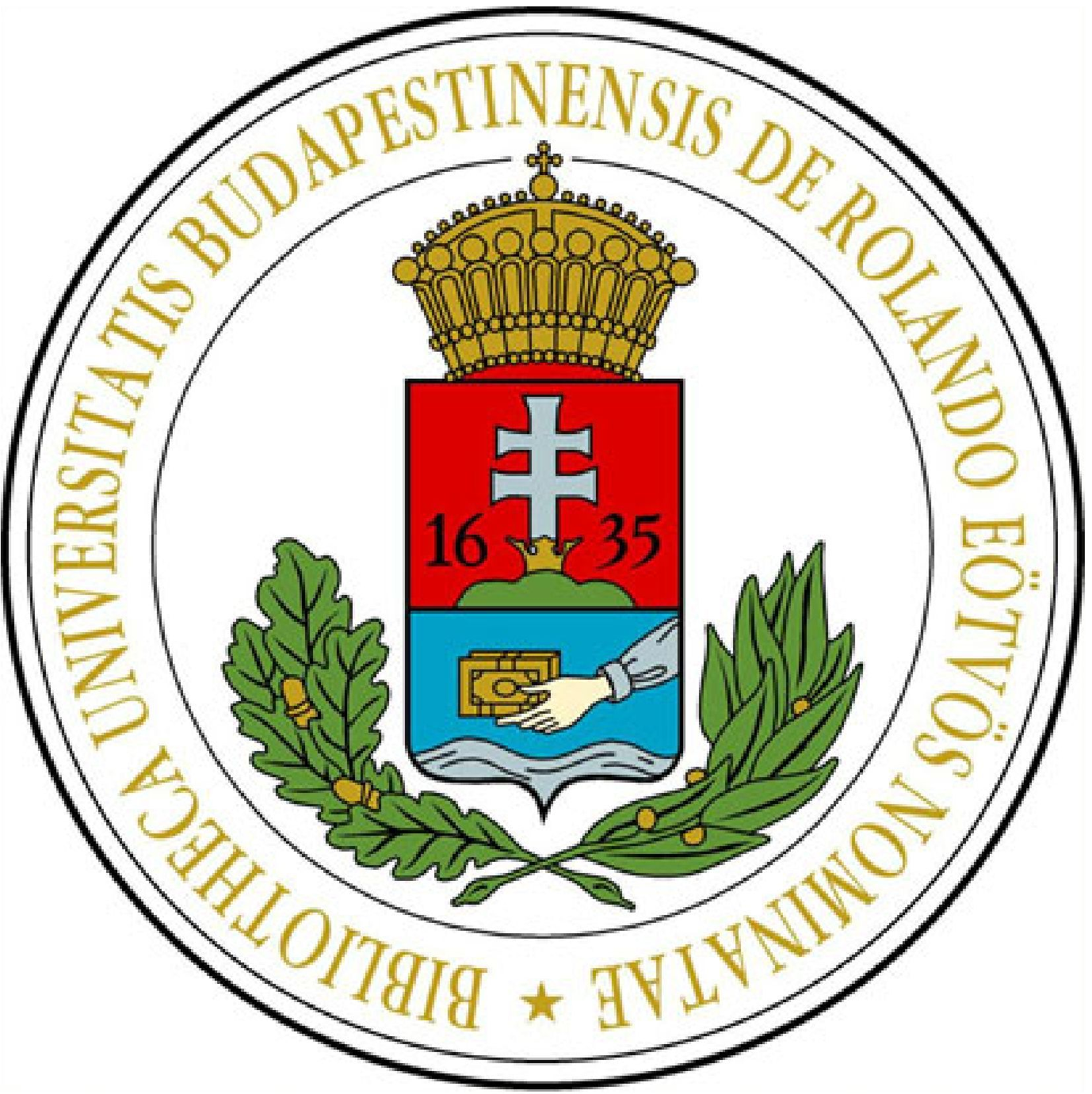}}
  \docprog{Particle Physics and Astronomy Program\\Program Leader: Dr. Ferenc Csikor\\[0.5cm]Budapest, 2011.}
\else
  \author{Gergely Endr\H odi}
  \university{Department of Theoretical Physics\\E\"otv\"os Lor\'and University, Faculty of Natural Sciences}
  \collegeordept{Doctoral School of Physics\\Head of Doctoral School: Dr. Zal\'an Horv\'ath}
  \docprog{Particle Physics and Astronomy Program\\Program Leader: Dr. Ferenc Csikor\\[0.5cm]Budapest, 2011.}
  \crest{\includegraphics[width=30mm]{Figs/eltelogo_nagy.ps}}
\fi
\advisor{Advisor: Dr. S\'andor Katz\\assistant professor}
\usepackage{StyleFiles/watermark}

\newcommand{\MeV}{\textmd{MeV}}
\renewcommand{\d}{\textmd{d}}
\newcommand{\Tr}{\textmd{Tr}}
\newcommand{\be}{\begin{equation}}
\newcommand{\ee}{\end{equation}}
\newcommand{\ba}{\begin{align}}
\newcommand{\ea}{\end{align}}

\newcommand{\Z}{\mathcal{Z}\!\!\!\raisebox{.31ex}{-}\,}

\renewcommand{\O}{\mathcal{O}}
\newcommand{\D}{{\cal D}}

\newcommand{\expv}[1]{\left \langle #1 \right \rangle}

\newcommand{\FIG}[6]{
	\begin{figure}[h!]
	\centering
	\includegraphics*[width=#1,angle=#2]{#3}
	\caption[#4]{#5}
	\label{#6}
	\end{figure}
}

\begin{document}

%\language{english}

\renewcommand\baselinestretch{1.2}
\baselineskip=18pt plus1pt

% A page with the abstract on including title and author etc may be
% required to be handed in separately. If this is not so, then comment
% the below 3 lines (between '\begin{abstractseparte}' and 
% 'end{abstractseparate}'), normally like a declaration ... needs some more
% work, mind as environment abstracts creates a new page!
% \begin{abstractseparate}
%   \input{Abstract/abstract}
% \end{abstractseparate}

% Using the watermark package which is in StyleFiles/
% and to remove DRAFT COPY ONLY appearing on the top of all pages comment out below line
%\watermark{DRAFT COPY ONLY}

\maketitle

%set the number of sectioning levels that get number and appear in the contents
\setcounter{secnumdepth}{3}
\setcounter{tocdepth}{3}

\thispagestyle{empty}
\frontmatter
%\vspace*{10cm}
%\quad
%\vspace*{10cm}
%\thispagestyle{empty}
%\newpage
%\thispagestyle{empty}
%\include{Dedication/dedication}
%\thispagestyle{empty}
%\vspace*{13cm}
%\include{Abstract/abstract}
%\vspace*{13cm}
%\thispagestyle{empty}

\tableofcontents
\listoffigures
%\phantomsection
%\addcontentsline{toc}{chapter}{List of figures}
\newpage
\newpage
%\printnomenclature  %% Print the nomenclature
%\phantomsection
%\addcontentsline{toc}{chapter}{Nomenclature}

\mainmatter

\chapter{Introduction}
The central objective of particle physics is to study the basic building blocks of matter, and to explore the way they bind together and interact with each other. Based on the interaction type one can distinguish between four different forces acting on the elementary particles: these are the gravitational, electromagnetic, weak and strong interactions. While a proper quantum description of gravity is not yet established, the latter three interactions are summarized in the structure which is called the Standard Model. The sector of the Standard Model that covers the strong force is described by a theory called Quantum Chromodynamics (QCD). The elementary particles of QCD notably differ from those of e.g. the electromagnetic interaction: they cannot be observed directly in nature. These particles -- quarks and gluons -- only show up as constituents of hadrons like the proton or the neutron. 

On the other hand, according to QCD, in certain situations quarks are no longer confined inside hadrons. One of the most important properties of QCD is asymptotic freedom, which implies that the interaction between quarks vanishes at asymptotically high energies. Due to asymptotic freedom, under extreme circumstances -- namely, at very high temperature or density -- quarks can be liberated from confinement. At this point a plasma of quarks and gluons comes to life that is substantially different as compared to the system of confined hadrons. This plasma state is referred to as the quark-gluon plasma (QGP).

\section{The quark-gluon plasma}
\label{sec:qgpintro}
There are two situations in which QGP is believed to exist (or have existed): first in the early Universe and second in heavy ion collisions. In both cases the temperature -- and with it the average kinetic energy of quarks -- is high enough to overcome the confining potential that is present inside hadrons and due to this the very different plasma state can be formed. For the case of the early Universe this plasma state was realized until about $10^{-5}$ seconds after the Big Bang, when the temperature sank below a critical temperature $T_c \sim 200\;\MeV$\footnote{This is equivalent to about $10^{12}$ degrees Celsius.}. The properties of the transition between the `hot', deconfined QGP and the `cold', confined hadrons play a very important role in the understanding of the early Universe~\cite{Schwarz:2003du}. The two above forms of strongly interacting matter can be thought of as phases in the sense that the dominant degrees of freedom in them are very different: colorless\footnote{The analogue of charge in QCD is called color, see section~\ref{sec:strong}.} hadrons in the cold, and colored objects in the hot phase. In accordance with this, the transition can be treated as a phase transition in the statistical physical sense.

One of the most important properties of the transition is its nature. We define a transition to be of {\it first order} if there is a discontinuity in the first derivative of the thermodynamic potential. For a {\it second-order} transition there is a jump in the second derivative, i.e. the first derivative is not continuously differentiable. For an analytic transition no such singularities occur, one may refer to such a process as being a {\it crossover}. In the case of a first-order phase 
\begin{wrapfigure}{r}{5.2cm}
\vspace*{-0.3cm}
\centering
  \includegraphics*[width=4.6cm]{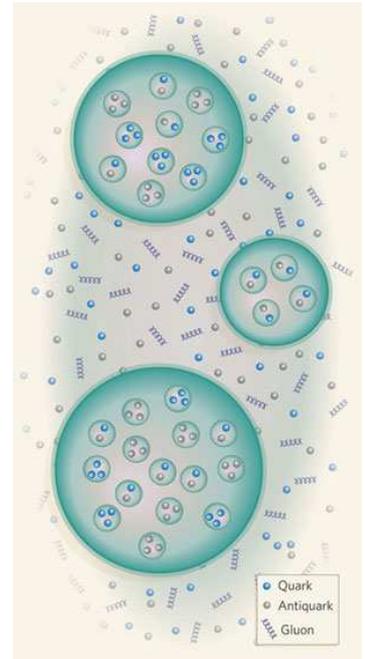}
\vspace*{-0.2cm}
  \caption[Illustration of a first-order transition.]{Illustration of a first-order transition~\cite{Wilczek:2006}.}
\vspace*{-0.5cm}
  \label{fig:firstorder}
\end{wrapfigure}
transition -- like the boiling of water -- the quark-gluon plasma would reach a super-cooled state in which smaller bubbles of the favored, cold phase can appear.
As the system aspires towards the minimum of the free energy, the large enough (supercritical) bubbles can further grow and after a while merge with each other (see illustration in figure~\ref{fig:firstorder}, taken from~\cite{Wilczek:2006}). This process can be treated as a jump through a potential barrier from a local minimum to a deeper minimum; from a so-called false vacuum to the real vacuum. On the other hand the transition can also be continuous (second order or crossover) -- in this case no such bubbles are created and the transition between the two phases occurs uniformly.

The cosmological significance of the above phenomenon is that if such bubbles indeed appeared then at the phase boundaries specific reactions can take place that one would be able to observe in the cosmic radiation. Such a transition may also have a strong effect on nucleosynthesis.
The boundaries of the bubbles can furthermore collide and as a result produce gravitational waves that may also be detected~\cite{Witten:1984rs}. According to lattice calculations of QCD however, the transition from QGP to confined, hadronic matter is much likely to be an analytic crossover~\cite{Aoki:2006we}. This means that the transition goes down continuously as illustrated in figure~\ref{fig:crossover}.

\FIG{8cm}{0}{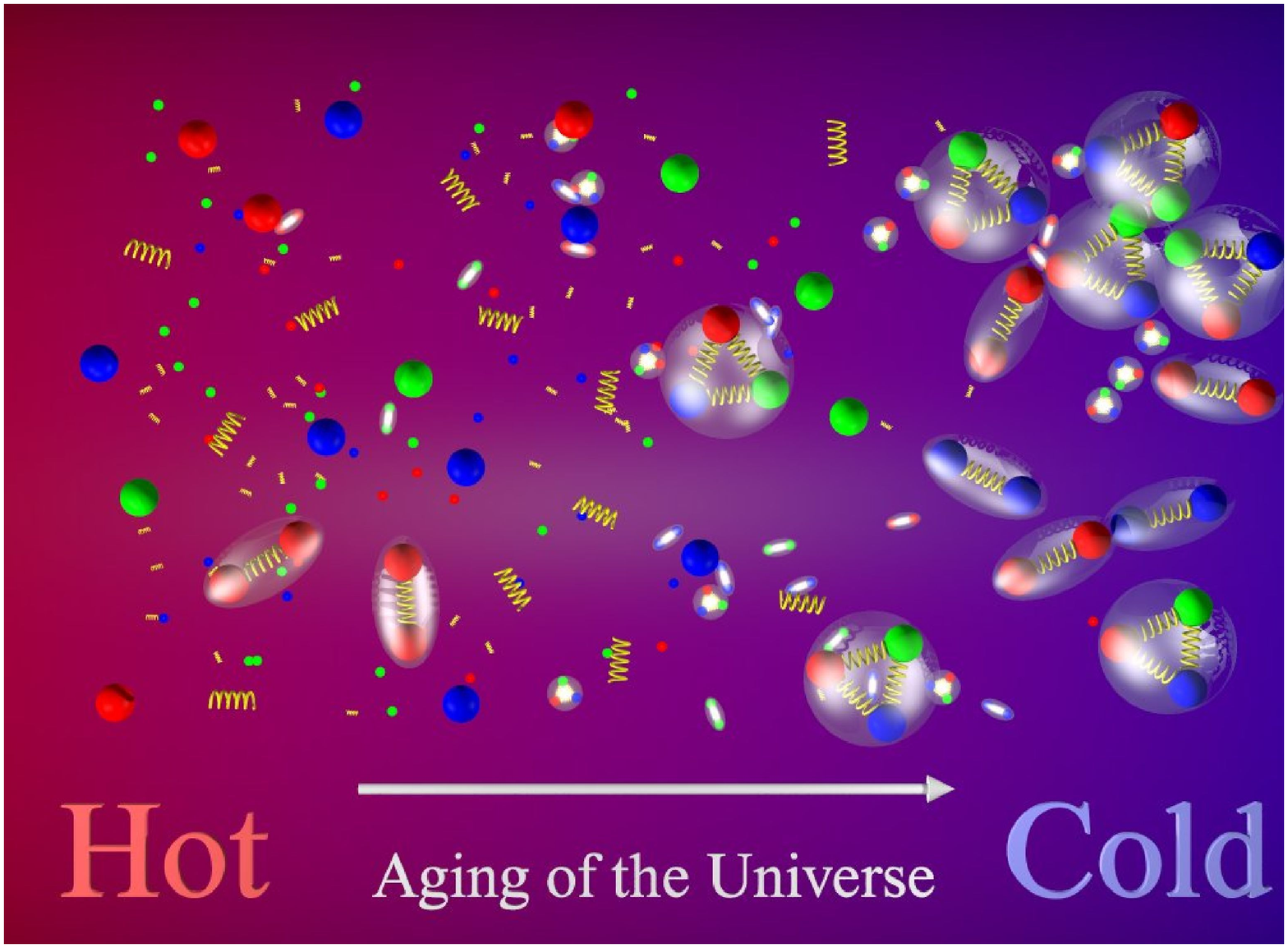}{Illustration of a crossover transition.}{Illustration of a crossover transition. Mesons (quark-antiquark bounded states) and hadrons (three-quark bounded states) appear continuously while the Universe cools down to its cold phase.}{fig:crossover}

The QCD transition plays a very relevant role also from the point of view of heavy ion collisions. It is now widely accepted that in high energy collisions of heavy ions conditions resembling the early Universe can be generated and the plasma phase of quarks can be recreated. Recently, in a collision of gold nuclei at the Relativistic Heavy Ion Collider (RHIC), an initial temperature beyond 200 MeV was reached~\cite{Adare:2009qk}. There are also further signals indicating that the QGP has indeed been created. One of these signals is the phenomenon of jet quenching. A jet is a beam of secondary particles that originates from the high-momentum quark that was broken out of the incoming protons or neutrons. Interactions between the jet and the hot, dense medium produced in the collision are expected to lead to a loss of the jet energy. Evidence for jet quenching has indeed been found at the Relativistic Heavy Ion Collider (RHIC)~\cite{Adler:2002tq}. 

\section{The phase diagram of QCD}
\begin{figure}[h!]
\centering
\includegraphics*[width=9.0cm, bb=6 4 757 501]{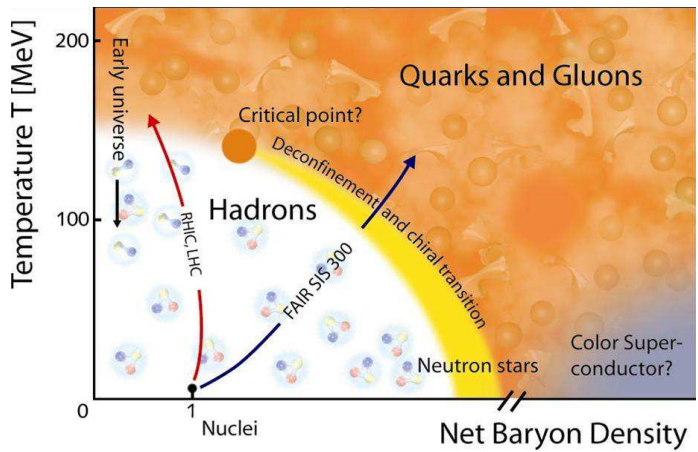}
\caption[A possible depiction of the phase diagram of QCD.]{A possible depiction of the phase diagram of QCD in the space of the state parameters: the temperature $T$ and the baryon density (which equals three times the quark density). Phases are separated by a crossover transition at zero chemical potential. At larger values of $\rho$ a first-order phase transition may take place, which is indicated by the yellow band. The crossover and first-order lines must then be separated by a critical endpoint. Regions at large densities and small temperatures are thought to describe the interior of dense neutron stars. At even higher densities exotic phases like a color superconductor are expected. Figure taken from~\cite{CBMbook}.}
\label{fig:pd}
\end{figure}

Just like at high temperature $T$, also at large quark densities $\rho$ we expect (in agreement with asymptotic freedom) the coupling between quarks to decrease and the QGP phase to be created. In the statistical physical approach to QCD thermodynamics, the net density of quarks\footnote{The density of quarks minus that of antiquarks.} can be controlled by a chemical potential $\mu$. Zero chemical potential corresponds to a situation where the density of quarks and antiquarks is the same.
The phases of strongly interacting matter can be represented on a $\mu-T$ or $\rho-T$ phase diagram (see a possible depiction on figure~\ref{fig:pd}, taken from~\cite{CBMbook}). On this diagram phases are separated by transition lines that can represent either first-order or second-order phase transitions or continuous transitions (crossovers). The $T,\mu$ parameters during the cooling down of the early Universe or a high energy collision also draw a trajectory on the phase diagram. These trajectories are contained in the small $\mu$ region of the phase diagram, since in both cases the number of quarks and antiquarks are roughly the same. Accordingly, this situation represents a thermodynamic system having zero or small chemical potential.

By increasing the density and keeping the temperature fixed -- i.e. compressing hadronic matter -- one can also move into the deconfined phase of quarks. We know far less about this region of the phase diagram, which is thought to exhibit phenomena like color flavor locking or color superconductivity. On the other hand, the low $\mu$ area is much better understood and theoretically more tractable. Besides phenomenological interest, the detailed structure of this area (like the transition temperature or the curvature of the transition line) is relevant for contemporary and upcoming heavy ion experiments. 
While most of the ongoing experiments like those at LHC or RHIC concentrate on achieving very high energies and thus small chemical potentials, there are
projects that aim for regions of the phase diagram with larger densities (RHIC II, FAIR)\footnote{In a heavy ion collision the density of the system is controlled by the center of mass energy $\sqrt{s_{NN}}$ and the centrality of the collision.}. Designing these next generation experiments can benefit greatly from developing theoretical understanding of the phase diagram.

According to lattice simulations, at zero chemical potential the transition is an analytic crossover~\cite{Aoki:2006we}. This is represented in figure~\ref{fig:pd} by the smooth transition between the white and the orange regions. A possible scenario about the $\mu>0$ region of the phase diagram is that a first-order transition emerges (yellow band), which also implies the existence of a critical endpoint (orange dot). Such a critical endpoint corresponds to a second-order phase transition that belongs to a given universality class. It could also happen that the transition is continuous also at larger $\mu$ values, and then no critical endpoint exists. There have been lattice indications favoring both scenarios, so this is still an open question. See e.g.~\cite{Fodor:2001pe} arguing for, and~\cite{deForcrand:2007rq} against the existence of a critical endpoint.

A further important characteristic of the system is its equation of state as a function of the temperature. The equation of state (EoS) is also sensitive to the transition between hadronic phase and the QGP and thus also plays a very important role in high energy particle physics. Moreover, recent results from RHIC imply that the high temperature quark-gluon plasma exhibits collective flow phenomena. It is also conjectured that the description of hot matter under these extreme circumstances can be given by relativistic hydrodynamic models. In turn, these models depend rather strongly on the relationship between thermodynamic observables, summarized by the equation of state. The EoS can be calculated using perturbative methods, but unfortunately, such expansions usually converge only at temperatures much higher than the transition temperature. Therefore the lattice approach (as a non-perturbative method) is a suitable candidate to study the EoS in the transition region $T\sim T_c$.

\section{Structure and overview}

In this thesis I concentrate on the low $\mu$, high $T$ region of the phase diagram, which -- according to the above remarks -- is interesting for both the context of the evolution of the early Universe and heavy ion collisions. 

The thesis is structured as follows. First I present a brief introduction to the theoretical study of the QGP. This includes the definition of the underlying theory, QCD, and the method with which QCD can be represented and studied using a finite discretization of the variables on a four-dimensional lattice (see chapter~\ref{chap:basic}). Using the lattice formulation one can study various thermodynamic properties of strongly interacting matter. This is investigated in detail for the case of vanishing quark density and also for the case where a positive net quark number is present. The analysis of the latter system entails a fundamental problem, so separate sections are devoted to this issue (see chapter~\ref{chap:thermo}).

After having discussed some of the basic elements of lattice QCD thermodynamics, I turn to present several new results regarding the transition between confined hadrons and the quark-gluon plasma. These results are divided into three separate chapters. First, in chapter~\ref{chap:phasediag} I present the study of the phase diagram at small chemical potentials.
In this project the pseudocritical temperature and the nature of the QCD transition are analyzed as a function of the quark density. With the help of these functions the phase diagram of QCD can be reconstructed. In particular, the curvature of the transition line lying between the two phases is determined, and the possibility of the existence of a critical endpoint is also addressed. Preliminary results regarding the curvature have been published in~\cite{Endrodi:2009sd}, while the full result has been published recently~\cite{Endrodi:2011gv}. This work was done in collaboration with Zolt\'an Fodor, S\'andor Katz and K\'alm\'an Szab\'o. My contributions to the project were the following:
\begin{itemize}
\item I have developed and implemented a method to define the curvature without the need to fit the $\mu>0$ data. This definition also gives information regarding the relative change in the strength of the transition.

\item I have performed all of the simulations and measured the Taylor-coefficients necessary for this definition. By means of a multifit to data measured at various lattice spacings I determined the curvature of the transition line separating the confined and deconfined phase.
\end{itemize}

Afterwards I turn to show results regarding the equation of state. The central quantity here is the pressure as a function of the temperature. After a brief overview of the literature and a discussion about how one can determine the pressure on the lattice I present the results. In chapter~\ref{chap:EoS} I study the EoS with dynamical quarks; this work has been recently published~\cite{Borsanyi:2010cj}. I participated in this project as member of the Budapest-Wuppertal collaboration. My contributions were:
\begin{itemize}
\item I have developed a multidimensional integration scheme that can be used to reconstruct a smooth hypersurface using scattered gradient data. I used this approach to determine the pressure of QCD in the two-dimensional parameter space spanned by the gauge coupling and the light quark mass. In this approach it is straightforward to study the quark mass dependence of the EoS and to estimate the systematic error in the pressure.

\item I have measured the charm condensate on part of the dynamical configurations and evaluated the charm contribution to the pressure and to other thermodynamic observables.

\end{itemize}
The multidimensional integration method is applicable on a general level and thus was also published individually~\cite{Endrodi:2010ai}. The method is summarized in more detail in appendix~\ref{appendix1}.

Finally, in chapter~\ref{chap:PG} I show results regarding the high temperature EoS in the pure gluonic theory. At these previously unreached temperatures it becomes possible to carry out a comparison to improved or resummed perturbation theory. Furthermore, the non-perturbative contribution to the trace anomaly is also quantified. This is also a project of the Budapest-Wuppertal collaboration, currently under publication~\cite{Szabolcs}, with preliminary results already published earlier~\cite{Endrodi:2007tq}. My contributions were the following:
\begin{itemize}

\item I have evaluated the pressure using a multi-spline fit to data measured at various lattice spacings and extracted the continuum extrapolated curve from this fit.

\item I have compared and matched lattice results with improved and resummed perturbative formulae. The free parameters were fitted to best reproduce the lattice results at high temperature. Furthermore, I quantified the non-perturbative contribution to the interaction measure with either a constant or a logarithmic ansatz.

\end{itemize}

\chapter{Quantum Chromodynamics and the lattice approach}
\label{chap:basic}

\section{A theory for the strong interactions}
\label{sec:strong}
Nowadays it is widely accepted that QCD is the appropriate tool to treat the interactions between quarks and gluons. For a long time it was however unclear what kind of theory could describe the strong forces~\cite{Muta:1997}. The electron-hadron collision experiments of the late sixties -- namely, the Bjorken-scaling of the structure functions in such scatterings -- suggested that electrons scatter off almost free, point-like constituents. The Bjorken-scaling implies that these constituents -- the quarks -- should interact weaker at shorter distances (at larger energies). Since for electromagnetism an appropriate describing theory was Quantum Electrodynamics (QED), it was reasonable to search for another quantum field theory (QFT) that succeeds to describe the dynamics of quarks. While most QFTs fail to fulfill the above requirement, it was proven in 1973 that non-Abelian gauge field theories on the other hand possess this property of exhibiting a weaker force at shorter distances. This property was named {\it asymptotic freedom}~\cite{Gross:1973id,Politzer:1973fx}.

QCD is a non-Abelian quantum field theory, which differs from its Abelian relative QED in the fact that here the symmetry transformations of the underlying theory cannot be interchanged. In other words, the corresponding symmetry consists of non-commutative generators: the gauge group here is SU(3) instead of U(1). Not much after the study of Gross, Wilczek and Politzer it was proven that non-Abelian gauge theories are not just a possible candidate to play the role of the theory of strong interactions; they represent the only class of theories in four space-time dimensions that exhibit asymptotic freedom.

The new symmetry (the {\it gauge symmetry}) that is generated by the non-commutative algebra describes a new type of property that quarks -- compared to electrons -- possess. This property was named color due to the apparent analogy of the system to color mixing: while quark fields carry a color quantum number, three of them can only build up a proton if the resulting combination possesses no color indices. In other words, three colored quarks can only be confined inside a proton if their `mixture' is colorless. If one assumes that only such colorless states can be realized, then it is obvious that quarks as isolated particles cannot exist. This property of QCD is called {\it confinement} -- referring to the fact that three quarks are confined inside a proton.

Just like in QED, in QCD there is also a mediating particle corresponding to the gauge field that transmits the interaction between color-charged quarks. The analogon of the QED photon is the gluon. While the photon does not have an electric charge (due to the non-commutative nature of the gauge group), the gluon possesses a color quantum number. Because of this, contrary to the photon, the gluon also couples to itself; this is the reason for the fact that instead of the $1/r$-like decaying potential of QED, in QCD a linearly rising potential appears between color-charged objects. The linear potential can be interpreted as a spring that binds quarks to each other inside a proton. This is just another way to describe the phenomenon of confinement.

In accordance with the above remarks, the Lagrangian density of QCD contains the fermion and antifermion fields $\psi$ and $\bar\psi$ (``quarks'' and ``antiquarks'') and the gauge field $A_\mu$ (``gluons''). The dynamics of the gauge field is governed by the field strength tensor
\be
F_{\mu\nu} = \partial_\mu A_\nu-\partial_\nu A_\mu - i g [A_\mu,A_\nu]
\ee
Furthermore, the interaction between gluons and quarks is determined by the minimal coupling, as contained in the covariant derivative
\be
D_\mu = \partial_\mu - i g A_\mu
\ee 
Putting all this together, the QCD Lagrangian describing $N_q$ number of different quark flavors (in Minkowski space-time) is:
\be
\mathcal{L} =-\frac{1}{2} \Tr \left(F_{\mu\nu} F^{\mu\nu}\right) + \sum\limits_{q=1}^{N_q} \bar \psi_q \left(i\gamma_\mu D^\mu - m_q\right)\psi_q
\label{eq:Lag}
\ee
The parameters of the theory are the gauge coupling $g$ and the masses of the quarks $m_q$. In nature there are six quark flavors (up, down, strange, charm, bottom and top) and thus six masses. However, the contribution of heavy quarks is usually negligible at the non-perturbative scale of $T_c\sim 200$ MeV and only the first three or four quark species play a significant role. Furthermore, the difference between the up and down masses is very small (compared to $T_c$) and thus it is a good approximation to take $m_u=m_d\equiv m_{ud}$.

QCD is an $SU(3)$ gauge theory, i.e. the Lagrangian~(\ref{eq:Lag}) is invariant under $SU(3)$ transformations. Quarks and antiquarks transform according to the fundamental representation of the gauge group, on the other hand gluons are placed in the adjoint representation. Accordingly, the fermionic fields have three color components and gluons contain eight degrees of freedom. Furthermore, in view of the Lorentz group quarks are bispinors and thus have four spin-components:
\begin{align}
\textmd{quarks: } &\psi^{c;\alpha}_q \quad c=1\ldots 3,\;\alpha=1\ldots4,\; q=u,d,s,c,t,b \\
\textmd{gluons: } &A_\mu=\sum\limits_{a=1}^8 A_\mu^a T_a
\end{align}
with $T_a$ being the infinitesimal generators of the gauge group, usually represented by the eight Gell-Mann matrices. The Lorentz- and color indices of the quark field will be suppressed in the following. Moreover, throughout the thesis the different quark flavors will be identified using their first letter in the index of the field e.g. $\psi_u$ will stand for an up quark.

\section{Perturbative and non-perturbative approaches}
According to the property of asymptotic freedom, for short-range reactions the interaction is weak and thus such processes can be safely analyzed by perturbation theory. On the other hand the temperature scale on which perturbative expansions converge is extremely high, and therefore in order to study e.g. the transition to QGP it is necessary to assess the dynamics of quarks using a non-perturbative approach. In the second half of the 20th century a new theory was constructed that is based on mathematical concepts and can be investigated through numerical simulations: lattice gauge theory. The main idea of this approach is to restrict the fields of the QCD Lagrangian to the points of a four-dimensional lattice. This theory is the only non-perturbative, systematically adjustable approach to QCD. Through lattice gauge theory we can gain information about quarks and gluons solely using first principles -- the Lagrangian of QCD.

The appearance of ultraviolet divergences, being a familiar property of quantum field theories, is also present in QCD. A possible way to deal with such infinities in QFT is to introduce some type of regularization (e.g. a cutoff) that makes the divergent Feynman-amplitudes mathematically tractable. After a proper renormalization of these amplitudes, the next step is the removal of the regularizing constraint. Necessarily, due to the redefinition of the renormalized quantities -- which will then converge to a finite value as the regularization is removed -- the bare parameters of the theory will also become cutoff-dependent. The renormalization program as implemented on the lattice will be discussed in details in section~\ref{sec:renorm}.

This procedure of regularization and renormalization can be realized using lattice gauge theories in an instructive manner. The lattice itself plays the role of the regulator, since the introduction of a finite lattice spacing $a$ is equivalent to setting a cutoff $2\pi/a$ in momentum-space. This way on a lattice of finite size obviously every amplitude is going to be finite. The removal of the finite lattice spacing is called the continuum limit: this is usually done through some kind of extrapolation to $a \rightarrow 0$.

The transition from confined hadrons to QGP is definitely a phenomenon that is only accessible by the lattice approach, since the relevant temperature scale of $T_c \sim 200$ MeV is still very far from the perturbative region. In the following sections I will address basic elements of lattice QCD, starting from the discretization of the variables, for the gauge fields and also for the fermion fields. The latter entails a fundamental problem that is stated in the so-called Nielsen-Ninomiya no-go theorem. Among the possible discretizations I will investigate the staggered version, since this was used to obtain all the results that are presented in this thesis.

\section{Quantum field theory on the lattice} 

In this section a very brief overview of lattice gauge theory is given. A full and detailed introduction can be found in e.g.~\cite{Montvay:1994},~\cite{Rothe:1992} or~\cite{Gattringer:2010zz}.

As mentioned above, the lattice can be thought of as a possible regulator of the divergent Feynman-amplitudes. The lattice approach can be formulated through the functional-integral formalism, which is a generalization of the quantum mechanical path integral of Feynman. This formalism readily shows how it becomes possible to treat the system non-perturbatively, without any use of perturbation theory.

According to the path integral in quantum mechanics, the Green's function of a particle propagating from $q_1$ to $q_2$ in the time interval $[t_1, t_2]$ can be written as an integral over various possible paths:
\be
G(q_2,t_2;q_1,t_1) \equiv \langle q_2| e^{-iH(t_2-t_1)}|q_1\rangle = \!\!\!\int\limits_{q_1\to q_2}\!\!\! Dq \; e^{iS[q]}
\label{eq:greensfunc}
\ee
where $S[q]$ is the action that belongs to the path given by $q$. This expression tells us to take every possible classical path that starts from $q_1$ and ends at $q_2$, and sum the corresponding phase factors in the integrand. These factors, however, oscillate very strongly, so the calculation has to be carried out in Euclidean space-time instead of the usual Minkowski space-time, which can be achieved using the $t\to -i\tau$ substitution. This latter change of variables is referred to as a Wick-rotation, after which time formally flows in the direction of the imaginary axis\footnote{In order to interpret results obtained after the Wick-rotation a continuation back to real time is of course due. However, for time-independent processes this is not necessary.}. Now in the above expression the exponent of the integrand changes to minus the Euclidean action $S_E$, which is the (imaginary) time integral of the Euclidean Lagrangian $\mathcal{L}_E$:
\be
G(q_2,t_2; q_1,t_1) = \!\!\!\int\limits_{q_1\to q_2}\!\!\! Dq \; e^{-S_E[q]}
\label{eq:pathintegral}
\ee
The above formula is mathematically well defined if one divides the time interval into $N$ pieces, calculates the finite sum for a given $N$ and then takes the $N\to\infty$ limit. Expression~\eqref{eq:pathintegral} should be interpreted in this sense.

The same procedure can be carried out in quantum field theories also. Here, instead of a finite number of degrees of freedom one deals with a field at each spacetime-point. While in quantum mechanics all information about the system is contained in the Green's function~(\ref{eq:greensfunc}), in a scalar QFT this role is played by an infinite set of ground state expectation values $\langle \varphi(x_1) \ldots \varphi(x_n) \rangle$. In the same manner as before one obtains
\be
\begin{split}
\langle \varphi(x_1) \ldots \varphi(x_n) \rangle &= \frac{1}{\Z} \int \D\varphi \; \varphi(x_1) \ldots \varphi(x_n) e^{-S_E[\varphi]} \\
\Z &= \int \D\varphi \; e^{-S_E[\varphi]}
\end{split}
\label{eq:scalarpartfunc}
\ee
Here, analogously to the quantum mechanical case, the $\D\varphi$ symbol indicates that the integrand has to be evaluated for each possible field configuration. Again, this expression is only well-defined for a finite number of degrees of freedom, so now not just time, but space also has to be discretized. This means that the field $\varphi$ has to be restricted to the sites of a four-dimensional lattice. The result for the above functional-integral is given by taking the limit where the discretization is taken infinitely fine.

More interesting from the quark-gluon point of view is the case of gauge fields and fermionic fields, since these appear in the Lagrangian of QCD shown in~(\ref{eq:Lag}). Let us consider the case of only one flavor of quarks with mass $m$ and coupling $g$. The latter usually enters the action in the combination $\beta \equiv 6/g^2$:
\be
\Z = \int \D A_\mu\, \D \bar\psi\, \D \psi\, e^{-S_E[A_\mu,\bar{\psi},\psi, m, \beta]}
\label{eq:partfunc}
\ee
Since quarks are fermions, according to the spin-statistics theorem the fields $\psi$ and $\bar\psi$ can be represented by anticommuting Grassmann-variables. With the functional integral given the next step is to put the theory on the lattice. However, the way the fields $A_\mu$, $\psi$ and $\bar\psi$ are discretized is not at all unique. Before I turn to the discretization definitions, it is instructive to interpret expression~\eqref{eq:partfunc} a bit more thoroughly.

\subsection{Statistical physical interpretation}

If the action $S_E$ is bounded from below, the expression~(\ref{eq:partfunc}) -- with its lattice discretized definition -- has the same form as the partition function of a classical statistical physical ensemble in four dimensions. This formal correspondence is valid at zero temperature. The quantum partition function $\Z=e^{-H/T}$ at a finite temperature $T$ is on the other hand represented by a functional integral in which the integral of $\mathcal{L}_E$ in the imaginary time direction is restricted to a finite interval of length $1/T$\footnote{The Boltzmann-constant is set here to unity: $k_B=1$.}. For bosonic fields periodic, for fermionic fields antiperiodic boundary conditions have to be prescribed in this direction. 

This interpretation justifies the lattice approach as a non-perturbative method as statistical physical methods can be used to calculate Green's functions like the one in~(\ref{eq:scalarpartfunc}). According to this analogy $\Z$ is called partition function and the Green's functions derived from $\Z$ are called correlation functions. The expectation value of an arbitrary operator $\phi$ can then be written as
\be
\label{eq:feynman2}
\langle \phi \rangle = \frac{1}{\Z} \int \D A_\mu\, \D \bar\psi\, \D \psi\, \phi\, e^{-S_E[A_\mu,\bar{\psi},\psi, m, \beta]}
\ee

An important remark to make here is that while in the quantum theory the temperature is determined according to the Boltzmann-factors $e^{-H/T}$, here $T$ is proportional to the inverse of the size of the system in the Euclidean time-direction. In particular, on a lattice with lattice spacing $a$ the temperature and the volume of the system are accordingly given as
\be
\label{eq:tempoflatt}
T=\frac{1}{N_t a}, \quad\quad V=(N_s a)^3
\ee
where $N_s$ ($N_t$) is the number of lattice sites in the spatial (temporal) direction.
In a usual computation one uses an $N_s^3 \times N_t$ lattice, so the sizes in the spatial directions are the same. It is important to note that lattices where $N_s\gg N_t$ correspond according to~(\ref{eq:tempoflatt}) to a system with finite nonzero temperature; on the other hand lattices with $N_t \ge N_s$ realize systems with roughly zero temperature. Also, the total volume of the system is given by $V_{4D}=V/T$.

\subsection{Gauge fields and fermionic fields on the lattice}

As part of the regularization process we have to discretize the action $S_E$, which is given by the four dimensional integral of the Euclidean Lagrangian density\footnote{In the following the subscript E is dropped.} of QCD. It can be proven that in order to preserve local gauge invariance -- which is of course indispensable to formulate a gauge theory -- gauge fields must be introduced on the links connecting the sites rather then on the sites themselves (as in the case of the scalar field). Only this way are we able to represent local gauge transformations in a manner that fits the definition of the continuum transformation. Assigning the $A_\mu$ gauge fields of the QCD Lagrangian to the sites of the lattice would make this compliance impossible. On the links the gauge fields can be represented with $U_\mu=e^{iagA_\mu}\;\in SU(3)$ matrices\footnote{This combination ensures that as $a\to0$ the correct continuum theory is approached.}. This also implies that the Hermitian conjugate (i.e. the inverse) of the matrix representing a given link equals the matrix corresponding to the link pointing to the opposite direction:
\be
U_\mu^{-1}(n) \equiv U_\mu^\dagger(n)=U_{-\mu}(n+\hat \mu)
\ee
Here $\hat{\mu}$ denotes the unit vector in the $\mu$ direction and $n$ is the lattice site. Due to the transformation properties of $U_\mu$, the simplest gauge invariant combination of gauge fields on the lattice can be constructed by taking the product of the links that build up a square (in the $\mu,\nu$ plane)
\be
U^{1\times1}_{\mu \nu}(n) = U_\mu (n) U_\nu (n+\hat \mu) U_\mu ^\dagger (n+\hat \nu) U_\nu ^\dagger (n)
\label{eq:plaquett}
\ee
and then calculating the trace of this expression. This trace is called the {\it plaquette}, based on which the action corresponding to pure gauge theory can be constructed. The resulting sum is the simplest real and gauge invariant expression that can be built using only gauge fields:
\be
S_G^{\rm Wilson}=\beta \sum\limits_{n,\mu<\nu} \left [ 1 - \frac{1}{3} \textmd{Re} \, \Tr \, U^{1\times1}_{\mu \nu}(n)  \right ]
\label{eq:wilsonaction}
\ee
The sum extends to every possible ($\mu - \nu$) square on the four-dimensional lattice. The combination~(\ref{eq:wilsonaction}) is called the Wilson gauge action. Choosing $\beta=6/g^2$ it is straightforward to show that in the $a\to 0$ limit the Wilson action approaches the continuum gauge action, namely the first term in~(\ref{eq:Lag}).\footnote{Note that the Wilson action can be extended by an arbitrary term that vanishes as $a\to 0$ and the resulting action still converges to the same continuum action. I get back to this in section~\ref{sec:impr}.}

To obtain the total action of QCD one also has to take into account the fermionic contribution. According to the transformation rules of the fermionic fields $\psi$ and $\bar\psi$ (which live on the sites of the lattice) other types of invariant combinations can also be composed. Since the QCD Lagrangian contains fermions quadratically, the general form of the action is written as
\be
S(U,\psi,\bar \psi) = S_G(U)-\bar \psi M(U)\psi
\ee
where $M(U)$ is the fermion matrix. The elements of this $12N\times 12N$ matrix (with $N$ being the number of lattice sites and $12=3\cdot4$ the number of colors times the number of Dirac-components) can be read off from the Lagrangian. The fermion matrix can be divided into a massless Dirac operator and a mass term: $M = \slashed D + m\mathds{1}$. In the partition function the integration over the fermions can be analytically performed and gives the following result\footnote{Recall that $\psi$ and $\bar\psi$ are Grassmann-variables.}:
\be
\label{eq:partfunc2}
\Z=\int \D U \D\bar\psi \D\psi \; e^{-S_G(U)-\bar\psi M(U) \psi} = \int \D U \; e^{-S_G(U)} \det M(U)
\ee
Here the integration measure for the fermions takes the simple product form
\be
\D\psi = \prod\limits_n d\psi(n)
\ee
the one over the gauge fields on the other hand depends on the 8 real parameters of the $SU(3)$ group, and the integral has to be performed over the whole group. If the parameters on the $\ell$th link are denoted by $\alpha_{\ell}^a$, the measure can be written as
\be
\D U = \prod\limits_{\ell}J(\alpha_{\ell})\prod\limits_{a=1}^8 d\alpha_{\ell}^a
\ee
where the structure of the $J(\alpha_{\ell})$ Jacobi-matrix can be determined by requiring gauge invariance. This integration measure is called the Haar-measure.

\subsection{Fermionic actions}
\label{sec:fermionactions}

The naive discretization of the fermionic part of the Lagrangian~(\ref{eq:Lag}) fails to give the correct continuum limit even in the free case. Namely, the naive fermionic action
\be
S_F^{\textmd{naive}}=\sum_n \left [am\bar{\psi}(n)\psi(n)+
\frac{1}{2}\sum_{\mu=1}^4
\left(\bar{\psi}(n)U_{\mu}(n)\gamma_\mu\psi(n+a\hat{\mu})-
\bar{\psi}(n+a\hat\mu)U^\dagger_{\mu}(n)\gamma_\mu\psi(n) \right) \right]
\label{eq:naiveferm}
\ee
gives a propagator for the free theory (where $U=1$) which possesses not 1 but 16 poles in the lattice Brillouin-zone $-\pi/a<p\le\pi/a$. Thus, the action~(\ref{eq:naiveferm}) describes 16 quarks and does not converge to the continuum action as $a\to 0$. This is a consequence of the Nielsen-Ninomiya theorem which states that this `doubling' problem cannot be solved without breaking the chiral symmetry of the QCD action in the $m\to 0$ limit. For the massless Dirac operator $\slashed D$, chiral symmetry means that
\be
\{ \slashed D,\gamma_5 \} \equiv \slashed D \gamma_5 + \gamma_5 \slashed D = 0
\label{eq:chiralsymm}
\ee
is satisfied. In the continuum theory, although this symmetry would imply the conservation of an axial-vector current, the corresponding current has an anomalous divergence due to quantum fluctuations. On a lattice with finite lattice spacing however, this current is indeed conserved, and the corresponding extra excitations are just the above mentioned `doublers'.

The two most popular methods to circumvent the doubling problem are the Wilson-type and the Kogut-Susskind (or staggered) type discretizations. In the former solution the mass of the 15 doublers is increased as compared to the original fermion. This is achieved by adding to the naive action $S_F^{\textmd{naive}}$ a term that contains a second derivative: $-r/2 \sum_n \bar\psi(n)\partial_\mu\partial_\mu \psi(n)$ with $r$ an arbitrary constant. This extra term is proportional to $a$, therefore it vanishes in the continuum limit. On the other hand it raises the masses of the unwanted doublers proportional to $1/a$. The action with Wilson-fermions then has the form
\be
\begin{split}
S_F^{\textmd{Wilson}} &=\sum_n (ma+4r)\bar\psi(n)\psi(n)\\
& - \frac{1}{2} \sum_{n,\mu} \left( \bar\psi(n) (r-\gamma_\mu) U_\mu(n) \psi(n+a\hat\mu) + \bar\psi(n+a\hat\mu) (r + \gamma_\mu) U^\dagger_\mu(n) \psi(n)\right) 
\label{eq:Wilsonfermaction}
\end{split}
\ee
This action breaks chiral symmetry for $r\ne 0$ even for zero quark masses on a lattice with finite lattice spacing. This implies that the quark mass will have an additive renormalization which makes it very difficult to study chiral symmetry breaking as for that a fine tuning of the parameter $m$ is required.

Another popular method to get rid of the doublers is to modify the naive action such that the Brilluoin zone reduces in effect. This is achieved by distributing the fermionic degrees of freedom $\psi_\alpha$ over the lattice such that the effective lattice spacing for each component is twice the original lattice spacing. We lay out the spinor components of $\psi(n)$ on the sites of the hypercube touching the site $n$. This way, in four dimensions the degrees of freedom reduces by $75\%$, so this formulation describes only 4 flavors of quarks. This discretization is referred to as the Kogut-Susskind or staggered fermionic action.

Applying an appropriate local transformation on the fields $\psi$ and $\bar\psi$ the naive action~(\ref{eq:naiveferm}) can be diagonalized in the spin-indices $\alpha$. This way the Dirac-matrices $(\gamma_\mu)_{\alpha \beta}$ are eliminated and with the new fields $\chi$ and $\bar\chi$ the staggered fermionic action is
\be
S_F^{\textmd{stag}}=\frac{1}{2} \sum_{n,\mu}\eta_{\mu}(n) \left[ \bar\chi(n) U_\mu(n) \chi(n+a\hat\mu) - \bar\chi(n+a\hat\mu) U^\dagger_\mu(n) \chi(n)\right] + ma \sum_n \bar \chi(n)\chi(n)
\label{eq:fermact_stag}
\ee
where the only remnants of the original Dirac structure are the phases $\eta_\mu(n) = (-1)^{n_1+n_2+\dots+n_{\mu-1}}$. A huge advantage of staggered fermions is that for zero quark mass an $U(1)_R \times U(1)_L$ symmetry (which is a remnant of the full chiral symmetry group) is preserved. Due to this there is no additive renormalization in the quark mass and thus no fine tuning -- as opposed to Wilson fermions -- is necessary. Consequently, using the staggered action it is possible to study the spontaneous breaking of this remnant symmetry and the corresponding Goldstone-boson. We remark furthermore, that the staggered action introduces discretization errors of $\mathcal{O}(a^2)$.

As mentioned above, the staggered discretization describes 4 flavors. Since (in the Wilson formulation) including a second quark flavor could be realized by inserting another determinant in~(\ref{eq:partfunc2}), one expects that taking the root of $\det M$ can be used to decrease the number of flavors. Thus for staggered fermions, the following partition function
\be
\label{eq:partfunc3}
\Z=\int \D U \; e^{-S_G(U)} \det M(U)^{N_q/4}
\ee
is expected to describe $N_q$ flavors. This ``rooting'' trick~\cite{Durr:2005ax} is theoretically not well established, since the above $\Z$ cannot be proven to correspond to a local theory (unlike that in~(\ref{eq:partfunc2})). Nevertheless numerical results seem to support the validity of the rooting procedure.

A further possibility to discretize fermions and simultaneously get around the Nielsen-Ninomiya theorem resides in defining a lattice version of chiral symmetry: $ \{ \slashed D,\gamma_5 \} = \O(a)$. This way the doublers are avoided and chiral symmetry in the continuum limit is recovered. These regularizations are called chiral fermions. Among solutions satisfying lattice chiral symmetry are the overlap and the fixed-point fermionic actions; the domain wall fermions on the other hand provide an approximation to such a Dirac operator.

\subsection{Positivity of the fermion determinant}
\label{sec:hermitic}

In this section an important property of the fermion matrix $M=\slashed D + m\mathds{1}$ is pointed out. It is straightforward to check that any of the above presented fermionic lattice discretizations satisfies the condition of $\gamma_5$-hermiticity\footnote{It is evident that equation~(\ref{eq:hermiticity}) also holds with $\slashed D$ replaced by $M$.}:
\be
\slashed D ^\dagger = \gamma_5 \slashed D \gamma_5
\label{eq:hermiticity}
\ee
For example for naive fermions~(\ref{eq:naiveferm}) the Dirac-operator takes the form (with the color and Dirac indices suppressed):
\be
\slashed D_{nm} = \frac{1}{2}\sum\limits_{\mu=1}^4 \left( U_\mu(n)\gamma_\mu\delta_{m,n+a\hat\mu} - U_\mu^\dagger(n-a\hat\mu)\gamma_\mu \delta_{m,n-a\hat\mu} \right)
\label{eq:naive_D}
\ee
Taking into account that for any $\mu$ the gamma-matrices satisfy $\gamma_5\gamma_\mu=-\gamma_\mu \gamma_5$ and $\gamma_5^2=1$, we have
\be
\begin{split}
\left(\gamma_5 \slashed D \gamma_5\right)_{nm} &= \frac{1}{2}\sum\limits_{\mu=1}^4 \left( -U_\mu(n)\gamma_\mu\delta_{m,n+a\hat\mu} + U_\mu^\dagger(n-a\hat\mu)\gamma_\mu \delta_{m,n-a\hat\mu} \right) \\
 &= \frac{1}{2}\sum\limits_{\mu=1}^4 \left( -U_\mu(m-a\hat\mu)\gamma_\mu\delta_{n,m-a\hat\mu} + U_\mu^\dagger(m)\gamma_\mu \delta_{n,m+a\hat\mu} \right)
\end{split}
\label{eq:naive_D_adj}
\ee
which is indeed the $nm$ matrix element of the adjoint of~(\ref{eq:naive_D}), $\slashed D_{nm}^\dagger$. Note that~(\ref{eq:naive_D_adj}) is the adjoint in color, Dirac, and coordinate space, since the gamma-matrices are self adjoint and $n$ and $m$ is interchanged.

Let us now consider the eigenvalue equation of $\slashed D$:
\be
\slashed D \chi_n = \lambda_n \chi_n
\ee
and define the characteristic polynomial of $\slashed D$ as $P(\lambda) = \det(\slashed D-\lambda \mathds{1})$. Then one obtains
\be
\begin{split}
P(\lambda) &= \det\left(\gamma_5^2(\slashed D-\lambda\mathds{1})\right) = \det\left(\gamma_5(\slashed D-\lambda\mathds{1})\gamma_5\right)\\
 &= \det\left(\slashed D^\dagger-\lambda\mathds{1}\right) = \det\left(\slashed D-\lambda^*\mathds{1}\right)^* = P^*(\lambda^*)
\end{split}
\ee
That is to say, if $\lambda$ is an eigenvalue ($P(\lambda)$=0), then $\lambda^*$ is also an eigenvalue, since $P(\lambda^*)=0$ also holds. This implies that the eigenvalues of $\slashed D$ are either real, or consist of complex conjugated pairs, i.e. the determinant of $\slashed D$ is real.

As will be discussed in section~\ref{sec:MonteCarlo}, the determinant $\det M$ is to be used as a probability weight and thus needs to be nonnegative. Combining~(\ref{eq:hermiticity}) with chiral symmetry~(\ref{eq:chiralsymm}) one observes that $\slashed D ^\dagger = -\slashed D$, i.e. $\slashed D$ is antihermitian: its eigenvalues are purely imaginary, $\lambda_n=i\eta_n$. Thus the eigenvalues of the fermion matrix are of the form $m\pm i\eta_n$ and the determinant of $M$ is indeed a nonnegative real number. 

It is straightforward to prove that~(\ref{eq:hermiticity}) holds also for~(\ref{eq:Wilsonfermaction}) and~(\ref{eq:fermact_stag}) and as a consequence the staggered fermion determinant is always nonnegative. Note however, that Wilson-fermions do not exhibit chiral symmetry, which means that there can be eigenvalues of $M$ that are real and negative which can spoil the positiveness of the Wilson-fermion determinant.

Also note that as a result of the inclusion of a $\theta$-term or a chemical potential, the Dirac-operator will no more be $\gamma_5$-hermitian. This in turn has serious consequences on the positivity of $\det M$, see section~\ref{sec:signproblem}.

\section{Continuum limit and improved actions}
\label{sec:impr}

After the field variables of~(\ref{eq:Lag}) have been discretized the continuum action is obtained by carrying out the $a\to 0$ limit. While the discretization procedure is not unique, different lattice actions have to give the same continuum limit. Accordingly, the expectation value of an arbitrary operator $\phi$ on the lattice can be written as
\be
\expv{\phi}_{lat} = \expv{\phi} + \O(a^p)
\label{eq:scaling}
\ee
with $\expv{\phi}$ being the expectation value of the operator in the continuum theory and the second term is the deviation or `lattice artefact' caused by the discretization. How fast the discretized action converges to the continuum one is determined by the exponent $p>0$ (and the coefficient of this term). For the Wilson gauge action this scaling is proportional to $a^2$ (i.e. $p=2$). For an improved action with larger $p$ the scaling is faster. Therefore with an improved action one may be able to approach the continuum limit faster, on the other hand, a complicated action can significantly slow down the simulation. The optimal choice may depend on the observable in question.

It is easy to see that the continuum limit of the lattice theory is equivalent to a second-order critical point of the underlying statistical physical system. Indeed, let us consider a particle with a finite mass $m$. This mass is a physical (constant) number, irrespective of how one measures it on the lattice. On the other hand, the mass measured in lattice units $\hat{m}=ma$ clearly has to vanish as $a\to0$, and therefore the corresponding correlation length has to diverge: $\xi\to\infty$. This is just the characteristic property of a critical point in statistical physics.

Near the critical point the statistical system exhibits the property of universality. This means that in this region the long-range behavior of the system depends only on the number of degrees of freedom, the space-time dimension and the symmetries of the theory. Consequently, the actual form of the action is less and less important; only the {\it relevant} operators matter. Nevertheless, irrelevant operators (which converge to zero as $a\to 0$) may modify the scaling~(\ref{eq:scaling}).

It was already mentioned that in general, as part of the renormalization program the bare parameters of the theory will depend on the regularization. On the lattice this means that these parameters will become a function of the lattice spacing, and as one approaches the continuum limit, they have to be tuned as a function of $a$.

\subsection{The line of constant physics}
\label{sec:lcp}

At a fixed temporal size $N_t$ one can change the lattice spacing by varying the bare parameters of the action:
the inverse gauge coupling $\beta$ and the quark masses $m_q$. The fact that towards the continuum limit the
lattice should reproduce the continuum physics, dictates the functional
relation between these parameters. This relation ensures that for each lattice spacing $a$ ``physics is the same''. 
A possible way to define this line of constant physics (LCP) is to fix ratios of experimentally measurable quantities 
to their physical value.

\begin{wrapfigure}{r}{8.5cm}
\centering
\vspace*{-1.8cm}
\includegraphics*[width=8.0cm]{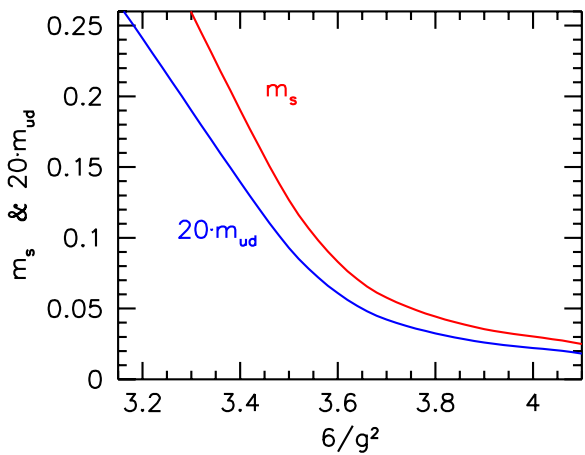}
\vspace*{-0.3cm}
\caption[The line of constant physics.]{The line of constant physics.}
\label{fig:lcp}
\vspace*{-0.7cm}
\end{wrapfigure}

In QCD with $2+1$ flavors we have three independent parameters: $\beta$, $m_{ud}$ and $m_s$. For the study of the phase diagram 
(chapter~\ref{chap:phasediag}) and the QCD equation of state (chapter~\ref{chap:EoS}) we fix the functions $m_s(\beta)$ and $m_{ud}(\beta)$ 
such that the ratios $f_K/m_\pi$ and $f_K/m_K$ are at their experimental value\footnote{Here $f_K$, $m_\pi$ and $m_K$ are the kaon decay constant, the pion mass and the kaon mass, respectively, which we take from~\cite{Amsler:2008zzb}.}. Through this procedure we get for 
the ratio of quark masses $m_s/m_{ud} = 28.15$. Note that different definitions may result in different functions $m_s(\beta)$, $m_{ud}(\beta)$, but these differences converge to zero as the continuum limit is approached.
The detailed determination of the line of constant physics can be found in~\cite{Aoki:2009sc,Aoki:2006br}. This definition of the LCP was used in the study of the phase diagram; the corresponding $m_s(\beta)$ and $m_{ud}(\beta)$ functions are shown in figure~\ref{fig:lcp}. For the study of the QCD EoS this relation was further improved~\cite{Borsanyi:2010cj}. 

\subsection{Scale setting on the lattice}
\label{sec:scalesetting}

On the lattice one can only measure dimensionless quantities. A measurement of e.g. the ratio of two particle masses can be compared to the physical value of this particular ratio, as it was used to define the LCP in the previous subsection. To determine the lattice spacing itself, one has to measure an experimentally accessible observable $A$ in lattice units, i.e. in units of a certain power $d$ (the mass dimension of the observable in question) of the lattice spacing $a$:
\be
A_{\rm latt} = A_{\rm exp} \cdot a^d
\ee
Then the lattice spacing can be calculated using the experimental value of the quantity $A_{\rm exp}$.

Arbitrary dimensionful observable can be used to define the lattice scale in this manner. A possible choice is one using the static quark-antiquark potential $V(r)$, which can be measured on the lattice using spatial-temporal loops constructed from the gauge field (the Wilson loops).  
In the confined phase the potential is linearly increasing with $r$. The coefficient of this term is given by the string tension $\sigma$:
\be
V(r) \xrightarrow{r\to \infty} \sigma r
\ee
The potential also contains a Coulomb-like repulsion which dominates at small distances. The shape of the potential as a function of $r$ can be used to implicitly define an intermediate distance $r_0$:
\be
\left.\left( r^2 \frac{\d V(r)}{\d r}\right)\right|_{r=r_0} = 1.65
\ee
The string tension and the parameter $r_0$ are only well defined in pure gauge theory. In the presence of dynamical quarks, at increasing distances mesons can be created from the vacuum and the string between the two color charges can break, making $\sigma$ and $r_0$ ill-defined. 

Therefore in dynamical simulations it is more practical to determine the lattice spacing in terms of a mass or a decay constant. For the study of the phase diagram (chapter~\ref{chap:phasediag}) and the QCD equation of state (chapter~\ref{chap:EoS}) we fixed the scale by measuring the kaon decay constant $f_K$. 

A further possibility is to use the critical temperature $T_c$ to fix the scale. To this end one has to determine the critical couplings $\beta_c$ on lattices with various temporal extent $N_t$. This scheme is particularly advantageous in pure gauge theory, where the phase transition is of first order~\cite{Celik:1983wz,Kogut:1982rt,Gottlieb:1985ug,Brown:1988qe,Fukugita:1989yb,Kogut:1987rz}, and therefore $T_c$ is sharply defined (as opposed to the case of full QCD, where a broad crossover separates the phases~\cite{Aoki:2006we}). Thus in the study of the pure gauge equation of state (chapter~\ref{chap:PG}) this approach was followed.

\subsection{Symanzik improvement in the gauge sector}
\label{sec:symanzik}

The scaling~(\ref{eq:scaling}) can be improved by inserting further gauge-invariant terms in the lattice action. It can be proven that the plaquette is the only relevant operator that can be built from purely gauge links. The second simplest combination is the $2\times1$ rectangle $U_{\mu\nu}^{2\times1}$, i.e. the ordered product of links along such a rectangle. The resulting improved action can be written as
\be
S_G^{\rm Symanzik}=-\beta \left[ c_0 \sum\limits_{n,\mu<\nu} \textmd{Re} \, \Tr \, U^{1\times1}_{\mu \nu}(n)  + c_1 \sum\limits_{n,\mu\ne \nu} \textmd{Re}\,\Tr \, U^{2\times1}_{\mu \nu}(n) \right]
\label{eq:symanzikact}
\ee
The lattice artefacts of the Wilson gauge action contain $\O(a^2)$ and $\O(g^2a^2)$ terms. 
If the coefficients in~(\ref{eq:symanzikact}) are set to $c_0=5/3$ and $c_1=-1/12$, the $\O(a^2)$ term is eliminated, thus the above combination will approach the continuum theory as $\O(a^4)$ on the tree level. This action is therefore called the tree-level improved Symanzik gauge action.

\subsection{Taste splitting and stout smearing}
\label{sec:stoutsmear}

As already mentioned, the staggered fermion discretization~(\ref{eq:fermact_stag}) describes (before applying the rooting trick) 4 flavors of quarks. The masses of these four fermion species (which are in this case called {\it tastes}) are however not the same. The $SU(4)$ flavor symmetry is violated by the staggered formulation, and as a result each
continuum hadron state has a corresponding multiplet of states on the lattice: due to the taste symmetry violation the masses of these states are split up~\cite{Ishizuka:1993mt}. This introduces a discretization error which is important mainly at
low energies~\cite{Huovinen:2009yb,Huovinen:2010tv,Borsanyi:2010bp}.

As an example, 16 lattice states correspond to each continuum pion state, each of
them contributing with a 1/16 weight. The following table lists the members of
the lattice pion multiplet with the taste structure (a $4\times4$ complex
matrix, $\Gamma_\alpha$) and the multiplicity ($n_\alpha$):
\begin{equation}
\begin{array}{|c||c|c|c|c|c|c|c|c|}
\hline
\alpha        & 0 & 1 & 2 & 3 & 4 & 5 & 6 & 7 \\
\hline
\Gamma_\alpha & \gamma_5 & \gamma_0\gamma_5 & \gamma_i\gamma_5 & \gamma_i\gamma_j & \gamma_i\gamma_0 & \gamma_i & \gamma_0 & 1 \\
\hline
16\cdot n_\alpha      & 1 & 1 & 3 & 3 & 3 & 3 & 1 & 1 \\
\hline
\end{array}
\nonumber
\end{equation}

Only $\alpha=0$ behaves like a Goldstone-boson, i.e. its mass vanishes in the chiral
limit. The other 15 states have masses of the order of several hundred MeVs for
sensible values of the lattice spacing. Though these mass differences vanish
in the continuum limit, it is very important to suppress them as much as
possible. The effect of the heavier ``pions'' on thermodynamic observables can be
significant: they can reduce the QCD pressure and can also shift the
transition temperature.

Strategies for the suppression have been studied extensively. 
An effective way to reduce splitting is to eliminate the ultraviolet noise from the gauge links (which appears as a result of the introduction of the finite lattice spacing), and ``smear'' the links.
During the smearing process each link is replaced by an appropriately defined average of the surrounding links. One possible way is to add to the gauge link the ``staples'' around it:
\be
U_\mu(n) \to Pr \left[ U_\mu(n) + \rho \sum\limits_{\nu \ne \mu} U_\nu(n) U_\mu(n+a\hat\nu) U_\nu^\dagger(n+a\hat\mu) \right]
\label{eq:stoutsmear}
\ee
with $\rho$ a constant parameter. As a sum of $SU(3)$ matrices the result in general will not be an element of the gauge group and thus a projection back to $SU(3)$ (denoted above by $Pr$) is necessary. 
This specific smearing method is called {\it stout smearing}~\cite{Morningstar:2003gk}. 

\begin{wrapfigure}{r}{8.5cm}
\centering
\vspace*{-0.3cm}
\includegraphics*[width=7.5cm]{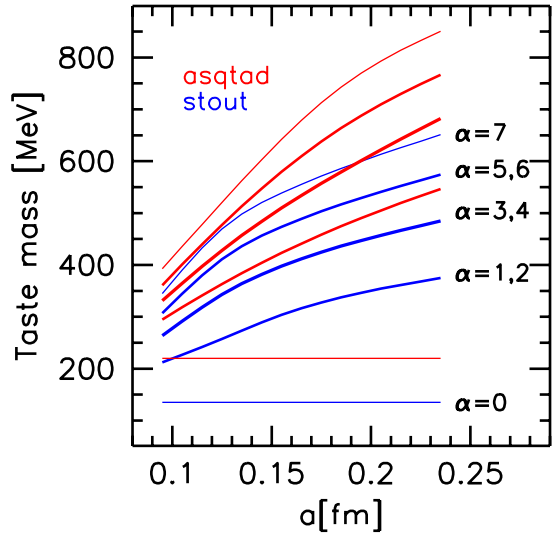}
\caption[Masses of lattice pion tastes as functions of the lattice spacing.]{Masses of lattice pion tastes as functions of the lattice spacing for the stout (blue lines) and the asqtad action (red lines). The pion states are labeled by $0\le \alpha \le 7$.}
\label{fig:splitting}
\vspace*{-1.7cm}
\end{wrapfigure}

The whole process can be repeated several times in order to increase the smoothness of the links. In the simulations that I present in this thesis $\rho=0.15$ was set and the smearing was carried out twice in a row. 
Stout smearing is proven to significantly reduce the lattice artefacts originating from taste splitting. The mass splitting in the pion
multiplet for the stout smeared action is
shown in figure~\ref{fig:splitting} as a function of the lattice spacing (blue lines). For physical quark masses the pion state with the lowest mass is
adjusted to the mass of the continuum pion. For comparison the splitting is also plotted for the asqtad improved action~\cite{Huovinen:2009yb} (red lines in the figure).

\subsection{Improved staggered actions}
\label{sec:impract}

The $\mathcal{O}(a^2)$ scaling of the staggered fermionic action can also be improved by considering a more complicated discretization for the derivative term in~(\ref{eq:fermact_stag}). Beside the 1-link term\footnote{The staggered fermionic field is denoted here and also in the following by $\psi$.}
\be
W_\mu^{(1,0)}(n) = \bar\psi(n) U_\mu(n) \psi(n+a\hat\mu)
\ee
one can also include higher order terms, like all the possible 3-link contributions. These are schematically written as the linear (3,0) and the bent (1,2) terms 
\be
\begin{split}
W_\mu^{(3,0)}(n) &= \bar\psi(n) U_\mu(n) U_\mu(n+a\hat\mu) U_\mu(n+2a\hat\mu) \psi(n+3a\hat\mu) \\
W_{\mu,\nu}^{(1,2)}(n) &= \bar\psi(n) U_\mu(n) U_\nu(n+a\hat\mu) U_\nu(n+a\hat\mu+a\hat\nu) \psi(n+a\hat\mu+2a\hat\nu) \\
                       &+ \bar\psi(n) U_\nu(n) U_\nu(n+a\hat\nu) U_\mu(n+2a\hat\nu) \psi(n+a\hat\mu+2a\hat\nu)
\end{split}
\ee
Furthermore, the 1-link terms can also be smeared with a method similar to the one presented in subsection~\ref{sec:stoutsmear}. By an appropriate setting of the coefficients of these improvement terms (such that the rotational symmetry of the quark propagator is improved~\cite{Heller:1999xz}) one can achieve a better scaling at the tree level (i.e. at zero gauge coupling). This implies that these actions (like the p4 or the asqtad action) approach the continuum action faster at very high temperatures. On the other hand this improvement does not suppress the taste splitting and therefore large lattice artefacts may be expected in the low temperature region (where the lattice spacing is large). 

Moreover, the splitting in the tastes can also produce $\O(a^2)$ errors through taste-exchange processes. By a further improvement these processes can also be suppressed; the resulting action is called the hisq discretization~\cite{Follana:2006rc}.
The hisq action together with the stout smeared action are proven to have significantly smaller splitting between the various tastes as compared to the asqtad or the p4 action. At low temperatures these actions are therefore expected to produce more reliable results.

\section{Monte-Carlo algorithms}
\label{sec:MonteCarlo}

In order to determine the expectation value of some observable (which is necessary to measure e.g. the thermodynamic quantities of chapter~\ref{chap:thermo}) one needs to calculate the functional integral of~(\ref{eq:partfunc2}). This integral, as discretized on a four dimensional lattice can have a dimension as high as $10^9$, which excludes usual numerical integration techniques. Such integrals can only be calculated by Monte-Carlo (MC) methods based on importance sampling. Furthermore, because of the Grassmann nature of the fermion fields standard MC methods are not applicable, and instead one integrates out the quark degrees of freedom to obtain the determinant of $M$ in the partition function, as in~(\ref{eq:partfunc2}).

Therefore, the expectation value of an arbitrary observable $\phi$ can be written as
\be
\expv{\phi} = \frac{1}{\Z} \int \D U \, \phi \, \det M \, e^{-S_G(U)}
\label{eq:foo1}
\ee
Importance sampling means that instead of selecting configurations in $U$ space randomly, we generate them according to the distribution
\be
\rho(U) = \frac{1}{\Z} \det M e^{-S_G(U)}
\ee
such that we have a set of configurations $\{U^{(i)}\}$ with $i=1\ldots N$. Then the expectation value of the observable is readily obtained as (assuming that the configurations are independent)
\be
\expv{\phi} = \lim\limits_{N\to \infty}\frac{1}{N} \sum\limits_{i=1}^N \phi(\{U^{(i)}\})
\ee
In practice the $N\to \infty$ limit cannot be carried out since one only has a finite sequence of configurations. The deviation from the exact expectation value in this case is given by terms of $\O(N^{-1/2})$ and can be estimated using the jackknife method~\cite{Montvay:1994}.

Note here that in order to interpret $\rho(U)$ as a probability measure and apply importance sampling, the fermion determinant has to be nonnegative. This constraint is fulfilled for the staggered lattice action~(\ref{eq:fermact_stag}), if a chemical potential is not present, see section~\ref{sec:hermitic}.

\subsection{Metropolis-method}

In order to generate configurations according to the desired distribution the only possible way is to construct a Markov chain, i.e. to generate the new configuration $\{U'\}$ from a previous one $\{U\}$ with a probability $P(U' \gets U)$. Markov chains in general converge to the distribution $\rho(U)$ if the above probability fulfills ergodicity (i.e. by successive steps the whole $U$ space can be covered) and detailed balance, which means
\be
\rho(U) P(U' \gets U) = \rho(U') P(U \gets U')
\ee
A simple Markov process is produced by the so-called Metropolis algorithm. Here, first one generates a new configuration $\{U'\}$ by a random change, and then accepts this according to the probability
\be
P_{\rm Met}(U' \gets U) = \min \left[ 1, e^{-\left( S_G(U')-S_G(U)\right )}\frac{\det M(U')}{\det M(U)} \right]
\ee
If the new configuration is not accepted, the original configuration remains for the next step\footnote{Here it can be explicitly seen that the positiveness of the determinant is necessary to obtain a probability for which $P\in[0,1]$.}. This procedure is however very inefficient, since it involves the calculation of the fermion determinant (i.e. $(N_s^3N_t)^3$ floating-point operations) in each step. Furthermore, the consecutive configurations are certainly not independent. 

In this context it is useful to introduce the notion of autocorrelation time, which is the number of steps after which the new configuration can be considered independent of the original one (i.e. when the correlation of the two falls below some small number). A further important quantity is the thermalization time, which can be identified with the number of steps necessary for the ensemble of the generated configurations to reach the equilibrium distribution $\rho$.

\subsection{The Hybrid Monte-Carlo method}

The Metropolis-algorithm can be improved in many aspects. A much more effective way to generate configurations is by means of the so-called Hybrid Monte Carlo (HMC) method~\cite{Duane:1987de,Gottlieb:1987mq}, which is a mixture of the Metropolis and the molecular dynamics method. First, we make the observation that the determinant of a hermitian matrix $H$ can be written as the (bosonic) integral of an exponential:
\be
\begin{split}
\det H &= \frac{1}{C} \int \D \varphi^\dagger \D \varphi e^{-\varphi^\dagger H^{-1} \varphi} \\
C &= \int \D \varphi^\dagger \D \varphi e^{-\varphi^\dagger \varphi}
\end{split}
\ee
The $\varphi$ fields are referred to as pseudofermions.
The fermion matrix $M$ itself is not hermitian, but the combination $M^\dagger M$ obviously is and thus can be used in the above formula in place of $H$ to obtain\footnote{One can further notice that the staggered action~(\ref{eq:fermact_stag}) connects only nearest neighbors and therefore in the matrix $M^\dagger M$ only the odd-odd and the even-even elements are nonzero. Furthermore, the determinant can be factorized as
\be
\det (M^\dagger M) = \det (M^\dagger M)_{\rm even} \cdot \det(M^\dagger M)_{\rm odd}
\nonumber
\ee
and using the actual form of the Dirac matrix it is also easy to see that the even and odd factors are equal. Therefore~(\ref{eq:detM}) indeed holds.
}
\be
\det M \equiv \sqrt{\det (M^\dagger M)} = \frac{1}{C} \int \D \varphi^\dagger \D \varphi e^{-\varphi^\dagger (M^\dagger M)_{\rm even}^{-1} \varphi}
\label{eq:detM}
\ee

Using this representation of the determinant the partition of~(\ref{eq:foo1}) can therefore be written in the following form:
\be
\Z = \frac{1}{C} \int \D U \D \varphi^\dagger \D \varphi \, e^{-S_G(U) - \varphi^\dagger (M^\dagger M)_{\rm even}^{-1} \varphi}
\ee
In the molecular dynamics method one introduces a new simulation time parameter $\tau$ and considers the time development of the system in this new variable, which can be obtained via the Hamiltonian formulation. The canonical variables\footnote{Here the index $n$ runs over all the links.} of this Hamiltonian are the gauge fields $U_n(\tau)$ and the corresponding conjugate momenta $\Pi_n(\tau)$, for a fixed value of the pseudofermion fields $\varphi$. Therefore we introduce the conjugate momenta and integrate over them also to rewrite the partition function as
\be
\begin{split}
\Z &= \frac{1}{CC'} \int \D\Pi \D U \D \varphi^\dagger \D \varphi \, e^{-\sum \Tr \Pi_n^2  -S_G(U) - \varphi^\dagger (M^\dagger M)_{\rm even}^{-1} \varphi} \\
C' &= \int \D\Pi \,e^{-\sum \Tr \Pi_n^2}
\end{split}
\ee
Thus the Hamiltonian of this system can be written as
\be
\mathcal{H} = \frac{1}{2} \sum\limits_{n} \Tr \Pi_n(\tau)^2 + S_G(U_n(\tau)) + \varphi^\dagger (M^\dagger(U_n(\tau)) M(U_n(\tau)))_{\rm even}^{-1} \varphi
\ee
The canonical equations of motion as derived from $\mathcal{H}$ can now be solved as a function of $\tau$. Along the solutions $\Pi_n(\tau)$, $U_n(\tau)$ the ``energy'' $\mathcal{H}$ of the system is constant. Thus, advancing along such a trajectory corresponds to a special Metropolis step for which the acceptance probability is 1. In this formulation the expectation value of an observable is obtained by averaging along the classical trajectory.

Since in practice the canonical equations can only be integrated approximately, in some discrete steps of $\delta\tau$, the conservation of energy will also be approximate. A possible prescription to carry out this numerical integration is the so-called leapfrog algorithm, which introduces errors of $\delta \mathcal{H}\sim \delta\tau^2$.
However, if a Metropolis acceptance test is inserted at the end of each trajectory, the systematic error caused by the finite $\delta\mathcal{H}$ can also be eliminated.

From the numerical point of view, the most demanding part of this algorithm is that in each step of the molecular dynamics trajectory (and also in the final Metropolis step), a matrix inversion has to be carried out to obtain the momenta $\Pi_n$ (which contains terms of the form $(M^\dagger M)^{-1} \varphi$). Equivalently, one has to exactly solve the system of linear equations
\be
\varphi = (M^\dagger M) \chi
\label{eq:foo2}
\ee
This can be solved by e.g. the conjugate gradient method. The time this algorithm needs for solving the above equation is proportional to the condition number of the matrix which, in turn, is related to the inverse quark mass. Due to this, simulations of systems with smaller quark masses are increasingly difficult.

\subsection{HMC with staggered fermions}
\label{sec:rhmc}

In order for the staggered lattice action to describe 1 (or 2) quark flavors, one needs to take the fourth (or square) root of the fermion determinant, as in~(\ref{eq:partfunc3}). Unfortunately, in this case the conjugate gradient method fails. However, one can use rational functions to approximate the root function as
\be
\sum\limits_{j=1}^{J} \frac{a_j+b_j x}{c_j + x} \simeq \frac{1}{\sqrt{x}}
\ee
For each term, the system of equations in~(\ref{eq:foo2}) can now be solved and using $J=10-15$ terms and appropriately tuned coefficients the exact solution for the inversion of the root of $M^\dagger M$ can be recovered. This algorithm is referred to as the Rational HMC (RHMC) method~\cite{Clark:2006fx}. This algorithm was used to obtain all of the results presented in this thesis.

\chapter{QCD thermodynamics on the lattice}
\label{chap:thermo}

After this brief introduction to the lattice approach of QCD, I will analyze the theory from the thermodynamic aspect. In this section I will identify the symmetries of the theory and consider the corresponding observables that one can use to extract the thermodynamic properties of the system. First of all, let us consider the partition function~(\ref{eq:partfunc3}) in the staggered formulation, generalized to the case of a higher number of flavors. The flavors are labeled by $q$, each having a mass of $m_q$, an assigned chemical potential $\mu_q$ and a degeneracy $N_q$ (thus, in the $2+1$-flavor system $N_u=2$ and $N_s=1$). Also, let us denote the total number of quarks as $N_Q=\sum_q N_q$. After integrating out the quark fields $\psi_q$, the partition function reads
\be
\Z = \int \D U e^{-S_G(U)} \prod_q \det M^{N_q/4}(U, m_q, \mu_q).
\label{eq:partfunc4}
\ee
where the dependence of the determinant on the chemical potential and the mass is explicitly written out. For each $q$, the chemical potential $\mu$ is treated in the grand canonical approach i.e. the action is complemented by a term $\mu N$ where $N$ is the number of quarks in the system. The lattice implementation of the chemical potential is studied in detail in section~\ref{sec:chempot}. Nevertheless, note already here that the chemical potential enters only the fermionic part of the action and is not present in $S_G$. The expectation value of an arbitrary observable $\phi$ based on the above partition function is written as
\be
\expv{\phi} = \frac{1}{\Z} \int \D U \, \phi \, e^{-S_G(U)} \prod_q \det M^{N_q/4}(U, \mu_q,m_q)
\label{eq:expvalO}
\ee

Here and in the following the fermionic determinant is calculated using the staggered discretization, which is used for obtaining all of the results in this thesis.

\section{Thermodynamic observables}
\label{sec:thermo_obs}

In lattice simulations the partition function~(\ref{eq:partfunc4}) itself is not directly accessible. There are on the other hand various observables one can measure using the partial derivatives of $\log \Z$. Such observables will in general be sensitive to the transition between hadronic matter and the QGP and thus play a very important role in thermodynamic studies. These are often referred to as {\it approximate order parameters}. The reason for this will be discussed in more detail in section~\ref{sec:nat}.

\subsection{Chiral quantities}
\label{sec:chiral}

It is well known that the QCD Lagrangian~(\ref{eq:Lag}) exhibits an $U(N_Q)_L\times U(N_Q)_R$ chiral symmetry in the limit $m_q\to 0$ where all flavors are massless. In particular, an axial $U(1)$ transformation on any field $\psi_q$ leaves the Lagrangian invariant:
\be
\psi_q \to e^{i\theta\gamma_5}\psi_q,\quad \bar\psi_q\to\bar\psi_q e^{i\theta\gamma_5}\; \Rightarrow \; \mathcal{L}(m_q=0)\to\mathcal{L}(m_q=0)
\ee
The staggered fermion formulation -- although not fully chirally symmetric -- is also invariant under such a transformation. This part of the group is therefore often referred to as the staggered remnant of the full chiral group. Note that Wilson-fermions do not preserve this symmetry and thus in this case the breakdown of chiral symmetry would be much more difficult to study.

The order parameter of this symmetry is the quark chiral condensate $\bar\psi_q\psi_q$. The chirally broken (low temperature) phase is characterized by a nonzero vacuum expectation value $\expv{\bar\psi_q\psi_q}>0$, while in the symmetric (high temperature) phase $\expv{\bar\psi_q\psi_q}=0$. The chiral condensate for the flavor $q$ can be written as the partial derivative of the partition function with respect to the quark mass $m_q$:
\be
\begin{split}
\expv{\bar\psi_q\psi_q} \equiv \frac{\partial \log \Z}{\partial m_q} &= \frac{N_q}{4}\frac{1}{\Z} \int \D U e^{-S_G(U)} \det M^{N_q/4-1} \frac{\partial \det M}{\partial m_q} \\
&= \frac{N_q}{4}\frac{1}{\Z} \int \D U e^{-S_G(U)} \det M^{N_q/4} \frac{\partial \log \det M}{\partial m_q}
\end{split}
\label{eq:psibarpsi}
\ee
Using the equality $\log \det = \Tr \log$ and taking into account that $\frac{\partial M}{\partial m_q} = \mathds{1}$, one obtains\footnote{Here the identity $(M^{-1})' = -M^{-1} M' M^{-1}$ is also used with the prime denoting differentiation with respect to an arbitrary variable. Also note that although suppressed, the Dirac operator of course flavor-dependent.}:
\be
\expv{\bar\psi_q\psi_q} = \frac{N_q}{4} \expv{ \Tr(M^{-1}) }
\label{eq:psibarpsi2}
\ee

The second derivative of $\log \Z$ with respect to the quark mass is also of interest; it is called the chiral susceptibility:
\be
\begin{split}
\expv{\chi_{\bar q q}} &= \frac{\partial^2 \log \Z}{\partial m_q^2}\\
&= -\frac{N_q^2}{16} \expv{ \Tr(M^{-1}) } ^2 + \frac{N_q^2}{16} \expv{ \Tr(M^{-1}) \Tr(M^{-1})} - \frac{N_q}{4} \expv{ \Tr(M^{-1}M^{-1} )} \\
&\equiv -\expv{\bar \psi_q \psi_q}^2 + \expv{\chi_{\bar q q}^{\textmd{disc.}} + \chi_{\bar q q}^{\textmd{conn.}}}
\end{split}
\label{eq:chi_pbp}
\ee
where the contribution originating from the single expectation value is divided into a disconnected and a connected part with
\be
\bar\psi_q\psi_q = \frac{N_q}{4} \Tr(M^{-1}), \quad \chi_{\bar q q}^{\textmd{disc.}} = (\bar\psi_q\psi_q)^2, \quad  \chi_{\bar q q}^{\textmd{conn.}} = \frac{\partial (\bar\psi_q\psi_q)}{\partial m_q}
\ee
Usually one studies the chiral condensate density and the chiral susceptibility density, which is obtained from the above combinations after a multiplication with $1/V_{4D}=T/V$.

\subsection{Quark number-related quantities}

A part of the full chiral group is the $U(1)$ vector symmetry. This symmetry of~(\ref{eq:Lag}) corresponds to the freedom of redefining the phases of the quark fields
\be
\psi_q \to e^{i\theta}\psi_q,\quad \bar\psi_q\to\bar\psi_q e^{-i\theta}\; \Rightarrow \; \mathcal{L}\to\mathcal{L}
\ee
This $U(1)$ symmetry -- which is valid for arbitrary $m_q$ -- is related to quark number conservation. Thus in this regard it is useful to study the quark number density $n_q$, which is proportional to the first derivative of $\log \Z$ with respect to the chemical potential $\mu_q$:
\be
\expv{n_q} \equiv \frac{\partial \log \Z}{\partial \mu_q} 
\ee
In the same manner as in~(\ref{eq:psibarpsi}) and~(\ref{eq:psibarpsi2}) one obtains:
\be
\expv{n_q} = \frac{N_q}{4} \expv{\Tr(M^{-1} M')}
\label{eq:qdensity}
\ee
where the prime indicates a derivative with respect to the chemical potential assigned to the quark labeled by $q$. Second derivatives with respect to the various chemical potentials can also be defined. For the study of the phase diagram we will only be interested in the diagonal susceptibilities, i.e. those that are twice differentiated with respect to the same $\mu_q$. These observables we will refer to as the quark number susceptibilities:
\be
\begin{split}
\expv{\chi_q} \equiv \frac{\partial^2 \log \Z}{\partial \mu_q^2} = &- \left(\frac{N_q}{4}\right)^2 \expv{ \Tr(M^{-1} M') }^2 + \left(\frac{N_q}{4}\right)^2  \expv{ \Tr(M^{-1}M')^2 } \\
&+ \frac{N_q}{4} \expv{ \Tr (M^{-1}M'' - M^{-1}M'M^{-1}M')}
\end{split}
\ee
which is compactly written as
\be
\expv{\chi_q} \equiv -\expv{n_q} ^2 + \expv{\chi_q^{\textmd{disc.}} + \chi_q^{\textmd{conn.}} }
\label{eq:chi_q}
\ee
where a disconnected and a connected term was once again defined:
\be
n_q = \frac{N_q}{4} \Tr(M^{-1} M'), \quad \chi_q^{\textmd{disc.}} = n_q^2, \quad \chi_q^{\textmd{conn.}} = \frac{\partial n_q}{\partial \mu_q}
\ee
To obtain the corresponding densities one should once again multiply by $T/V$. Moreover, it is customary to study the combination $\expv{\chi_q}/T^2$ for reasons discussed in section~\ref{sec:mudepobs}.

Note that despite its name, the quark number susceptibility does not exhibit the peak-like structure that is usual for a susceptibility in statistical physics. This is due to the fact that $\chi_q$ can also be written as the first derivative (with respect to $\mu_q^2$) of the thermodynamic potential $\log\Z$.

\subsection{Confinement-related quantities}
\label{sec:confine}

In the limit $m_q\to\infty$ of infinitely heavy quarks the QCD Lagrangian possesses an additional symmetry. In this limit quarks decouple from the theory and one is left with a purely gluonic system described by $S_G$. This system is invariant under a center transformation of the temporal links, i.e. a transformation where each temporal link $U_4(n)$ is multiplied by a $Z\in Z(3)$ center element (and the adjoint links by $Z^+$). Since by definition $Z$ commutes with every link variable, any closed loop of links is invariant under this transformation, except for loops that wind around the temporal direction. Therefore there is an observable that explicitly breaks this $Z(3)$ symmetry, which can be constructed by multiplying the links along timelines:
\be
P = \frac{1}{V} \sum\limits_{n_1,n_2,n_3} \Tr \prod\limits_{n_4=0}^{N_t-1} U_4(n)
\label{eq:polyakovloop}
\ee
This observable is called the Polyakov loop. Its expectation value is connected to the free energy of a static quark-antiquark pair taken infinitely far apart $F_{\bar q q}(r\to \infty)$:
\be
\left| \expv{P} \right| = e^{-F_{\bar q q}(r\to\infty)/2T}
\label{eq:ploop2}
\ee
In the low temperature phase of pure gauge theory $F_{\bar q q}(r)$ diverges as $r\to\infty$ and as a consequence $\expv{P}=0$. This is just the phenomenon of confinement: it takes infinitely large energy to separate a quark from an antiquark (this can be thought of as the presence of a string that connects the quark and the antiquark). At high temperatures quarks are no longer confined and thus $F_{\bar q q}(r\to\infty)$ is finite, producing a nonzero expectation value for the Polyakov loop. In view of the observation that~(\ref{eq:polyakovloop}) is not invariant under $Z(3)$ transformations, this means that the Polyakov loop acts as the order parameter of center symmetry: at low temperature the symmetry is intact, while at high temperatures it is spontaneously broken.

In full QCD with finite quark masses $Z(3)$ is not a valid symmetry anymore, as the fermion determinant contains terms that transform nontrivially. This non-invariance can pictorially be described by the fact that quark-antiquark pairs can be created from the vacuum and the ``string'' formed between strong charges can break. Nevertheless, the Polyakov loop still signals the transition from hadronic matter to the QGP by increasing from almost zero to a larger value.

For the scale setting procedure of chapter~\ref{chap:PG} we will also use the susceptibility of the Polyakov loop, which is defined as
\be
\chi_P = V \left( \expv{P^2} - \expv{P}^2 \right)
\ee

\subsection{Equation of state-related quantities}
\label{sec:eosquant}

The partition function also serves to define observables that can be used to establish the equation of state of the theory. Such observables play an important role in describing the thermodynamic properties of the system; their definition is given in this section. These definitions will be applied in chapter~\ref{chap:EoS} for the determination of the equation of state both in pure gauge theory and in full QCD.

The free energy density is related to the logarithm of the partition function as
\be
f = -\frac{T}{V} \log \Z
\ee
The pressure is given by the derivative of $T\log\Z$ with respect to the volume. Assuming that we have a large, homogeneous system, differentiation with respect to $V$ is equivalent to dividing by the volume. Therefore in the thermodynamic limit the pressure can be written as minus the free energy density:
\be
p = -\lim_{V\to\infty}f.
\label{eq:press}
\ee
In lattice simulations the validity of this assumption has to be checked. This will also be elaborated on in chapter~\ref{chap:EoS} and chapter~\ref{chap:PG}. Having calculated the pressure as a function of the temperature $p(T)$, all other thermodynamic observables can also be reconstructed. The trace anomaly $I$ is a straightforward derivative of the normalized pressure:
\be
I\equiv\epsilon-3p=T^5 \frac{\partial}{\partial T} \frac{p(T)}{T^4}
\label{eq:Idef}
\ee
This combination is often called interaction measure as it measures the deviation from the equation of state of an ideal gas $\epsilon=3p$. The inverse relation can easily be written as
\be
\frac{p(T)}{T^4} = \int\limits_0^T {\frac{I(T')}{T'^5} \d T'}
\label{eq:pfromI}
\ee
Using the pressure and the trace anomaly the energy density $\epsilon$, the entropy density $s$ and the speed of sound $c_s$ can be calculated as
\be
\epsilon= I + 3p,\quad\quad s = \frac{\epsilon+p}{T},\quad\quad c_s^2=\left.\frac{\partial p}{\partial \epsilon}\right|_s
\label{eq:eosq}
\ee
Note that in the absence of a chemical potential all the above observable are functions of only one variable, namely the temperature $T$. Therefore varying the pressure or the energy density inevitably modifies the entropy density also. Nevertheless, the speed of sound remains a well-defined quantity, since it can be rewritten as a ratio of two partial derivatives at constant volume~\cite{Gavai:2004se}:
\be
c_s^2 = \left. \frac{\partial p}{\partial T}\right|_{V} \Big/ \left.\frac{\partial \epsilon }{ \partial T}\right|_{V}
\ee

\section{Renormalization}
\label{sec:renorm}

In the previous sections I presented the most important observables that one can use to study the thermodynamic properties of the system. These are all constructed using the free energy density, which is however an ultraviolet divergent quantity. Thus, in order to have a meaningful continuum limit, a proper renormalization to these observables has to be applied. Based on dimensional reasoning, the free energy itself contains additive divergences in the following form~\cite{smilga:1992}:
\be
f=f_r + \O(a^{-4}) + \O(m^2a^{-2}) + \O(m^4 \log(a))
\label{eq:divergences}
\ee
with $f_r$ being the renormalized free energy density. Note that a term $T^2 a^{-2}$ -- despite having the correct dimension -- is not present in the above expression. The absence of this contribution expresses the fact that divergences are in general independent of the temperature. Once one calculated the divergences at $T=0$ and carried out a proper renormalization, the effect of heating up the system to a finite temperature is just equivalent to assigning different weights to the states according to the Boltzmann-factors. This of course should not bring in new divergences.

One can eliminate the additive divergences by subtracting the $T=0$ contribution from the free energy. Thus the following expression is ultraviolet finite:
\be
f_r(\alpha_i, T) = f(\alpha_i, T) - f(\alpha_i, T=0)
\label{eq:f_renorm}
\ee
where $\alpha_i$ are the bare parameters of the theory, on which the dependence of $f$ is explicitly written out in order to emphasize that the subtraction has to be carried out at the same value for each bare parameter. Remember that the temperature can be set according to the number of lattice sites $N_t$ in the temporal extent, see~(\ref{eq:tempoflatt}). Therefore, on a finite lattice, zero temperature can never be realized. Usually, however, a large enough $N_t$ (such as $N_t \gtrsim N_s$) represents a system that is so deep in the low-temperature phase that can it can safely be considered as having effectively zero $T$. Accordingly, in the following, $T=0$ should be thought of in this sense.

\subsection{Renormalization of the approximate order parameters}

The chiral condensate -- as a result of being a derivative with respect to the bare mass -- also contains a multiplicative divergence besides the additive one. This can be cancelled if one multiplies with the bare mass such that the resulting combination is a renormalization group invariant. In order to obtain a dimensionless expression a simple normalization by some mass scale $Q^4$ can be carried out\footnote{The chiral condensate $\bar\psi_q\psi_q$ is abbreviated here as $\bar\psi\psi$; the renormalization procedure applies to any of the quark flavors. Also, dependence on the bare parameters is suppressed from now on.}:
\be
\bar\psi\psi_r(T) = \frac{m}{Q^4}\cdot \left (\bar\psi\psi(T) - \bar\psi\psi(T=0) \right)
\label{eq:pbp_renorm}
\ee
in the same manner, the chiral susceptibility is renormalized with the square of $m$:
\be
\chi_{\bar\psi\psi,r}(T) = \frac{m^2}{Q^4}\cdot \left (\chi_{\bar\psi\psi}(T) - \chi_{\bar\psi\psi}(T=0) \right)
\ee
For the normalization $Q^4$ one can use the fourth power of the $T=0$ pion mass $m_\pi^4$, or any other (possibly temperature dependent) combination, e.g. $T^4$ or $m_\pi^2T^2$. This will be elaborated on in more detail in section~\ref{sec:mudepobs}. We remark here that this renormalization procedure leads to a somewhat unusual chiral condensate which vanishes at $T=0$ and reaches a negative value at $T\to\infty$.

The quark number density and the quark number susceptibility inherit no divergent contributions from the free energy density and thus remain finite as the continuum limit is approached. The origin of this is the absence of a $\mu^2a^{-2}$ term in~(\ref{eq:divergences}), which follows from the fact that quark number is a conserved quantity and thus needs no renormalization. Conserved currents associated with non-Abelian symmetries are in general not renormalized since they obey nonlinear commutation relations and thus their overall normalization is already fixed. For Abelian symmetries like the $U(1)$ symmetry of quark number conservation nonrenormalization can be deduced from the corresponding Ward identity.

The divergences of $f$ also show up in the Polyakov loop. These can be eliminated by the renormalization of the static quark-antiquark potential $V(r)$ at $T=0$~\cite{Kaczmarek:2002mc}, namely, prescribing the condition $V_r(r_0)=0$ for the renormalized potential. This can be interpreted for the Polyakov loop to satisfy the renormalization condition
\be
\left| \expv{P_r} \right| = \left| \expv{P} \right| e^{V(r_0)/2T}
\ee
The potential $V(r)$ on the other hand can be determined at $T=0$ from Wilson loops.

\subsection{Renormalization of the pressure}
\label{sec:renorm_pres}

On the lattice the renormalization~(\ref{eq:f_renorm}) can be written in the form
\be
f^{\rm lat}_r(\alpha_i, N_s, N_t) = f^{\rm lat}(\alpha_i, N_s, N_t) - f^{\rm lat}(\alpha_i, N_s, N_t^{\rm sub})
\label{eq:f_renorm_lat}
\ee
where the temperature (to which the left hand side corresponds to) is determined through $T=(N_t a(\alpha_i) )^{-1}$ with the lattice spacing treated as a function of the bare parameters. Furthermore it is useful to define the integer parameter $\xi>1$ by
\be
N_t^{\rm sub}=\xi N_t
\label{eq:xi_def}
\ee
that is to say, the lattice used for the subtraction corresponds to a temperature of $T/\xi$.
Instead of using a large $\xi$ (i.e. a large $N_t^{\rm sub}$, which is computationally very demanding), one can also calculate the renormalized observables using a reasonably small $\xi$ (due to the fact that divergences are in general independent of the temperature). This approach is presented here for the case of the pressure, but it can also be generalized to any other quantity that is additively renormalized.

Let us introduce the following combination~\cite{Endrodi:2007tq}:
\be
\bar p_\xi(T) = p(T)-p(T/\xi)
\ee
with $\xi$ being an integer larger than one. Using this difference the renormalized pressure can be built up as
\be
\begin{split}
p_r(T) &=p(T)-p(T/\xi) + p(T/\xi) - p(T/\xi^2) + p(T/\xi^2) - p(T/\xi^3) + \ldots = \\
 &= \bar p_\xi(T) + \bar p_\xi(T/\xi) + \bar p_\xi(T/\xi^2) + \ldots
\end{split}
\ee
Furthermore, for the dimensionless combination this means
\be
\frac{p_r(T)}{T^4} = \left.\frac{\bar p_\xi}{T^4}\right|_T + \frac{1}{\xi^4} \cdot \left.\frac{\bar p_\xi}{T^4}\right|_{T/\xi} + \frac{1}{\xi^8} \cdot \left.\frac{\bar p_\xi}{T^4}\right|_{T/\xi^2} + \ldots
\label{eq:xi_expand}
\ee
Since the forthcoming terms are suppressed by increasing powers of $\xi^{-4}$ (and since the pressure itself gives smaller contributions at smaller temperatures) in practice one can safely truncate the series after a few terms. Simply using the first term in the series with $\xi=3$ causes a relative error of somewhat more than one percent. For the study of the EoS in section~\ref{sec:dyneos} this ratio is below the typical statistical error and thus can safely be ignored at the level of the present statistics.

The renormalization prescription~(\ref{eq:xi_expand}) can also be applied to the trace anomaly $I$, the energy density $\epsilon$ and the entropy density $s$. However, since these can all be derived from the pressure, if once the renormalization of the pressure is carried out, the other observables are also going to be finite in the continuum limit.

\section{Chemical potential on the lattice}
\label{sec:chempot}

In heavy ion collisions usually there is a nonzero net quark density in the system resulting from the initial excess of quarks over antiquarks. In the grand canonical approach to statistical physics the density of quarks can be controlled by a chemical potential $\mu$, which can be introduced by including in the action a term $\mu N$ with $N$ being the number of quarks. In the Euclidean formulation the equivalent of $N$ is the four-volume integral of $j_4\equiv\bar\psi\gamma_4\psi$. A naive inclusion of the chemical potential by an additional $\mu j_4$ term in the action would however cause quadratic divergences in the energy density~\cite{Hasenfratz:1983ba}.

However, the term $\bar\psi i g \gamma_\nu A_\nu \psi$ in the QCD Lagrangian shows that the chemical potential actually acts like the fourth component of a purely imaginary, constant vector potential ($\mu \sim i g A_4$). Consequently, a straightforward way to introduce $\mu$ on the lattice is to complement the fourth component of the gauge field with it in the fermionic part of the action\footnote{Remember that the gauge field is defined as $U_\nu=e^{igaA_\nu}$.}. This amounts to multiplying the links in the forward Euclidean time direction with $e^{a\mu}$, and the backward links with $e^{-a\mu}$.
As a result the fermionic action will be (using the staggered formulation):
\be
\begin{split}
S_F^{\textmd{stag}}(\mu) &= \frac{1}{2} \sum_{n} \bigg[ \sum_{\nu=1}^3 \eta_{\nu}(n) \left( \bar\psi(n) U_\nu(n) \psi(n+a\hat\nu) - \bar\psi(n+a\hat\nu) U^\dagger_\nu(n) \psi(n)\right) \\
	&\hspace*{1.83cm}+ \eta_{4}(n) \left( \bar\psi(n) U_4(n)e^{a\mu} \psi(n+a\hat 4) - \bar\psi(n+a\hat 4) U^\dagger_4(n)e^{-a\mu} \psi(n) \right) \bigg]\\  &+ ma \sum_n \bar \psi(n)\psi(n)
\end{split}
\label{eq:fermact_stag_mu}
\ee
On the other hand it is easy to prove that the fermion determinant can be written as a sum over closed loops. In such loops factors of $e^{\pm a\mu}$ once again cancel, unless the loop winds around the Euclidean time direction. For a loop with $n_w$ number of windings the total contribution will therefore be $\left(e^{\pm a\mu}\right)^{n_w N_t}$. This however implies that implementing the chemical potential on the lattice can be done by multiplying the time-like links by $e^{a\mu\cdot N_t} =e^{\mu/T}$ on a single timeslice (and the links in the opposite direction in that timeslice by $e^{-\mu/T}$). This implementation is connected to that in~(\ref{eq:fermact_stag_mu}) by a transformation of the links, which leaves the fermion determinant unchanged.

The action~(\ref{eq:fermact_stag_mu}) describes one quark flavor. For the case of more flavors a different chemical potential has to be assigned to each quark field. As mentioned before, in the energy range under study it is useful to take into account the three lightest flavors $u$, $d$ and $s$. In heavy ion collisions the strangeness of the initial states is zero, and -- due to the strangeness-conserving nature of the strong interactions -- strange quarks $s$ can only be produced together with their antiquarks $\bar s$. This implies that the net strange density throughout the whole process is zero, thus $\mu_s=0$. Furthermore, it is also realistic to set the chemical potential assigned to $u$ and $d$ the same: $\mu_u = \mu_d\equiv\mu_{Q}$. This quark chemical potential is then equal to one third of the baryonic chemical potential: $\mu_{Q}=\mu_B/3$. In the following, if not stated otherwise, $\mu$ will always denote the chemical potential assigned to the light quarks $\mu_Q$.

\subsection{The sign problem}
\label{sec:signproblem}

In section~\ref{sec:hermitic} it was shown that the fermion matrix satisfies the condition~(\ref{eq:hermiticity}). However, if a chemical potential is present, the important property of $\gamma_5$-hermiticity is lost, and consequently the determinant of $M$ will be complex. This implies -- as can be explicitly checked using the discretization~(\ref{eq:fermact_stag_mu}) -- that in such cases the fermion matrix satisfies (for any $\mu\in\mathbb{C}$):
\be
\gamma_5 M(\mu) \gamma_5 = M^\dagger(-\mu^*)
\label{eq:hermiticityfinmu}
\ee
This of course gives the original condition for $\mu=0$. Note however, that $\gamma_5$-hermiticity also holds for purely imaginary values of the chemical potential.

For a real chemical potential on the other hand the determinant will be some complex function and thus cannot be used as a probability weight in~(\ref{eq:partfunc2}). Since physical observables have real expectation values, it is instructive to use the real part $\textmd{Re} \det(M) e^{-S_G}$ as the weight. This can however still be negative and as a consequence can cause large cancellations when averaged over different configurations. This is called the {\it sign problem}. 
Note that a nonzero imaginary part of the determinant is necessary in order to describe a system with nonzero baryon number~\cite{deForcrand:2010ys}. We remark furthermore, that the sign problem is a general characteristic of Monte-Carlo studies of fermionic systems and is not particular to the lattice approach.

Recently several methods were developed to circumvent the sign problem and thus access the region of small
chemical potentials. They are all based on simulations at zero or
purely imaginary chemical potentials where as was argued for according to~(\ref{eq:hermiticityfinmu}) the sign problem is absent.

In the {\it reweighting method} one generates configurations with the action with zero (or purely imaginary) $\mu$, and then assigns new weights to them in a way that they describe a $\mu>0$ ensemble~\cite{Fodor:2001au,Fodor:2001pe,Fodor:2002km,Csikor:2004ik,Fodor:2004nz}. Although this approach is exact in the infinite statistics limit, on a finite number of configurations the weights oscillate strongly, resulting in a large cancellations. Furthermore, since it requires the evaluation of the fermionic determinant, the reweighting method is unfortunately restrained to rather small lattices. Another method is to carry out measurements at various values of the purely imaginary $\mu$ and then {\it analytically continue} to a real chemical potential according to some ansatz function~\cite{deForcrand:2002ci,D'Elia:2002gd,deForcrand:2003hx, D'Elia:2004at,deForcrand:2006pv, Wu:2006su,D'Elia:2007ke,Conradi:2007be, deForcrand:2008vr,D'Elia:2009tm}. One needs to carefully choose the actual form of the ansatz function since the continued results depend very strongly on it. A further approach is the use of the {\it canonical ensemble}, where one works in sectors with fixed baryon number~\cite{Alexandru:2005ix,Kratochvila:2005mk, Ejiri:2008xt} and once again the determinant of the fermion matrix needs to be calculated. Finally, a possible method to analyze the system with nonzero $\mu$ is to carry out a {\it Taylor expansion} around zero (or purely imaginary) chemical potential. Such studies can be improved by measuring higher order coefficients in the expansion. 
This approach can be shown to be just the expanded version of the reweighting method. Furthermore, the Taylor-expansion is not restricted to small lattices, which allows for the systematic study of finite size and lattice discretization errors.
In principle the expansion is expected to converge up to the singularity on the complex $\mu$-plane that is closest to the origin.

In this thesis I present results regarding the phase diagram that were obtained using the Taylor expansion technique. Therefore the next section is devoted to the detailed study of this approach.

\subsection{Taylor expansion in \texorpdfstring{$\mu$}{mu}}
\label{sec:tech}

It is instructive to begin with two general remarks. First, time reversal symmetry ensures that $\Z(\mu) = \Z(-\mu)$ i.e. the partition function -- and thus e.g. the chiral susceptibility also -- is an even function of the chemical potential. As a consequence of this, derivatives with respect to $\mu^2$ and second derivatives with respect to $\mu$ are proportional to each other at zero chemical potential:
\be
\left.\frac{\partial^2 \expv{\chi_{\bar\psi\psi}}}{\partial \mu^2}\right|_{\mu=0} = 2 \left.\frac{\partial \expv{\chi_{\bar\psi\psi}}}{\partial (\mu^2)}\right|_{\mu=0}
\label{eq:depmusq}
\ee
Second, let us expand the fermion determinant in the chemical potential as
\be
\det M(\mu) = a_0 + a_1 \mu + a_2 \mu^2 + a_3 \mu^3 + \ldots
\ee
Equation~(\ref{eq:hermiticityfinmu}) implies that for the determinant $\det M(\mu) = \det M^\dagger(-\mu^*)$ holds. For the coefficients $a_i\in\mathbb{C}$ this accounts to
\be
a_0=a_0^*; \quad a_1=-a_1^*; \quad a_2=a_2^*; \quad \cdots
\label{eq:det}
\ee
which means that even coefficients $a_{2n}$ are real, while odd coefficients $a_{2n+1}$ are purely imaginary (for any $n \in \mathbb{N}$).

By means of the Taylor expansion approach one reconstructs observables at finite chemical potential using their derivatives at zero chemical potential. Instead of $\mu=0$, the expansion could also be carried out around a purely imaginary $\mu=i\mu_I$. However, for the curvature of the pseudocritical line, which will be the primary objective of the next chapter of the thesis, the behavior of observables around $\mu=0$ is of interest.

Suppose now that we have an observable $\phi$, which may depend explicitly on $\mu$. Remember that $\mu$ now denotes the light chemical potential. As a straightforward calculation shows, the derivative of the expectation value $\langle \phi \rangle$ (see~(\ref{eq:expvalO})) with respect to $\mu$ can be written as:
\be
\frac{\partial \langle \phi \rangle}{\partial \mu} = \expv{ n \phi } - \expv{ n } \expv{ \phi } + \expv{ \phi' }
\ee
where $n$ is the light quark number density as in~(\ref{eq:qdensity}) and the prime stands for the derivative with respect to the light chemical potential. The second derivative is similarly given by:
\be
\begin{split}
\frac{\partial^2 \langle \phi \rangle}{\partial \mu^2} &= \expv{ \phi (n^2 + n') } - \expv{ n } \expv{ n \phi } - \expv{ \phi } \expv{ n^2 + n' } - \expv{ n \phi } \expv{ n } + 2 \expv{ n }^2 \expv{ \phi } +\\
&+ \expv{ n \phi' } - \expv{ \phi' } \expv{ n } + \expv{ \phi'' } + \expv{ n \phi' } - \expv{ \phi' } \expv{ n }
\end{split}
\ee
where the combination $n^2+n' = \chi^{\rm disc.}+\chi^{\rm conn.}$ is the disconnected plus the connected part of the light quark number susceptibility (see definition in~(\ref{eq:chi_q})).
Now we can take into account that according to (\ref{eq:det}), odd derivatives of $\det M$ -- such as $n$ -- are purely imaginary. Therefore, expectation values of such operators necessarily vanish at zero $\mu$. Using this fact we can rewrite the above two expressions as:
\begin{align}
\label{eq:muder1}
\left.\frac{\partial \langle \phi \rangle}{\partial \mu}\right|_{\mu=0} &= \expv{ n \phi } + \expv{ \phi' } \\
\left.\frac{\partial^2 \langle \phi \rangle}{\partial \mu^2}\right|_{\mu=0} &= \expv{ \phi (n^2+n') } - \expv{ \phi } \expv{ n^2+n' } + 2\expv{ n \phi' } + \expv{ \phi'' }
\label{eq:muderivative}
\end{align}
For observables that do not depend explicitly on the (light) chemical potential, like $\phi=P$ or $\phi=n_s$ (the latter only depends on $\mu_s$), the last two terms in~(\ref{eq:muderivative}) vanish. However, for $\mu$-dependent observables, like $\phi = \bar\psi\psi \equiv \bar\psi\psi_{ud}$ this explicit $\mu$-dependence shows up and we have a nonzero contribution.

For the study of the phase diagram in chapter~\ref{chap:phasediag} two observables will be addressed: the renormalized chiral condensate and the strange quark number susceptibility. For $\phi=\bar\psi\psi_r$ the formula~(\ref{eq:muderivative}) is directly applicable. Note that the additive renormalization (see section~\ref{sec:renorm}) of the condensate does not influence the derivative in question. For the strange quark number susceptibility (as in~(\ref{eq:chi_q}) with $q=s$) -- even though at zero chemical potential $\expv{n_s}=0$ -- there is an additional term that contributes to the second $\mu$-derivative. Finally, at $\mu=0$ one obtains for both observables
\begin{align}
\label{eq:pbpmud}
\frac{T}{V}\left.\frac{\partial^2 \langle \bar\psi\psi_r \rangle}{\partial \mu^2}\right|_{\mu=0} &= \frac{T}{V} \left[ \expv{ \bar\psi\psi_r (n^2+n') } - \expv{ \bar\psi\psi_r } \expv{ n^2+n' } + 2\expv{ n \bar\psi\psi_r' } + \expv{ \bar\psi\psi_r'' } \right]\\
\label{eq:chismud}
\frac{T}{V}\left. \frac{\partial^2 \expv{\chi_s/T^2}}{\partial \mu^2}\right|_{\mu=0} &= \frac{T}{V}\frac{1}{T^2} \left[ \expv{\chi_s (n^2+n')} - \expv{\chi_s}\expv{n^2+n'} - 2\expv{n_s n}^2 \right]
\end{align}

\chapter{The phase diagram}
\label{chap:phasediag}

As it was pointed out in the introduction, the phase diagram of QCD plays a very important role in high energy physics, both from the theoretical and the experimental point of view. The diagram displays the phases of strongly interacting matter as a function of the state parameters. As argued for in section~\ref{sec:chempot}, the density of quarks is controlled by a chemical potential and the relevant parameter (in e.g. heavy ion collisions) is the light baryonic chemical potential $\mu_B$. Consequently, it is instructive to study the phase diagram on the $\mu_B-T$ plane. In the following the light baryonic chemical potential is denoted simply by $\mu$.

For $\mu=0$ various lattice results can be found in the literature. In particular the nature and the temperature of the transition were studied in detail. Before I present results regarding the phase diagram it is necessary to get acquainted with these $\mu=0$ studies, therefore an overview of the topic is presented in section~\ref{sec:nat}. Based on this grounding next I will address the case of nonvanishing chemical potentials. Unfortunately, here the situation is much more difficult, since the theory suffers from a {\it sign problem}, see section~\ref{sec:signproblem}. In section~\ref{sec:curv} I report results on the curvature of the transition line, as the first $\mu>0$ study in which a continuum extrapolation was carried out.

For this study we used the Symanzik improved gauge action (see section~\ref{sec:symanzik}), while the fermions are discretized using the one-link staggered action with stout-smeared gauge links (see section~\ref{sec:stoutsmear}).

\section{The finite temperature transition}
\label{sec:nat}

For the study of the finite temperature transition the first task is to identify the transition itself. To this end one has to explore the transition region and measure some observable that is sensitive to the transition. The singularity or analytic jump in such observables indicates that the system transforms from one phase to another. We know that there are two distinctive symmetries that are connected to the finite temperature transition: chiral symmetry and $Z(3)$ symmetry (see detailed discussion in section~\ref{sec:chiral} and~\ref{sec:confine}). The breakdown of the $Z(3)$ symmetry and the restoration of chiral symmetry are the two phenomena that signal the finite temperature transition. On the other hand, in nature these symmetries are explicitly broken. While chiral symmetry is only valid when the quark masses are zero, $Z(3)$ symmetry is present only in the absence of dynamical quarks, i.e. when the quarks are infinitely heavy.

One sees that the transition depends very strongly on the actual values of the quark masses. In nature quarks are neither massless nor infinitely heavy, thus the two observables related to the two symmetries -- the chiral condensate and the Polyakov loop -- are not real order parameters. Nevertheless, they still signal the transition from one phase to the other and thus are useful to study as being {\it approximate order parameters}.

\begin{wrapfigure}{l}{6.8cm}
\centering
\includegraphics*[height=5.6cm]{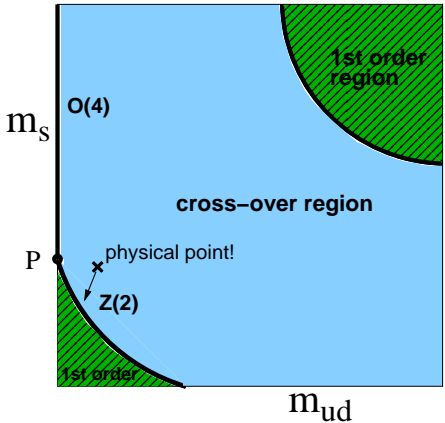}
\caption[The Columbia-plot.]{The Columbia-plot: the nature of the finite temperature transition as a function of $m_{ud}$ and $m_s$. First-order transitions (green, dashed regions) are separated from crossovers (blue region) by second-order transitions (solid lines).}
\label{fig:Columbia}
\vspace*{-0.4cm}
\end{wrapfigure}
It is instructive to analyze the transition from the point of the quark masses a bit more in detail. One may can summarize the dependence of the nature of the transition on the light and strange masses in a compact form on the so-called Columbia-plot (see figure~\ref{fig:Columbia}).
In the upper right corner, which corresponds to $m_s\to\infty$, $m_{ud}\to\infty$, lattice results unambiguously identified a first-order transition~\cite{Celik:1983wz,Kogut:1982rt,Gottlieb:1985ug,Brown:1988qe,Fukugita:1989yb,Kogut:1987rz}. For the opposite limit one is restricted to model calculations. Such models have the same symmetries as QCD, and universality guarantees that near the critical points their behavior essentially coincides with that of QCD. At the origin ($m_s=m_{ud}=0$) we expect to have a first-order transition, while along the vertical axis a second-order transition is predicted, with a universality class of $O(4)$. The physical point $m_{ud}^{phys}$, $m_s^{phys}$ on the other hand lies somewhere in the crossover region, as shown by lattice calculations. There are also indications that the second-order line in the lower left corner of the plot belongs to the $Z(2)$ universality class~\cite{Endrodi:2007gc}. The position of this transition line relative to the physical point is also of phenomenological interest~\cite{deForcrand:2007rq,deForcrand:2006pv}. Lattice results predict that the crossover transition may persist down to $\sim 0.1 m^{phys}$~\cite{Endrodi:2007gc} (pointed to by the arrow on the plot). Note that the scales on the plot are exaggerated. It is useful to remark here that the strength of the transition is a non-monotonic function of the quark masses. A similar non-monotonic behavior as a function of the chemical potential is therefore not to be excluded.

\begin{figure}[h!]
\centerline{
\includegraphics*[height=5.2cm]{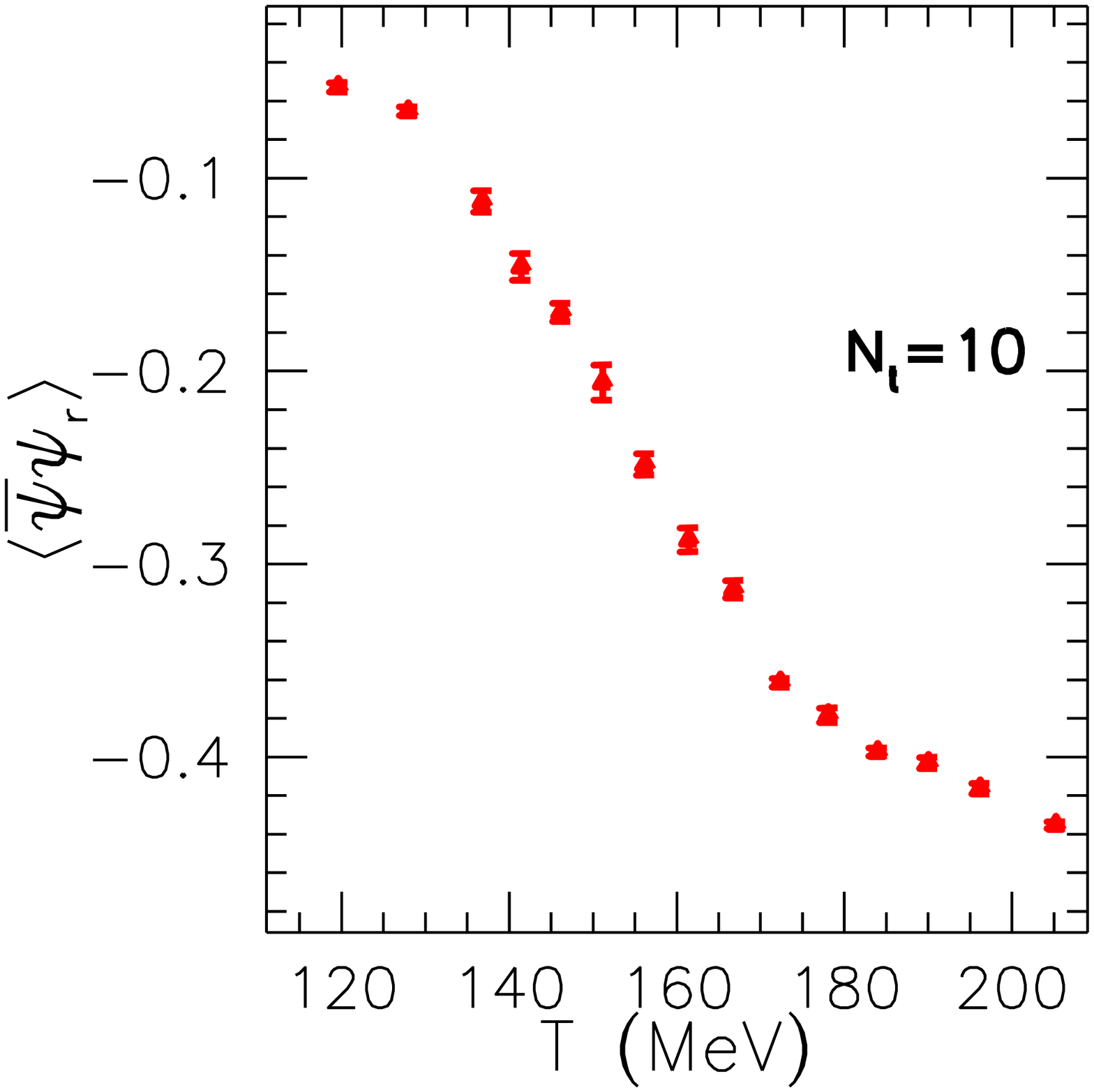}\,
\includegraphics*[height=5.2cm]{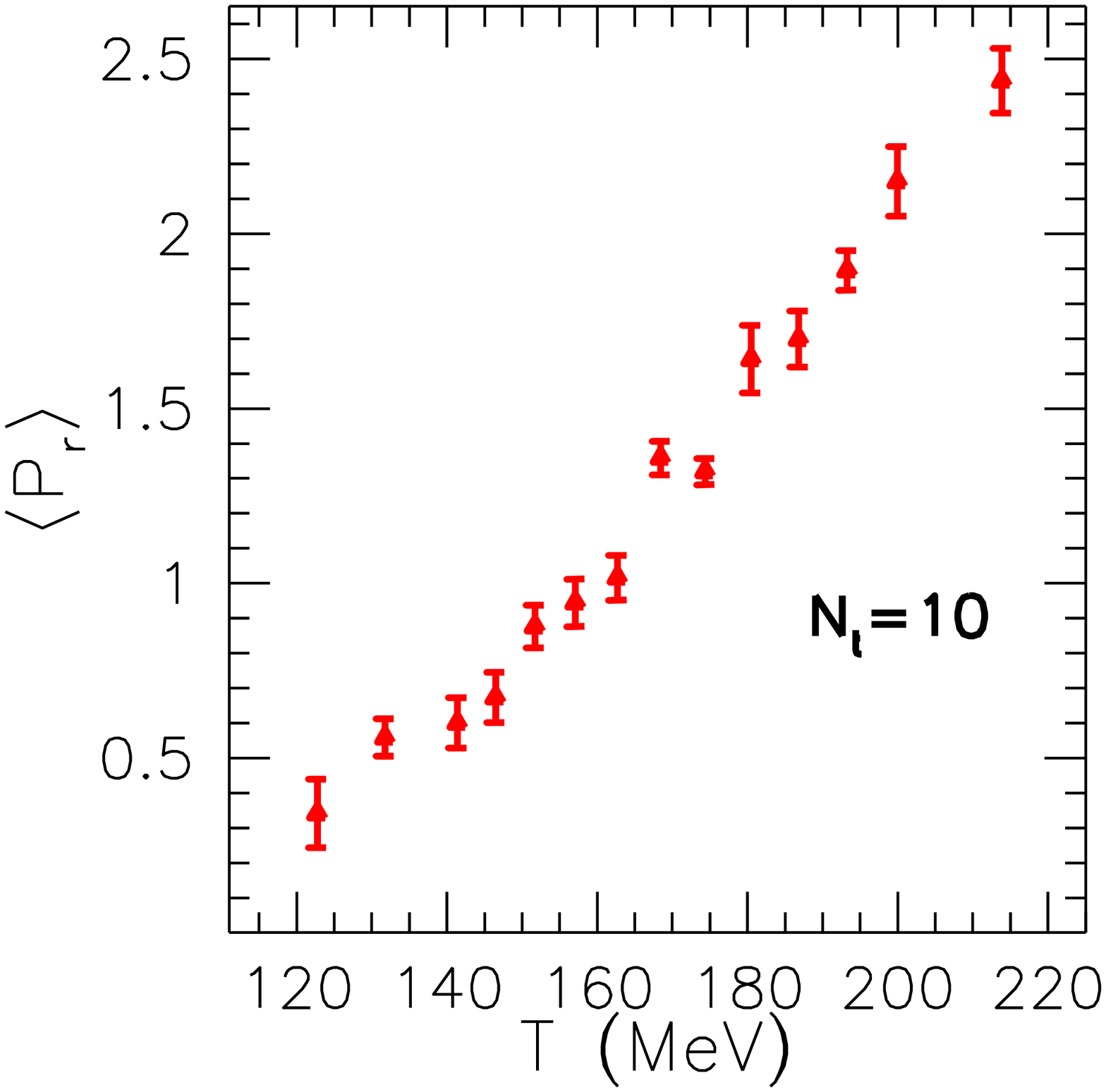}\,
\includegraphics*[height=5.2cm]{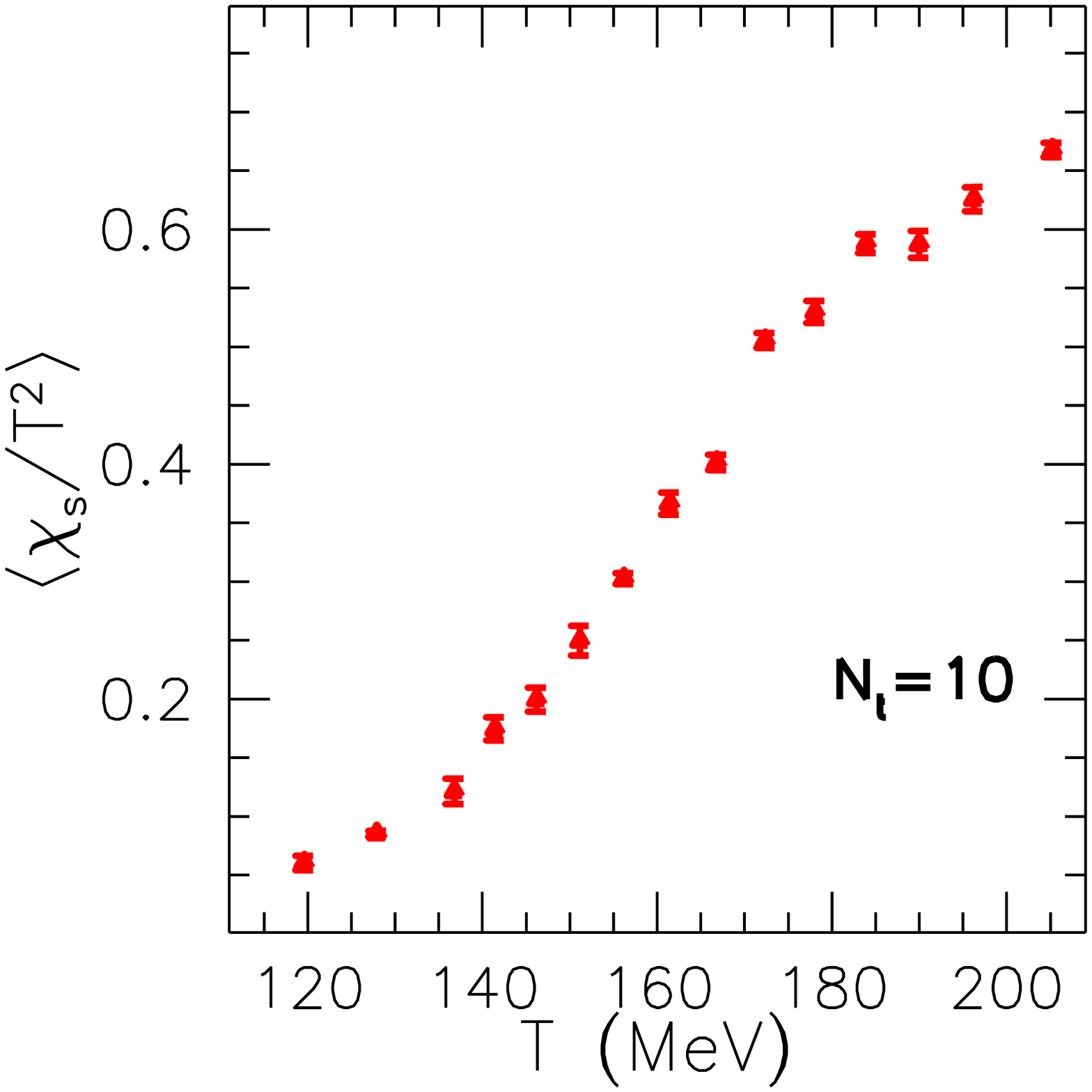}
}
\caption[The renormalized chiral condensate, Polyakov loop and the strange quark number susceptibility at zero chemical potential.]{The renormalized chiral condensate, Polyakov loop and the strange quark number susceptibility as functions of the temperature at $\mu=0$.}
\label{fig:obs_illustr}
\end{figure}

In the following I concentrate on the physical point, that is to say, the quark masses are set to their physical values (see section~\ref{sec:lcp}). 
The observables of interest are shown as a function of the temperature in figure~\ref{fig:obs_illustr}. From left to right the renormalized chiral condensate, Polyakov loop, and the normalized strange quark number susceptibility are plotted in the transition region, measured on $N_t=10$ lattices. While $\bar\psi\psi_r$ decreases from zero to a negative value (the details of the renormalization of the condensate is discussed in section~\ref{sec:renorm}), $P$ and $\chi_s/T^2$ increase from almost zero to a nonzero value. Apparently there is no singularity in the observables shown in figure~\ref{fig:obs_illustr}. This is no surprise, since on a finite lattice no real phase transition can be realized, due to the fact that the partition function $\Z$ is the finite dimensional integral of a positive integrand. A real phase transition can only be recovered in the infinite volume limit $V\to\infty$. It is therefore instructive to carry out measurements on various lattices with increasing (three dimensional) volume, and study the finite size scaling of the observables. Such a careful analysis was carried out recently and the results show that the continuum extrapolated observables show no singularities even in the infinite volume limit~\cite{Aoki:2006we}. This reflects that the transition is an analytic crossover, as opposed to being a first-order phase transition.

Such a crossover transition -- just like the melting of butter -- does not exhibit a unique critical temperature. On the contrary, one can characterize the finite temperature transition of QCD by a {\it pseudocritical temperature} and a (nonzero) width\footnote{In the following the term transition temperature will refer to the pseudocritical temperature.}. In particular, the transition temperature and the width can be determined by means of e.g. a cubic fit: the former is related to the inflection point of the curve, while the latter to the inverse slope at the inflection point. Note that the Polyakov loop and the strange susceptibility tend to give a somewhat higher transition temperature as compared to the chiral condensate (see figure~\ref{fig:obs_illustr}). This difference also persists in results extrapolated to the continuum limit~\cite{Aoki:2009zzc,Borsanyi:2010bp}. This behavior is not unexpected and its reason lies in the crossover nature of the transition. Different definitions may give different results for the transition temperature, simply due to the nonsingular behavior of the logarithm of the partition function. 

Throughout this chapter the stout smeared staggered fermionic action is used. This discretization was applied in the above cited studies also. The transition temperature was however also studied using various other formulations, including improved staggered actions or further fermionic implementations, see e.g. Refs.~\cite{Bazavov:2010bx,Cheng:2009be}.

\section{The curvature of the \texorpdfstring{$T_c(\mu)$}{Tcmu} line}
\label{sec:curv}

\begin{figure}[h!]
\centering
\includegraphics*[height=7cm]{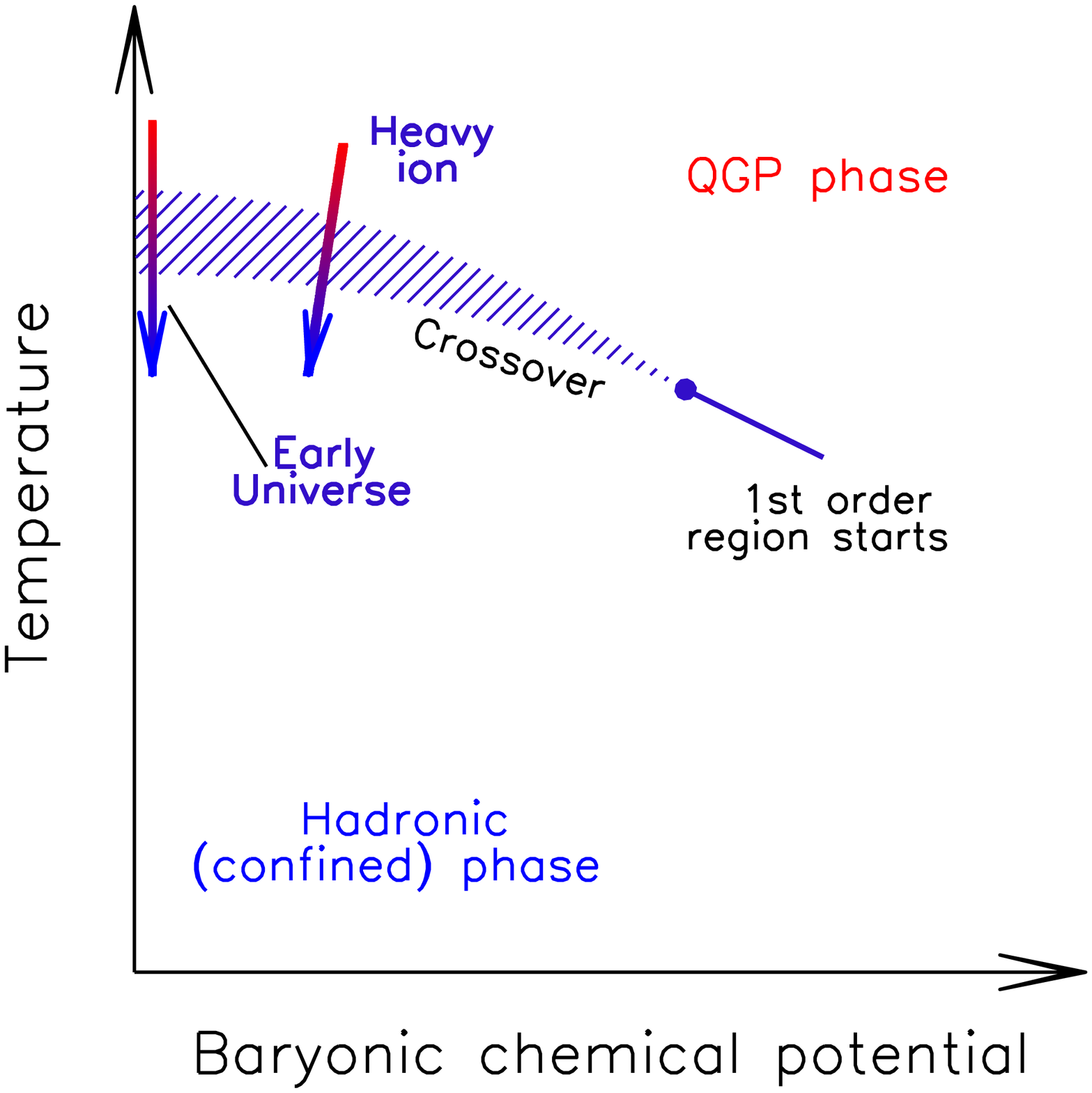} \quad
\includegraphics*[height=7cm]{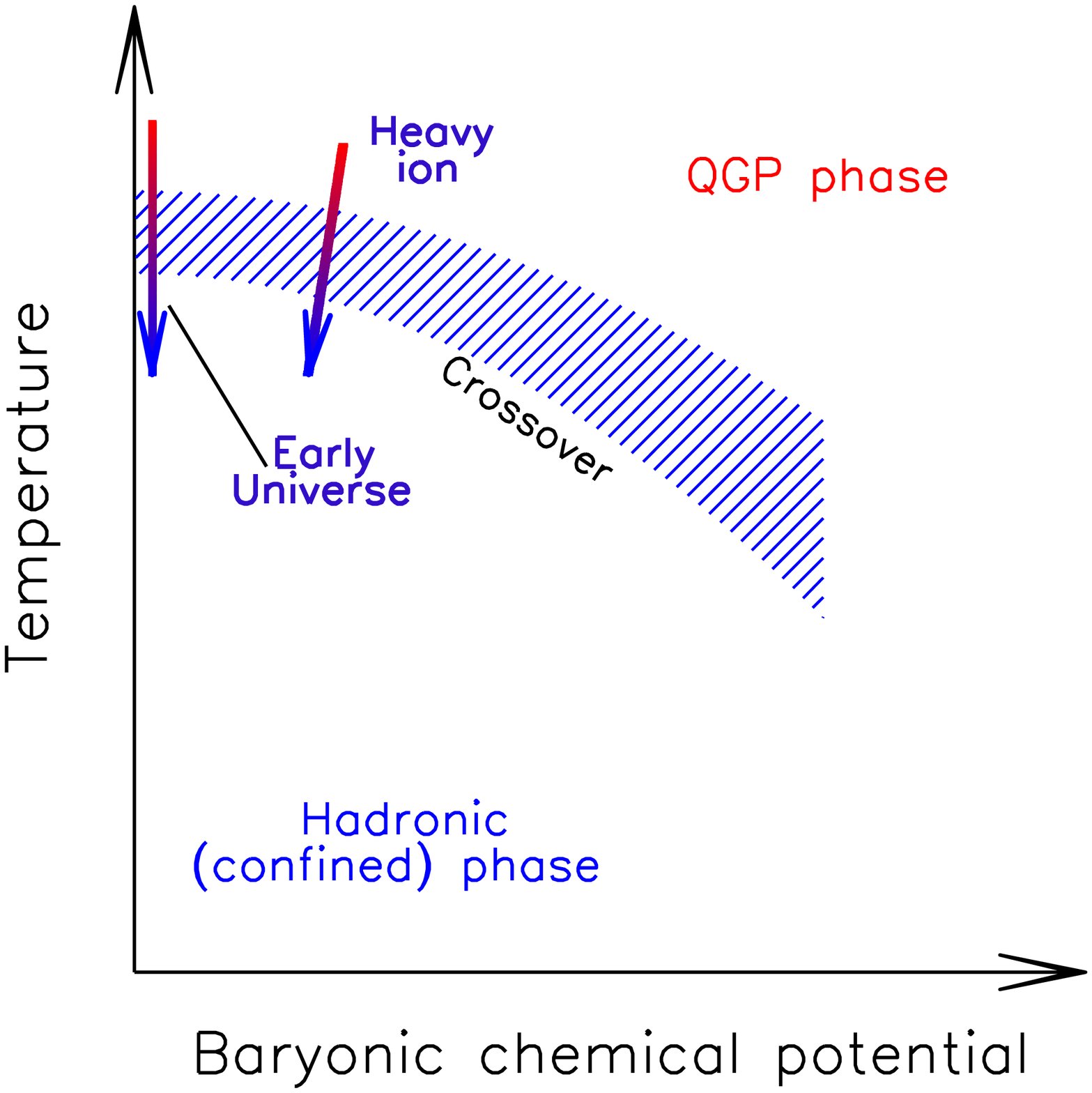}
\caption[Two scenarios for the QCD phase diagram on the $\mu-T$ plane.]{Two possible scenarios for the QCD phase diagram on the $\mu-T$ plane. The left panel shows a phase diagram with a transition growing stronger and possibly even turning into a real, first-order phase transition at a critical endpoint. The right panel on the other hand corresponds to a scenario with a weakening transition and no critical endpoint. The paths representing systems that describe the early Universe and a heavy ion collision are
also shown by the arrows.
}
\label{fig:scenarios}
\end{figure}

While at zero chemical potential lattice calculations provide reliable and accurate results~\cite{Cheng:2006qk,Aoki:2006br,Aoki:2009sc,Bazavov:2009zn}, there are various questions about the $\mu>0$ region of the phase diagram still to be answered. Two possible scenarios are illustrated in figure~\ref{fig:scenarios}. As discussed above, the transition at $\mu=0$ is a crossover~\cite{Aoki:2006we} and the transition temperature $T_c$ is expected to decrease as we increase $\mu$ (one might think that a system with high density favors the deconfined phase already ay smaller temperatures). Apart from the actual shape of the $T_c(\mu)$ function, a particularly interesting question is whether the transition becomes weaker or stronger as $\mu$ grows.

A strengthening of the transition could lead to the existence of a critical point, where the crossover transforms into a true phase transition (see left side of figure~\ref{fig:scenarios}). Indications for the existence of such a critical endpoint are present in the literature, see e.g.~\cite{Fodor:2001pe}.

Another possibility is that the transition weakens with increasing $\mu$ (see right side of figure~\ref{fig:scenarios}). Recently it was found on relatively coarse lattices that the transition at small $\mu$ is weaker as compared to the $\mu=0$ situation~\cite{deForcrand:2006pv,deForcrand:2008vr}. The existence of the critical point is not ruled out by this result, but it requires a non-monotonic behavior. Note, that the strength of the transition is also non-monotonic as a function of quark masses as it was demonstrated in section~\ref{sec:nat}. A similar non-monotonic behavior can be observed at finite $\mu$ in a linear $\sigma$ model with two quarks~\cite{Kapusta:2009ac}. 

The transition temperature $T_c(\mu)$ is determined in the following as a function of the chemical potential through a Taylor-expansion technique. A particularly important property of this function is its curvature $\kappa$ -- i.e. its second derivative with respect to $\mu$ at $\mu=0$. Due to the crossover nature of the transition, different definitions may give different results for $T_c(\mu)$ -- and thus also for $\kappa$. To extract this quantity I will show the analysis of two different observables that are sensitive to the transition: the chiral condensate and the strange quark number susceptibility. Furthermore, the method also reveals information about the change in the slope of these observables as a function of $T$ near the transition temperature as $\mu$ is increased. This is closely related to the change in the width -- or, in other words, the strength -- of the transition. This information can indicate whether the transition grows and turns into a real, first-order phase transition or remains an analytic crossover.

\subsection{The transition temperature at nonzero \texorpdfstring{$\mu$}{mu}} 

Due to the fact that the partition function $\Z$ is an even function of the chemical potential, the transition temperature also depends only on $\mu^2$. Accordingly, let us parameterize the transition line in the vicinity of the vertical $\mu = 0$ axis as
\be
T_c(\mu^2)=T_c \left( 1 - \kappa \cdot \mu^2/T_c^2 \right)
\label{eq:kappadef}
\ee
with $T_c$ being short for $T_c(0)$. This implies that the curvature can be written as 
\be
\kappa= - T_c \left. \frac{\d T_c(\mu^2)}{\d (\mu^2)} \right|_{\mu=0}
\label{eq:curvdef}
\ee
In order to determine the curvature we have to measure $T_c$ as a function
of $\mu$ for small chemical potentials. 
A straightforward way to do this would be to calculate the value of some observable as a function of $T$ at a nonzero $\Delta\mu$ using its value at $\mu=0$ and the derivative~(\ref{eq:muderivative}). Then one can analyze the behavior of this observable in the transition region by means of e.g. a cubic fit to find the inflection point (like shown in section~\ref{sec:nat}). However, this approach turns out to be somewhat ineffective since then a fitting of the reweighted data is required.
Instead of this, we use a definition of $T_c$ which is more suitable for determining the curvature.

Let us consider a quantity $\phi(T,\mu^2)$ that is monotonic in $T$ in the transition region, and fulfills the following 
constraints\footnote{In order not to complicate the notation, in the text we simply write $\phi$ instead of $\expv{\phi}$.}:
\be
\lim_{T\rightarrow 0} \frac{\partial}{\partial \mu^2} \phi(T,\mu^2) = 0, \quad\quad \lim_{T\rightarrow \infty} \frac{\partial}{\partial \mu^2} \phi(T,\mu^2) = 0 
\label{eq:constraints}
\ee
that is to say, $\phi$ does not depend on the chemical potential in 
the limiting cases $T\rightarrow 0$ and $T\rightarrow \infty$. For any fixed $\mu$ we can implicitly define a
transition temperature $T_c(\mu^2)$ as the temperature
at which $\phi(T,\mu^2)$ takes the predefined constant 
value $C$:
\be
\left.\phi(T,\mu^2)\right|_{T=T_c(\mu^2)}=C.
\ee
We will choose a $C$ that corresponds to the inflection point 
of $\phi(T,0)$\footnote{The dependence of $T_c$ on the constant $C$ is suppressed in the following.}.

\subsection{Definition of the curvature}

Now let us determine the curvature using the definition of $T_c(\mu^2)$. 
The total derivative of the observable $\phi(T,\mu^2)$ may be written as
\be
\d\phi=\left. \frac{\partial \expv{\phi}}{\partial T}\right|_{\mu=0} \cdot \d T + \left. \frac{\partial \expv{\phi}}{\partial (\mu^2)}\right|_{\mu=0}\cdot \d \mu^2
\ee
Along the $T_c(\mu^2)$ line, $\phi$ is constant by definition, thus 
$\d\phi=0$. One obtains
\be
\frac{\d T_c}{\d \mu^2}=  \underbrace{ - \left.\left( \frac{\partial \expv{\phi}}{\partial \mu^2} \right)\right|_{\substack{T=T_c\\\hspace*{-0.1cm}\mu=0}} \Big / \left.\left(\frac{\partial \expv{\phi}}{\partial T} \right)\right|_{\substack{T=T_c\\\hspace*{-0.1cm}\mu=0}} }_{R(T)}
\label{eq:tcdef}
\ee
Thus, for every $C$ we can define a curvature. Since the $T_c(C)$ function is
invertible for the whole $C$ range, we can also write the right hand side of
(\ref{eq:tcdef}) as a function of the temperature, $R(T)$.

\begin{wrapfigure}{r}{7.5cm}
\centering
\vspace*{-0.5cm}
\includegraphics*[height=6.0cm]{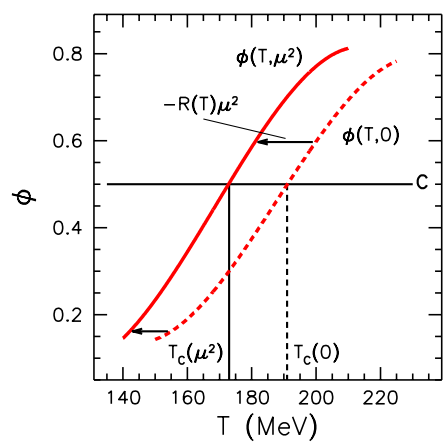}
\vspace*{-0.2cm}
\caption[Illustration of the behavior of the observable at $\mu=0$ and $\mu>0$.]{Illustration of the behavior of the observable $\phi$ at $\mu=0$ and $\mu>0$. The function $T_c(\mu^2)$ is defined as the temperature where $\phi(T,\mu^2)$ crosses a constant value $C$ (see definition in text).}
\vspace*{-0.26cm}
\label{fig:illustr}
\end{wrapfigure}
The function $R(T)$ is related to the distance that the $\phi(T)$ curve 
shifts along the $T$ axis as the chemical potential is varied. Given
$\phi(T)$ and $R(T)$ at zero chemical potential, the shift for non-zero
$\mu$ at leading order is $R(T) \cdot \mu^2$ (the curve moves to the left if $R(T)$ is negative and to
the right otherwise). This behavior is illustrated in figure~\ref{fig:illustr}.
Using $R(T)$ we can define a temperature dependent curvature according to (\ref{eq:curvdef})
as $\kappa(T)=-T_c \cdot R(T)$. The meaning of $\kappa(T)$ is again simple:
it gives the curvature of the $\phi=\rm{const.}$ curve which starts from
$T$ at $\mu=0$.

We use the value of $\kappa(T)$ at $T=T_c$ to define the curvature
for a given observable. The shape of the $\kappa(T)$ function
also has important consequences. The slope of $\kappa(T)$ around $T_c$
is related to the width of the transition as follows:
if the slope is zero, i.e. $\kappa(T)$ is constant around $T_c$, then
all points shift the same amount along the $T$ axis when a small chemical potential
is switched on. This means that to leading order in $\mu$ the shape of
the $\phi(T)$ function (and thus, the width of the transition) does 
not change. If the slope is positive, then points with larger $T$ shift
more than the ones with smaller $T$ resulting in a compression of points, i.e. a 
narrower transition. Similarly, a negative slope indicates a broadening 
of the transition.

Putting all this together, the expression $\partial \kappa / \partial T$ is related to the relative change in the width $W(\mu)$ of the transition as the chemical potential increases:
\[
\frac{1}{W} \frac{\partial W}{\partial (\mu^2)} =- \left . \frac{1}{T_c}\frac{\partial \kappa}{\partial T} \right |_{T=T_c}
\]
where we assume that $W$ is proportional to the inverse slope of the quantity in question: $W \sim \left | \left.(\partial \phi / \partial T\right |_{T=T_c})^{-1} \right |$.

\subsection{The \texorpdfstring{$\mu$}{mu}-dependence of the measured observables}
\label{sec:mudepobs}

The two quantities we used to play the role of $\phi$ are the renormalized chiral condensate and the strange 
quark number susceptibility. The definition and renormalization of these observables is detailed in section~\ref{sec:thermo_obs}. 
For the case of the strange susceptibility it is useful to study the combination $\chi_s/T^2$, since it obeys the conditions of~(\ref{eq:constraints}). It is easy to see that at $T=0$ one gets $\chi_s/T^2=0$ and at $T\to\infty$ the normalized quark number susceptibility $\chi_s/T^2$ reaches its  $\mu$ independent Stefan-Boltzmann limit of 1.

The situation with the chiral condensate is more delicate. The final normalization of this observable was carried out by $Q^4=m_\pi^4$ (see renormalization condition~(\ref{eq:pbp_renorm})). Note that a division by $T^4$ which would also render the condensate dimensionless, changes the temperature-dependence to be non-monotonic, which would be disadvantageous in the present context. Furthermore, as it was already pointed out in section~\ref{sec:renorm}, $\bar\psi\psi_r$ is zero at $T=0$ and approaches a negative value as the temperature is increased. A more conventional condensate which is positive at $T=0$ and goes to zero at large temperatures can be obtained by a constant shift which is irrelevant for our present study.

Finally we need to show that the renormalized chiral condensate is independent of $\mu$ at $T=0$ and $T\to\infty$. At $T=0$ the partition function is independent of $\mu$ as long as $\mu$ is smaller than a $\mu_c$ critical value (the approximate baryon mass) and no baryons can be created from the vacuum. Only for $\mu>\mu_c$ does the partition function have a non-trivial $\mu$-dependence. Therefore, for $\mu<\mu_c$ all derivatives of $\Z$ (thus $\bar\psi\psi_r$) are independent of $\mu$. The chemical potential regime of interest lies in this region. In the Stefan-Boltzmann limit ($T\to\infty$) the $\mu$-independence can be proven by a straightforward calculation.

Now, according to~(\ref{eq:kappadef}) and~(\ref{eq:tcdef}), in order to measure the curvature two derivatives are necessary. The derivative $\partial \expv{\phi}/\partial T$ is calculated numerically, using the $\mu=0$ data as a function of the temperature. The combination $\partial \expv{\phi}/\partial (\mu^2)$ on the other hand is determined using the technique detailed in section~\ref{sec:tech}, in particular, see equations~(\ref{eq:pbpmud}) and~(\ref{eq:chismud}). The derivatives of the explicit dependence of the chiral condensate on $\mu$ (see last two terms of~(\ref{eq:pbpmud})) were calculated numerically, using a purely imaginary chemical potential $\Delta \mu_i$. The value of $\Delta \mu_i$ was varied in the
range $0.01 \ldots 0.0005$, and it was checked that the finite
differences converge fast enough to the $\Delta \mu_i \rightarrow 0$
values and the error coming from this approximation is negligible
compared to statistical errors. Taking into account these
considerations $\Delta \mu_i=0.001$ was used.

\subsection{Finite size effects}

\begin{wrapfigure}{r}{8.0cm}
\centering
\vspace*{-0.6cm}
\includegraphics*[height=7.0cm]{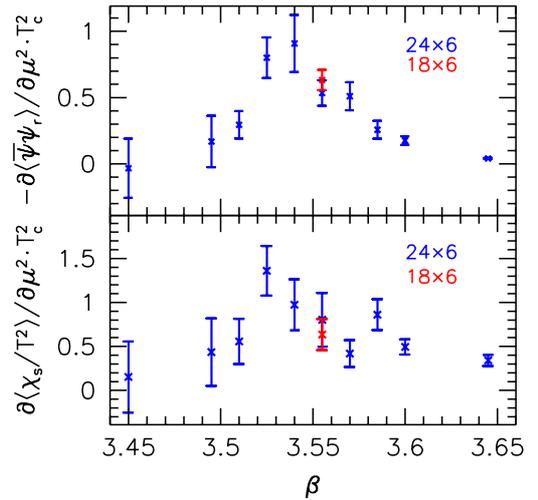}
\vspace*{-0.2cm}
\caption[Finite size analysis of the observables.]{The derivative of the observables with respect to $\mu^2$ measured on $N_t=6$ lattices with $N_s=24$ (blue) and $N_s=18$ (red).}
\vspace*{-0.4cm}
\label{fig:finV}
\end{wrapfigure}
In order to check for finite size effects the measurement of the derivatives $\partial \expv{\phi}/\partial (\mu^2)$ was also performed on lattices with different physical spatial extent for $N_t=6$. In particular we compared results at $\beta=3.555$ obtained on $24^3\times6$ and on $18^3\times6$ lattices. This value of $\beta$ corresponds to about $155$ MeV, i.e. is near the pseudocritical temperature. The larger box is of physical size $\sim5$ fm. We observe a good agreement as the results for $\partial \phi/ \partial (\mu^2)$ agree within statistical errors for both the renormalized chiral condensate $\phi=\bar\psi\psi_r$ and the strange quark number susceptibility $\phi=\chi_s/T^2$, as can be seen in figure~\ref{fig:finV}. Thus we conclude that finite size errors can be neglected at the present statistical accuracy. Note that this volume-independence is a consequence of the crossover nature of the transition.

\section{Results}

\begin{figure}[h!]
\centering
\mbox{
\includegraphics*[height=6.2cm]{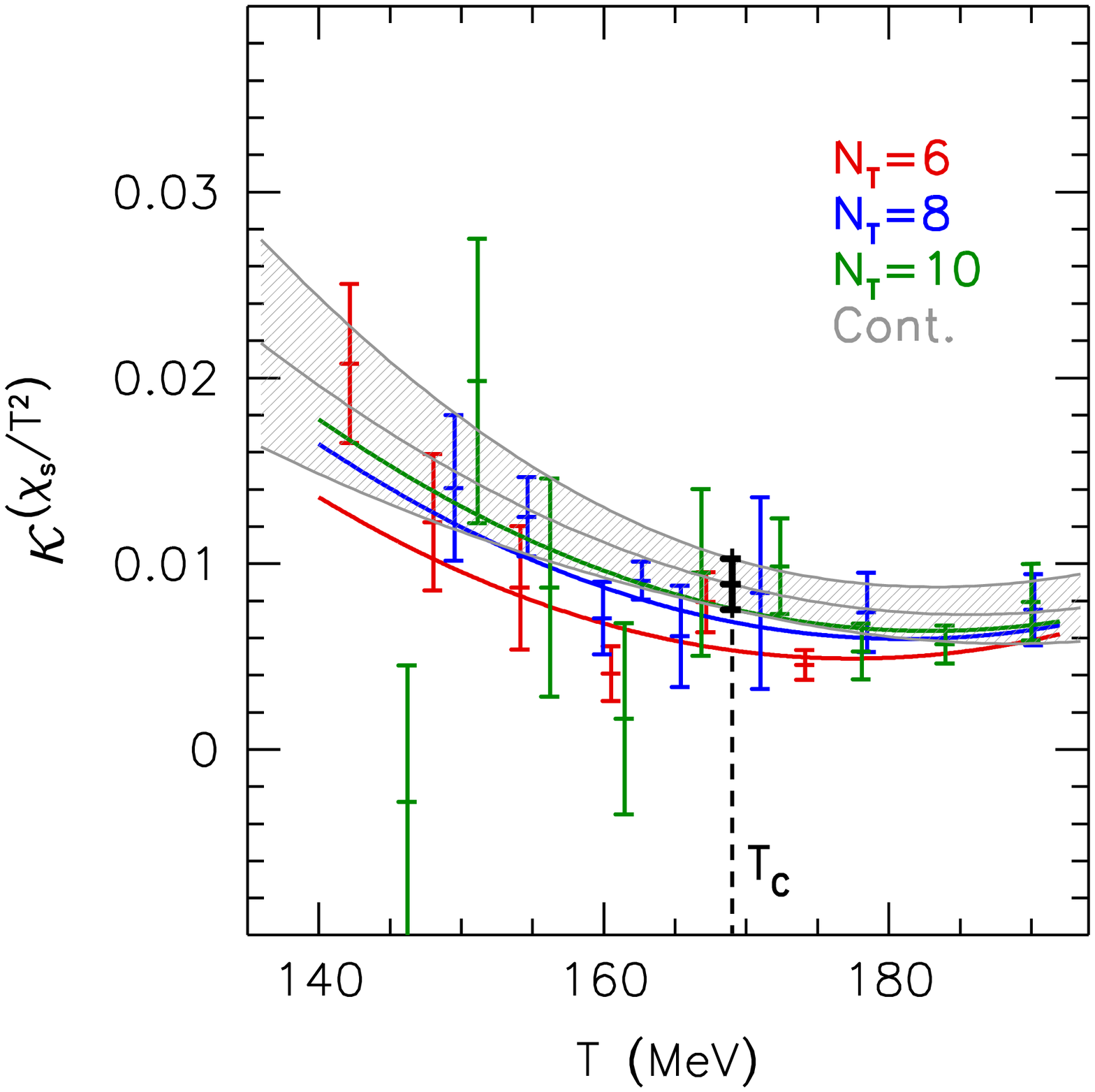}
\includegraphics*[height=6.2cm]{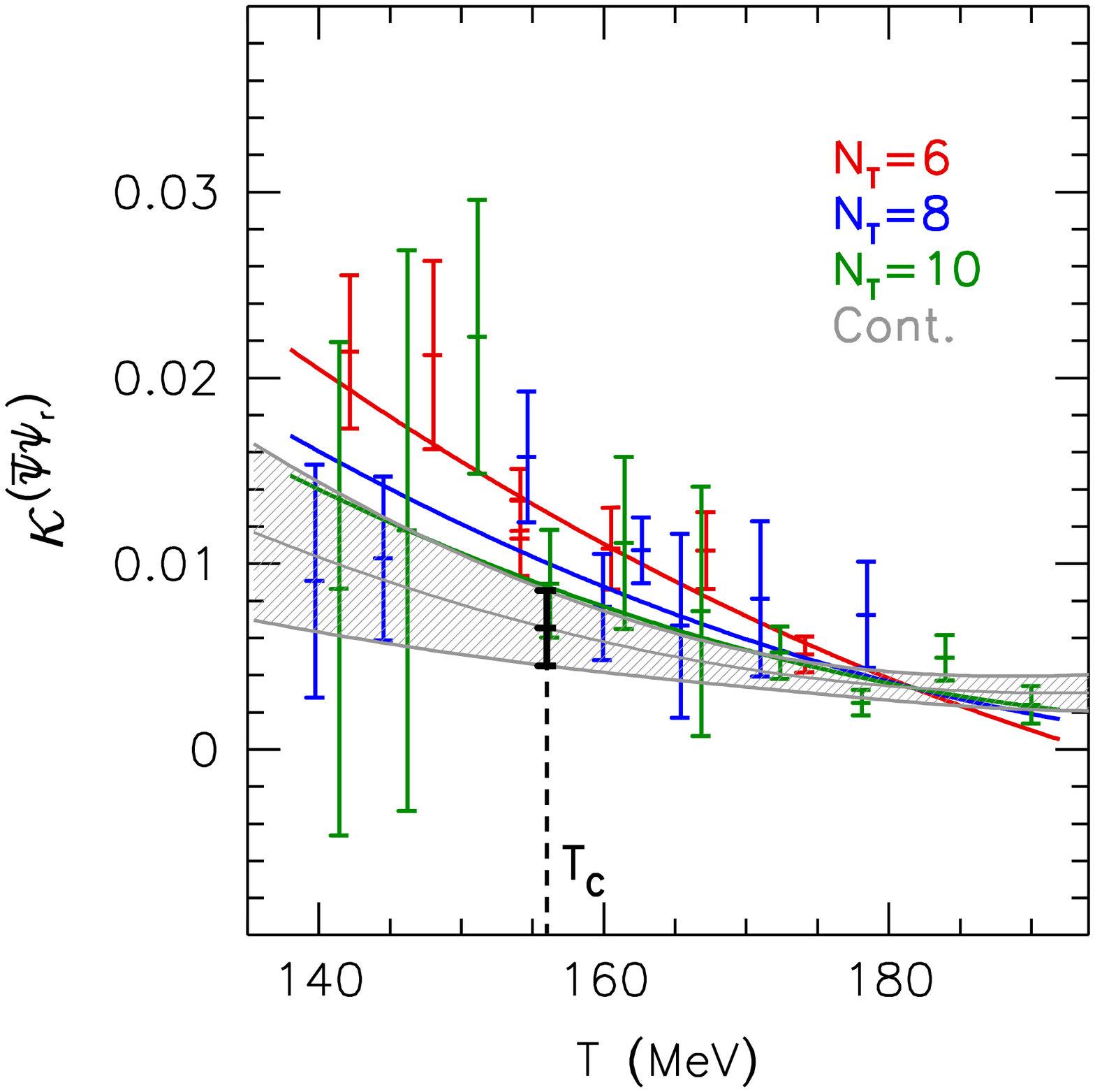}
}
\caption[The temperature dependent curvature for the chiral condensate and the strange susceptibility.]{The curvature $\kappa(T)$ determined using 
the strange quark number susceptibility (left) and the renormalized chiral 
condensate (right), respectively. A result of the combined fit (see detailed description in 
text) is shown by the gray band. The fit results for the 
individual $N_t=6,8,10$ lattices are shown by the red, blue 
and green curves, respectively.}
\label{fig:avg}
\end{figure}

Since the actual shape of the $\kappa(T)$ function is unknown it is instructive to carry
out a Taylor expansion around $T_c$ in the $t=(T-T_c)/T_c$
dimensionless variable:
\be
\kappa(T)=\kappa(T_c)+c_0\cdot t+c_1\cdot t^2
\ee
For each lattice spacing (i.e. each $N_t$) we have several simulation
points, corresponding to different temperatures. In order to fit all
of our points at once, we allow a lattice spacing-dependence for the
constant and linear terms (having a lattice spacing-dependence
 of the quadratic term is also possible, but it does not improve the quality of the fits). Therefore we fit all of our simulation points with the following function:
\be
\kappa(T;N_t)=\kappa(T_c;\textmd{cont})+c_0\cdot t+c_1\cdot t^2+c_2/N_t^2+c_3\cdot t/N_t^2\ee
with fit parameters
$\kappa(T_c;\textmd{cont}),c_0,c_1,c_2$ and $c_3$. The independent data points as well as the fitted curves (for each
$N_t$ and in the continuum) are shown in figure~\ref{fig:avg}.

The $\chi^2/\textmd{d.o.f.}$ values of the two fits are 1.08 and 1.29, respectively, indicating good fit qualities. The continuum curvatures are given by the $\kappa(T_c;\textmd{cont})$ fit parameter. The final results concerning the curvature and the relative change in the width of the transition are 
\begin{align*}
\kappa^{(\chi_s/T^2)} &= 0.0089(14), & \kappa^{(\bar\psi\psi_r)} &= 0.0065(20) \\
\Delta W/W^{(\chi_s/T^2)}  &= 0.033(16), & \Delta W/W^{(\bar\psi\psi_r)} &= 0.030(18)
\label{eq:result}
\end{align*}
with the statistical error of the appropriate quantities in parentheses. 
The curvatures obtained from the two observables are consistent with each other within errors.
Using the $\kappa$ values the transition lines
defined by any of the observables can be given as
\be
T_c(\mu)=T_c[1-\kappa\cdot\mu^2/T_c^2]
\ee
with $T_c$ once again being short for $T_c(\mu=0)$.

The results also suggest that the transition remains a weak crossover
with essentially constant strength for small to moderate chemical potentials.
Actually, there is a slight increase in the width of the 
transition determined from
both quantities. This effect is, however, very weak: the width only changes by a few percent up to $\mu\approx T_c$. 

\begin{figure}[h!]
\centering
\vspace*{-1.1cm}
\includegraphics*[height=9.0cm]{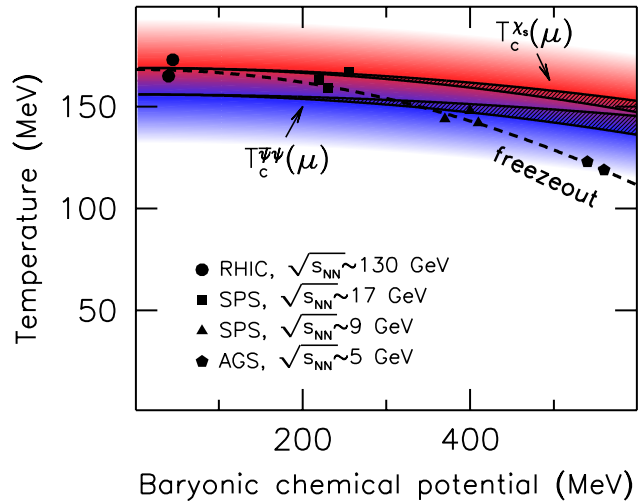}
\vspace*{-0.9cm}
\caption[The phase diagram of QCD for small chemical potentials.]{The phase diagram of QCD for small chemical potentials. 
The crossover transition between the `cold' and `hot' phases is 
represented by the colored area. The lower solid band shows the result 
for $T_c(\mu)$ defined through the chiral condensate and the upper one 
through the strange susceptibility. The width of the bands represent
the statistical error of $T_c(\mu)$ for the given $\mu$ coming from the error of the curvature $\kappa$
for both observables.
The dashed line is the freeze-out curve 
from heavy ion experiments~\cite{Cleymans:1998fq}. Also indicated are with different symbols the individual 
measurements of the chemical
freeze-out from RHIC, SPS (Super Proton Synchrotron) and AGS (Alternating Gradient Synchrotron), respectively. The center of mass energies $\sqrt{s_{NN}}$ for each are shown in the legend.}
\label{fig:finalresult}
\vspace*{-0.2cm}
\end{figure}

The validity range of our result is hard to estimate from the present study alone. 
A conservative estimate
for the limit where the result obtained through the Taylor-expansion is still reliable
is $\mu\approx T_c$ i.e. where the expansion parameter exceeds unity. In the baryonic 
chemical potential this corresponds to about 500 MeV. Beyond this limit higher-order corrections are by all means expected to be important. 
Earlier experience with the exact reweighting method~\cite{Fodor:2004nz}
also shows that the leading-order quadratic behavior of the $T_c(\mu)$ function
dominates up to the above mentioned limit in the baryonic chemical potential.
To investigate whether higher order terms may lead to a critical point
one must carry out a similar analysis with full reweighting,
beyond the reach of present computational resources.

The final result is shown in figure~\ref{fig:finalresult}. The crossover region's extent changes little as the chemical potential increases, and within it two
definitions give different curves for $T_c(\mu)$. It is useful to
compare the whole picture to the freeze-out curve~\cite{Cleymans:1998fq} which summarizes experimental results on the $\{T,\mu\}$ points where hadronization of the quark-gluon plasma was observed. This curve is expected to lie in the interior of the crossover region, as is indicated by our results as well.

\chapter{The equation of state with dynamical quarks}
\label{chap:EoS}

As it was stressed throughout the introduction, the study of the finite temperature transition of QCD is relevant for both the case of the early Universe and that of heavy ion collisions. In both situations strongly interacting matter is thought to enter the high temperature QGP phase of quarks and gluons. While the nature and the temperature scale of the transition imply interesting cosmological consequences (see section~\ref{sec:qgpintro}), they also have a direct impact on contemporary high energy experiments in particle accelerators.
In particular, the deconfined phase of QCD is claimed to have been produced in the
ultrarelativistic heavy ion collision experiments at CERN SPS, RHIC at
Brookhaven National Laboratory, ALICE at the LHC and the future FAIR at the
GSI.
The experimental results available so far show that the hot QCD matter produced
experimentally exhibits robust collective flow phenomena, which are well and
consistently described by near-ideal relativistic hydrodynamics
\cite{Teaney:2000cw,Teaney:2001av,Kolb:2003dz}. These hydrodynamic models
need as an input an Equation of State (EoS) which relates the local
thermodynamic quantities. Therefore, an accurate determination of the QCD EoS
is an essential ingredient to understand the nature of matter created in
heavy ion collisions, as well as to model the behavior of the QGP in the early universe.

QCD thermodynamics and in particular the EoS have received increasing attention in the past years. These include among others calculations in the quenched approximation \cite{Boyd:1996bx,Okamoto:1999hi}, two-flavor simulations
\cite{Bernard:1996cs,AliKhan:2001ek}, studies with heavier-than-physical
\cite{Karsch:2000ps,Kanaya:2009nq}, almost physical \cite{Bernard:2006nj,Cheng:2007jq,Bazavov:2009zn,Cheng:2009zi} and physical quark masses~\cite{Aoki:2005vt}. However, results for an EoS
involving physical masses together with a careful continuum limit are so far
missing; their relevance for the physics of the hot deconfined matter is
obvious. In section~\ref{sec:dyneos} I present results on various thermodynamic observables for a system of $N_q=2+1$ flavors of dynamical quarks. The findings have recently been published, see Ref.~\cite{Borsanyi:2010cj}. Two important aspects of this study are to be emphasized. First, the quark masses are set to their physical values, which is very important due to the strong dependence of thermodynamic quantities on these parameters. Second, a continuum limit estimate is given using lattices of temporal extent $N_t=6,~8,~10$ and a few checkpoints at $N_t=12$. Just as for the study of the phase diagram, once again the Symanzik improved gauge and the stout smeared staggered fermionic action is used.

\section{Determination of the pressure}

First it is instructive to review the standard integral method, which can be used to determine the equation of state on the lattice~\cite{Engels:1990vr}. This technique is then further developed by the multidimensional spline method presented in section~\ref{sec:spline_overview}.

\subsection{The integral method}

The starting point is the (dimensionless) pressure as defined on the lattice
\be
p^{\rm lat}(\beta,m_q)=(N_tN_s^3)^{-1}\log\Z(\beta,m_q)
\ee
as a function of the bare parameters, i.e. the inverse gauge coupling and the quark masses. This combination in itself is unfortunately not accessible using conventional algorithms, only its partial derivatives
with respect to the bare parameters of the action are measurable. Using these
partial derivatives the pressure can be rewritten as a
multidimensional integral along a path in the space of bare parameters:
\be
p^{\rm lat}(\beta,m_q)-p^{\rm lat}(\beta_0,m_{q0})=
(N_tN_s^3)^{-1} \int^{(\beta,m_q)}_{(\beta_0,m_{q0})}
\left(
d\beta \frac{\partial \log \Z}{\partial \beta} +
\sum_q d m_q \frac{\partial \log \Z}{\partial m_q}\right),
\label{eq:multiint}
\ee
where the index `0' is used to denote the starting point of the integration.
The value of the pressure at parameters `0' has to be handled with
care: one either chooses the starting point so deeply in the strong coupling
regime, so that the $p^{\rm lat}(\beta_0,m_{q0})$ can already be neglected or
it can be taken from model computations (see section~\ref{sec:adjustint}). The 
derivatives in~(\ref{eq:multiint}) are the gauge action density and
the quark chiral condensate densities (as defined in~(\ref{eq:psibarpsi})), measured in lattice units:
\be
\langle -s_g\rangle= (N_tN_s^3)^{-1}\frac{\partial \log\Z}{\partial \beta},
\quad \quad \langle\bar\psi_q\psi_q\rangle=(N_tN_s^3)^{-1}\frac{\partial \log\Z}{\partial m_q}.
\label{eq:ders}
\ee

As discussed in section~\ref{sec:renorm_pres}, the renormalization of the pressure can be carried out by subtracting its value measured on a lattice with the same bare parameters, but at a different temperature value, ie. with a different temporal extent $N_t^{\rm sub}$. In the following for these `cold' lattices we use $N_t^{\rm sub}=3 N_t$, i.e. $\xi=3$ in the notation of~(\ref{eq:xi_def}) and~(\ref{eq:xi_expand}). 
The finiteness of $N_t^{\rm sub}$ causes an error of the order $\xi^{-4}$, which is smaller than the typical size of the statistical error of the results, see section~\ref{sec:dyneos}.

Using the subtracted observables
\begin{align}
\langle s_g\rangle^{\rm sub} &= \langle s_g\rangle_{N_t,N_s} - \langle s_g\rangle_{N_t^{\rm sub},N_s} \\
\langle\bar\psi_q\psi_q\rangle^{\rm sub} &= 
\langle\bar\psi_q\psi_q\rangle_{N_t, N_s} - \langle\bar\psi_q\psi_q\rangle_{N_t^{\rm sub}, N_s}
\end{align}
one can express the renormalized pressure, divided by $T^4$ as:
\be
\label{eq:intp}
\frac{p(T)}{T^4}-\frac{p(T_0)}{T_0^4}=
N_t^4 \int^{(\beta,m_q)}_{(\beta_0,m_{q0})}
\left(d\beta \langle -s_g \rangle^{\rm sub} +
\sum_q d m_q \langle \bar\psi_q\psi_q\rangle^{\rm sub}\right)
\ee
where $T$ and $T_0$ are the
temperature values corresponding to the lattice spacing at the bare parameters
$(\beta,m_q)$ and $(\beta_0,m_{q0})$. Using~(\ref{eq:Idef})
one can also relate the integrand to the trace anomaly:
\be
\frac{I(T)}{T^4} \frac{dT}{T}=
N_t^4\left(d\beta \langle -s_g \rangle^{\rm sub} +
\sum_q d m_q \langle \bar\psi_q\psi_q\rangle^{\rm sub}\right).
\label{eq:tracealat}
\ee
Let us make an observation here: due to the $N_t^4$ prefactor in the
trace anomaly the subtracted observables decrease with the lattice
spacing (at a fixed temperature this means an increasing $N_t$). While
$\langle -s_g \rangle^{\rm sub}$ decreases as $N_t^{-4}$, the chiral condensate
behaves substantially different. Due to chiral symmetry\footnote{Staggered
fermions have only a remnant chiral symmetry, but this does not affect the
argument.} $\langle \bar\psi_q\psi_q \rangle$ is proportional to the bare quark
mass, moreover it is also multiplied by another power of the bare quark mass
(through the $dm_q$ line-element) in the trace anomaly. Since the bare mass (in lattice units) also
decreases with the lattice spacing, the subtracted condensate
only decreases as $N_t^{-2}$. These scalings are directly translated into different precisions
for the two terms in~(\ref{eq:tracealat}). The term with the gauge
action density has $N_t^2$ times larger errors than the term with the chiral
condensate, if they are evaluated with the same statistics. This is also reflected by the fact that in $s_g$ all of the divergences of the free energy density remain, while in $\bar\psi_q\psi_q$ the highest order term is $\mathcal{O}(a^{-2})$. Therefore the subtracted observable $\langle s_g\rangle ^{\rm sub}$ has larger cancellations as compared to $\langle \bar\psi_q\psi_q\rangle ^{\rm sub}$.

The standard integral method proceeds as follows: first one calculates the
trace anomaly for several temperatures along the lines of constant physics (LCP, see definition in section~\ref{sec:lcp})
and then integrates it to get the pressure up to an integration constant (which can
be either neglected or taken from a model calculation). This path was used in
several lattice studies, e.g. \cite{Cheng:2007jq,Bazavov:2009zn,Cheng:2009zi,Aoki:2005vt}. In the following a generalized version of the integral technique is presented.

\subsection{Multidimensional spline method}
\label{sec:spline_overview}

In order to compare the pressure to results obtained with other fermionic formulations is is useful to study the light quark dependence of $p$. A major goal of the study presented in section~\ref{sec:dyneos} was therefore to determine the EoS for
several (heavier than physical) different values of the light quark mass $m_{ud}\equiv m_u=m_d$.
To this end simulations were carried out covering an extended region in the
$\{\beta,m_s,m_{ud}\}$ parameter space. The strange mass $m_s$ was always set to its physical
value $m_s^{\rm phys}(\beta)$ along the line of constant physics.
On the other hand, $m_{ud}$ was set as
$m_{ud}(\beta)=m_s(\beta)/R$ with the quark mass ratio $R$ ranging from $1$ to
$R^{\rm phys}=28.15$. The value $R=1$ corresponds to the three degenerate
flavor case ($N_q=3$), whereas $R^{\rm phys}$ is the ``real world'' value of the quark
mass ratio ($N_q=2+1$). At these parameters we can consider the pressure as a function of
only two variables: $\beta$ and $R$. The respective derivatives can be measured
on the lattice, they are given by the following combinations:
\begin{eqnarray}
\label{eq:dbeta}
D_\beta= \left.\frac{\partial }{\partial \beta}\right|_{R}\frac{p(T)}{T^4}&=& N_t^4 \left[
\langle -s_g \rangle^{\rm sub} + 
\frac{\partial m_s^{\rm phys}}{\partial \beta}(\langle \bar\psi_{s}\psi_{s}\rangle^{\rm sub} +
\frac{1}{R}\langle \bar\psi_{ud}\psi_{ud}\rangle^{\rm sub})\right] \\
\label{eq:dR}
D_R= \left.\frac{\partial }{\partial R}\right|_{\beta}\frac{p(T)}{T^4}&=& -N_t^4
\frac{m_{s}^{\rm phys}}{R^2}\langle \bar\psi_{ud}\psi_{ud}\rangle^{\rm sub}
\end{eqnarray}

\begin{wrapfigure}{r}{7.1cm}
\centering
\vspace*{-0.8cm}
\includegraphics*[width=7.0cm]{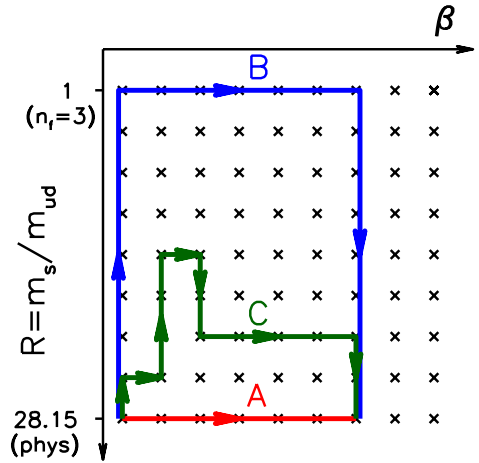}
\vspace*{-0.9cm}
\caption[Illustration of possible integration paths in the space of the bare parameters.]{Illustration of possible integration paths.}
\vspace*{-0.6cm}
\label{fig:int1}
\end{wrapfigure}

These derivatives have to be integrated to obtain the pressure. Since the
integrand is in itself -- by construction -- a total derivative, the result has to be independent of the
chosen integration path in the two-dimensional space of the parameters $\beta$ and $R$.
A few possible paths are illustrated in figure~\ref{fig:int1}. The simulation
points are denoted by crosses. A straightforward way would be to perform a
one-dimensional integral in $\beta$ at a fixed value of the quark mass ratio
$R$, this corresponds to path A.  However one can also imagine zigzagging
routes, like path B or C on the figure. Averaging several of such integrals,
one can increase statistics and furthermore also provide an estimate of the systematic
error related to the integration path.

Here I propose a generalized method that takes into account every possible
integration path at the same time. The main idea is -- instead of
parameterizing the derivatives of the pressure and then integrate -- to
parameterize the pressure itself. A straightforward parameterization is one
using the actual values $p_{kl}$ of the pressure at some node points
$\{\beta_k,R_l\}$ that build up a two-dimensional spline function -- i.e. that
unambiguously determine the whole pressure surface as a function of $\beta$ and
$R$. The spline parameters $p_{kl}$ can be set to minimize the deviation between the
derivatives of this surface and the measured values $D_\beta$ and $D_R$. This
minimum condition leads to a system of linear equations that can be solved for
$p_{kl}$. The solution for the spline parameters in turn determines the surface $P(\beta,R)$.

This method gives the pressure directly using the information
contained in the derivatives $D_\beta$ and $D_R$ without the need to carry out
an actual integration. Derivatives of the surface are then straightforward to compute, e.g. the trace anomaly can be obtained using the relation~(\ref{eq:Idef}).  
However, in order to ensure the smoothness of further derived quantities ($\epsilon$, $s$ and $c_s$), a 4-degree spline fit to the trace anomaly was carried out to give the results in section~\ref{sec:dyneos}.
Using this method we are also capable of estimating the systematic error of the result (as opposed to the conventional integral technique) by the variation of the nodepoints
of the spline interpolation. Details of the spline fitting and the determination of the systematic errors can be found in appendix~\ref{appendix1} and in~\cite{Endrodi:2010ai}.

Both the integral technique and the spline method can easily be generalized to the case of more quark flavors -- of particular interest is the contribution of the charm quark.
In order to obtain this contribution, the $\beta$ derivative of the
pressure in~(\ref{eq:dbeta}) has to be complemented by the charm
condensate. With the charm mass set according to $m_c=Q\cdot m_s$ this derivative is modified as
\be
D_{\beta} \to D_{\beta} + \frac{\partial m_s}{\partial \beta}\cdot Q \cdot \langle \bar\psi_c \psi_c \rangle^{\rm sub}
\ee

\subsection{Adjusting the integration constant}
\label{sec:adjustint}

The multidimensional spline method developed in the previous section determines the pressure only up to an overall constant factor,
which corresponds to an integration constant (this uncertainty is of course also present in the conventional integral method). 
This originates from the fact that
a constant shift of the surface leaves its gradient unchanged - i.e. a shifted solution is an equivalently good fit to the measured derivatives. 
This integration constant unfortunately cannot be determined on the lattice since
the derivatives $D_\beta$ and $D_R$ are measured at finite values of the
temperature (where the actual value of the pressure is a priori unknown).
In order to obtain a final result for the pressure this constant shift needs to be estimated; here
I present two possible approaches to do so.

\subsubsection*{Set $p$ to zero at small $T$ in $N_q=3$ theory}
\label{sec:setzero}

A straightforward way to set the integration constant is to set the pressure to zero at small temperature, that is to say, somewhere deep in the hadronic phase where contributions to $p$ are suppressed. This suppression is exponential in the hadron masses $m_H$ and the temperature: $\sim e^{-m_H T}$. Therefore it is reasonable to set the zero at larger quark masses, where hadrons are also heavier and thus the contribution to the pressure is smaller. Taking into account these considerations we set the integration constant such that at the smallest temperature for three degenerate flavors (the largest $m_{ud}$) the pressure is set to zero:
\be
\label{eq:nullpt}
p(T=100\,{\rm MeV}, R=1)= 0
\ee
The error of this choice is hard to estimate from the lattice alone. According to the HRG
model (see next subsection) however, the pressure at this point $p^{\rm HRG}(T)/T^4= 0.02$ is much smaller than the typical statistical errors on the lattice.

\subsubsection*{Set $p$ according to Hadron Resonance Gas model}
\label{sec:hrg}

Though it is possible to calculate the pressure solely using lattice methods, it is instructive to compare this result to model calculations, especially at low temperature, where lattice discretization errors are expected to be large. Moreover, there is an additional effect that may influence the pressure in the hadronic phase. This is due to the fact that the staggered formulation does not preserve the flavor
symmetry of continuum QCD, and as a consequence, the spectrum of low lying hadron
states is distorted (see section~\ref{sec:stoutsmear}). Recent analyses performed by various collaborations~\cite{Huovinen:2009yb,Huovinen:2010tv,Borsanyi:2010bp} have pointed out that
this distortion can have a dramatic impact on the thermodynamic quantities. In
order to quantify this effect, one can compare the low temperature behavior of
the observables obtained on the lattice, to the predictions of the Hadron
Resonance Gas (HRG) model.

The HRG model has been widely used to study the hadronic phase of QCD in comparison with lattice data~\cite{Karsch:2003vd,Karsch:2003zq,Tawfik:2004sw}. The model builds on the fact that the low temperature phase of QCD is dominated by pions. 
Goldstone's theorem implies weak interactions between pions at low energies, which allows to study them within chiral perturbation theory~\cite{Gasser:1983yg}.
As the temperature $T$ increases, heavier states
become more relevant and need to be taken into account. In the HRG approximation~\cite{Dashen:1969ep}, the microcanonical partition function of the interacting system can be calculated in the thermodynamic limit $V\rightarrow\infty$, assuming that it is a gas of
non-interacting free hadrons and resonances~\cite{Venugopalan:1992hy}. 
 
The pressure of the HRG model can therefore be written as the sum of
independent contributions coming from non-interacting resonances
\be
\frac{p^{\rm HRG}}{T^4} = \frac{1}{VT^3}\sum_{i\in\;mesons}\hspace{-3mm} 
\log{\Z}^{M}(T,V,\mu_{X^a},m_i) + \frac{1}{VT^3}
\sum_{i\in\;baryons}\hspace{-3mm} \log{\Z}^{B}(T,V,\mu_{X^a},m_i)
\label{eq:ZHRG}
\ee
where
\be
\begin{split}
\log{\Z}^{M}(T,V,\mu_{X^a},m_i) &=- \frac{V {d_i}}{{2\pi^2}} \int_0^\infty dk k^2
\log(1- z_ie^{- \varepsilon_i/T}) \\
 \log{\Z}^{B}(T,V,\mu_{X^a},m_i) &= \frac{V {d_i}}{{2\pi^2}} \int_0^\infty dk\, k^2
\log(1+ z_ie^{- \varepsilon_i/T})
\end{split}
\label{eq:ZMB}
\ee
with energies
$\varepsilon_i=\sqrt{k^2+m_i^2}$, degeneracy 
factors $d_i$ and fugacities
\be
z_i=\exp\left(\frac{1}{T}\sum\limits_a X_i^a\mu_{X^a} \right)
\label{eq:fuga}
\ee
In the above equation, $X^a$ are all possible conserved charges, including the
baryon number $B$, electric charge $Q$ and strangeness $S$. The sums in~(\ref{eq:ZHRG}) include all known baryons and mesons up to 2.5 GeV, as
listed in the latest edition of the Particle Data Book~\cite{Amsler:2008zzb}. The trace anomaly of the HRG model can also be calculated from~(\ref{eq:ZHRG}) using the relation~(\ref{eq:Idef}).

\begin{figure}[h!]
\centering
\includegraphics*[width=7.2cm]{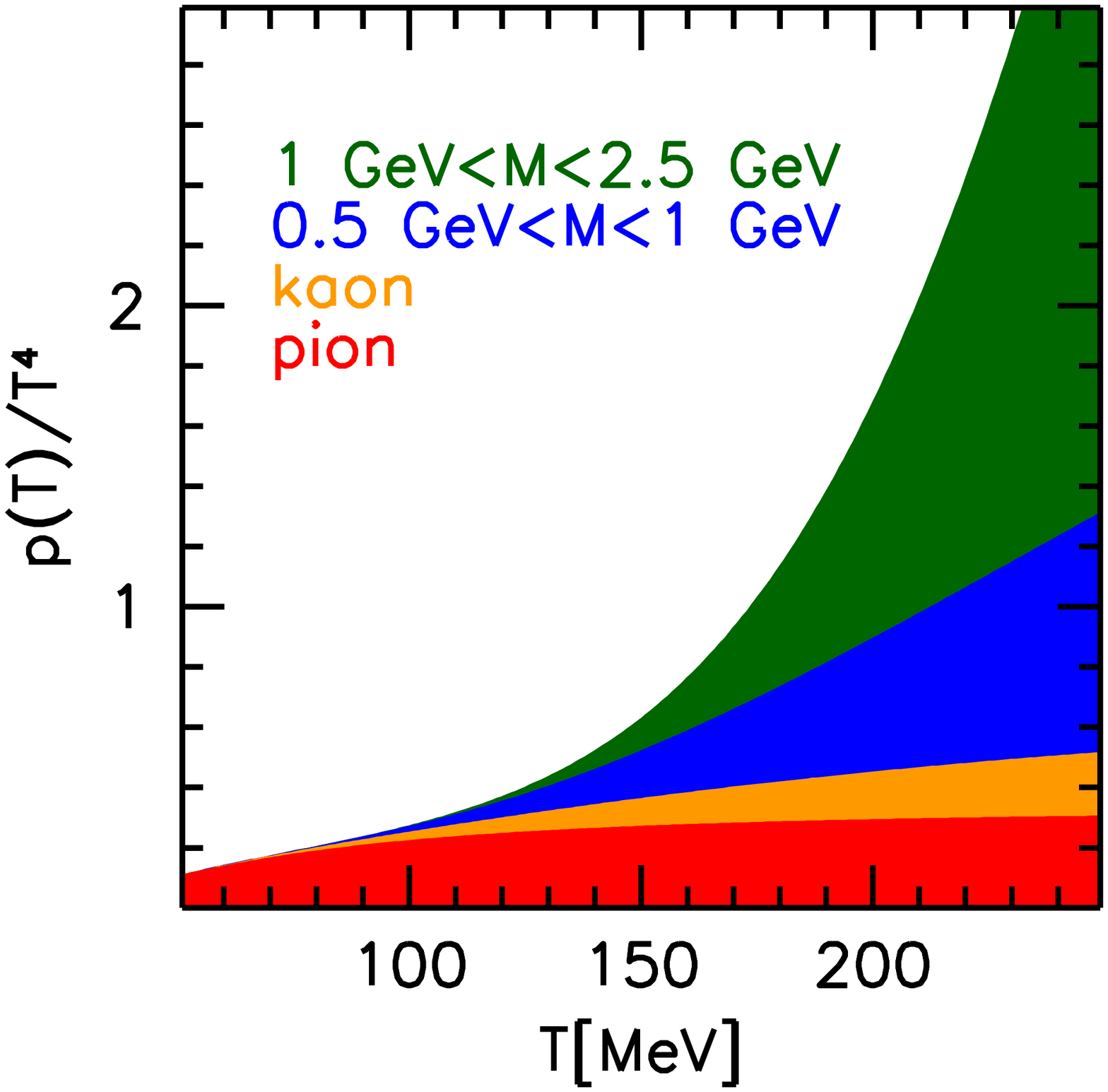} \;
\includegraphics*[width=7.2cm]{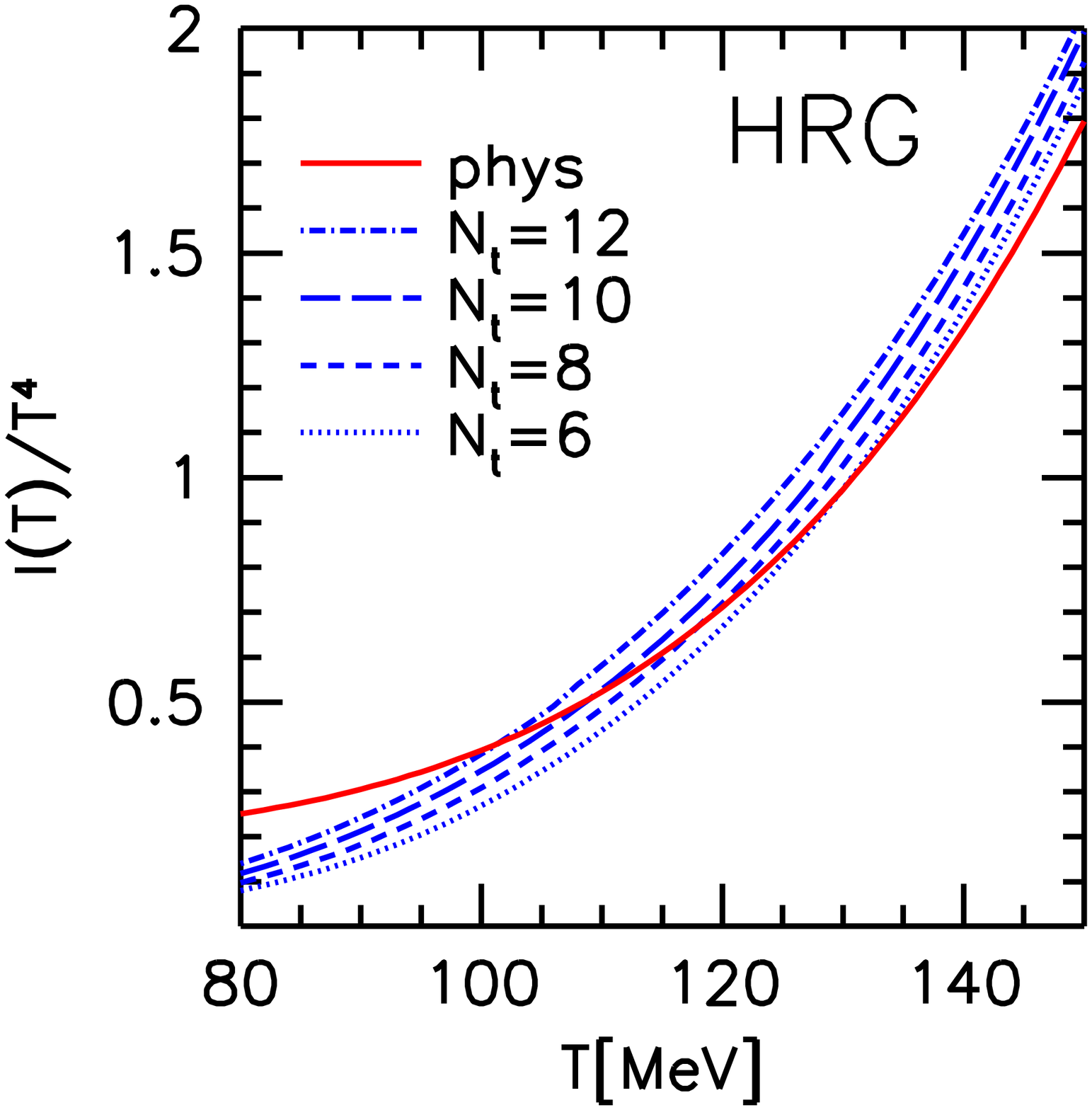}
\caption[The pressure and the trace anomaly according to the HRG and the ``lattice HRG'' model.]{Contributions to the pressure at low temperatures according to the HRG model (left side). The normalized trace anomaly in the physical HRG model (solid red line) and in the HRG models with the lattice hadron spectrum (dashed blue lines) (right side).}
\label{fig:phrgillustr}
\end{figure}

As expected, for temperatures $T\lesssim$ 60-100 MeV the contribution of the pions dominate the pressure. On the other hand, for larger temperatures the kaon contribution becomes sizable and slowly, heavier states also become relevant. The contributions coming from these sectors are illustrated by the colored bands in the left side of figure~\ref{fig:phrgillustr}. One sees that at $T\sim100$ MeV (where a comparison to lattice results is desired) the pressure is almost completely due to the pionic contribution and resonances with large widths do not play a significant role. 

As already mentioned the staggered lattice discretization has a considerable impact on 
the hadron spectrum. In
order to investigate these errors, we define a ``lattice HRG'' model, where in
the hadron masses lattice discretization effects are taken into account.  We
consider only the taste splitting effects for the pions and the kaons. For
example the contribution of the pions to the normalized pressure is now given
by a sum, which runs over the different pion tastes:
\begin{equation}
\frac{1}{VT^3}\sum_{\alpha=0}^{7} n_\alpha
\log{\Z}^{M}(T,V,\mu_{X^a},m_\alpha),
\label{eq:ps}
\end{equation}
where $n_\alpha$ are the multiplicities of the various taste sectors, and $m_\alpha$ the masses of the pion tastes (see section~\ref{sec:stoutsmear}).

In the right side of figure~\ref{fig:phrgillustr} we plot the normalized trace anomaly of the HRG model
with the physical and with the lattice distorted spectrum for the four
different lattice temporal extensions used in our investigations.  The
difference between the physical and lattice curves is a first estimate of the
lattice discretization errors arising from the taste violation. As the
temperature decreases at a fixed $N_t$, the lattice spacing gets larger and so
do the taste violation effects. The model calculation suggests, that the lattice
results may have sizeable systematic errors in the low temperature region.
Above $T\sim 100$ MeV, this error estimate for the interaction measure is
smaller than the typical magnitude of other errors in lattice QCD calculations.

\section{Systematic effects}

As it was pointed out in chapter~\ref{chap:basic}, there are two main sources of systematic error that one has to handle with care. One stems from the lattice artefacts and the other from finite size effects. In the next two subsections I will present results that support that neither of the above two systematic errors are present in our analysis of the EoS.

For high temperatures the thermodynamic quantities approach the corresponding values of the
non-interacting massless relativistic gas, i.e. the Stefan-Boltzmann (SB)
limit. For the three flavor pressure the SB value is $p_{SB}/T^4\approx 5.209$. Using~(\ref{eq:eosq}) one can also calculate the limit for the energy density $\epsilon_{SB}=3p_{SB}$, the entropy density $s_{SB}=4p_{SB}/T$ and the square of the speed of sound $c_{s,SB}^2=1/3$.

In order to decrease lattice artefacts -- besides using an improved action -- we also carry out an improvement on the level of the observables.
In lattice thermodynamics the cutoff effects in general depend on $N_t$, however, for $N_t\to\infty$ they disappear. In the following we use a tree-level improvement for the pressure: we divide the lattice results by the appropriate improvement coefficients $\kappa(N_t)$. At the tree-level these coefficients are proportional to the value of the pressure as measured in the noninteracting system on a lattice with finite temporal extent $N_t$:
\be
\kappa(N_t)=\frac{p_{SB}(N_t)}{p_{SB}}
\ee
For the action of our choice the $\kappa(N_t)$ factors are calculated to be:
\be
\begin{array}{|c|c|c|c|}
\hline
N_t=6&N_t=8&N_t=10&N_t=12\\
\hline
1.517&1.283&1.159&1.099\\
\hline
\end{array}
\nonumber
\ee
We use the same factor for the two different spatial volumes in our finite
volume study ($6\times 18^3$ and $6\times36^3$). Using thermodynamic
relations one can obtain these improvement coefficients for the energy density,
trace anomaly and entropy, too. The speed of sound receives no improvement
factor at tree level.

\begin{figure}[h!]
\centering
\vspace*{-0.1cm}
\includegraphics*[width=7.0cm]{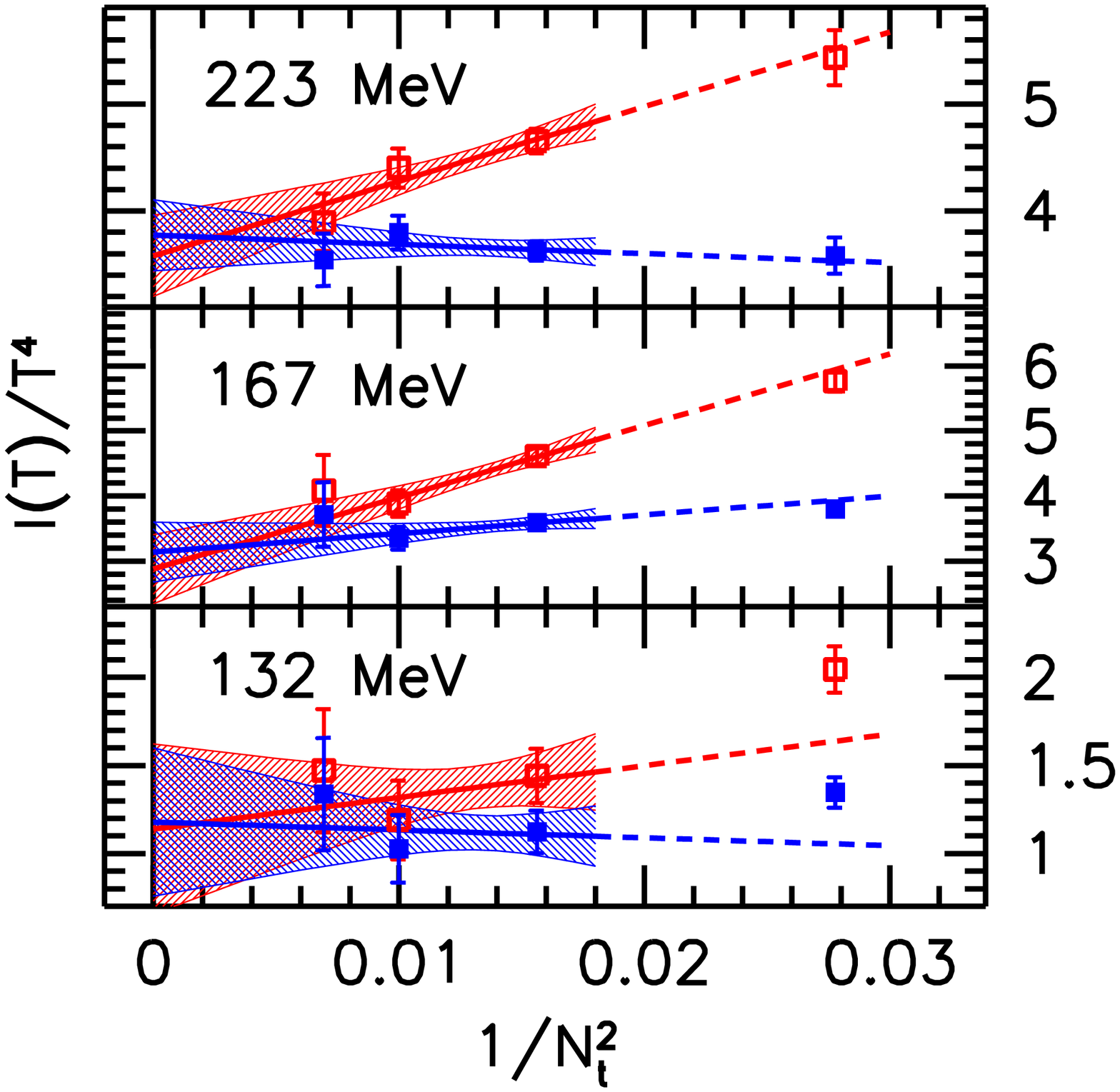} \quad
\includegraphics*[width=7.0cm]{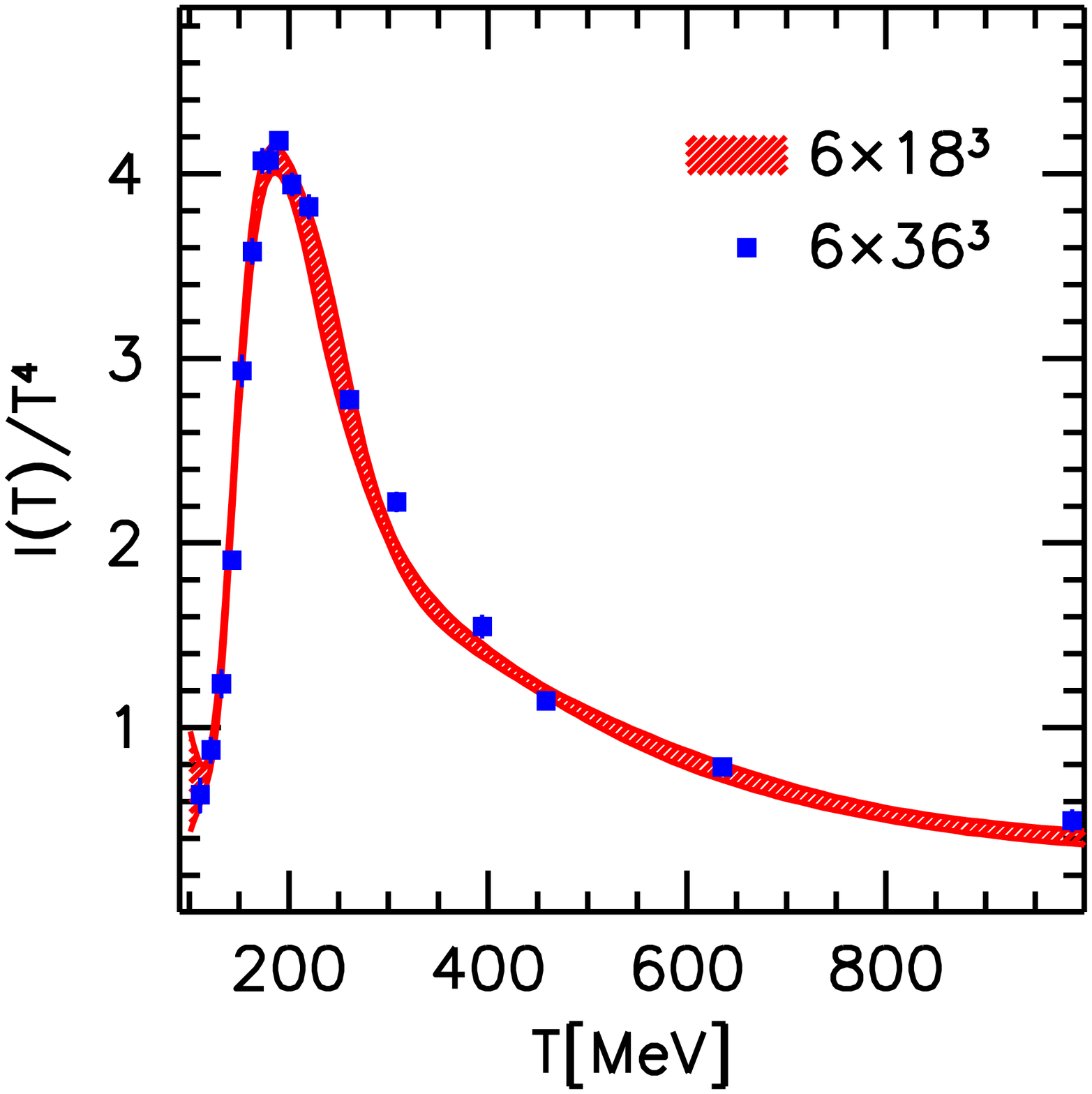}
\vspace*{-0.1cm}
\caption[Finite volume errors in the trace anomaly and continuum scaling of the pressure.]{The normalized trace anomaly at three different temperatures as a function of $1/N_t^2$. Filled red symbols represent the results within the lattice tree-level improvement framework, blue opened symbols show the results without this improvement. The error of the continuum extrapolated value is about $0.4$ for all three temperatures (left panel). The trace anomaly on lattices with different spatial volumes:
red band for $N_s/N_t=3$ and blue points for $N_s/N_t=6$ (right panel).}
\label{fig:finV2}
\end{figure}

In the left side of figure~\ref{fig:finV2} we illustrate at three temperature values ($T=132$, $167$ and $223$ MeV) the effectiveness of this improvement procedure. We show both the
unimproved and the improved values of the normalized trace anomaly $I/T^4$ for $N_t=6,8,10$ and $12$ as a function of $1/N_t^2$. The lines are linear continuum extrapolations\footnote{Note that the staggered fermionic action approaches the continuum action as $\sim a^2$. This scaling motivates that the extrapolation is linear in $1/N_t^2\sim a^2$.}
using the three smallest lattice spacings. As visible in the figure, the continuum
limit of both the unimproved and the improved observables agree well. Thus, our results confirm the expectations, that the lattice tree-level improvement effectively reduces cutoff effects. At all three temperatures the unimproved observables have larger cutoff effects (i.e. larger deviation from the continuum value) than the improved ones.

Note that the improvement coefficients $\kappa(N_t)$ are exact only
at tree-level, thus in the infinitely high temperature, non-interacting case.
As we decrease the temperature, corrections to these improvement coefficients
appear, which have the form $1+b_2(T)/N_t^2+...$. Empirically one finds that
the $b_2(T)$ coefficient, which describes the size of lattice artefacts of the
tree-level improved quantities, is tiny not only at very high temperatures,
but throughout the deconfined phase. 
In particular, for the case of the three temperatures considered in the left panel of figure~\ref{fig:finV2}, the $b_2(T)$ coefficients are found to differ from zero with less than one standard deviation.

The second possible systematic error is that connected to the finite lattice size.
In order to verify that there are no significant finite size effects present in
the lattice data with the aspect ratio $N_s/N_t=3$, we checked our $N_t=6$ data
against a set of high precision $N_s/N_t=6$ simulations. The latter lattice geometry
corresponds to about 7~fm box size at the transition temperature. The right panel of figure~\ref{fig:finV2} shows the comparison between the two volumes for the
normalized trace anomaly $I/T^4$. From this result we conclude that it is
acceptable to perform the more expensive simulations throughout with
$N_s/N_t=3$.  Let us note here, that the volume-independence in the
transition region is an unambiguous evidence for the crossover type of the
transition.

\section{Results}
\label{sec:dyneos}

In this section I present results on the equation of state of $N_q=2+1$ QCD. The pressure, the interaction measure, the energy and entropy
density as well as the speed of sound are shown as a function of the temperature.
The characteristic points of these observables are identified and a parametrization for the trace anomaly is presented. Afterwards I study the dependence of the pressure on the light quark mass $m_{ud}$. Furthermore, the contribution of the charm quark to the pressure is discussed within the partially quenched framework. Finally a comparison to results obtained with other fermionic discretizations is carried out.

The multidimensional spline method (see section~\ref{sec:spline_overview} and appendix~\ref{appendix1}) was used to determine the equation of state on
$N_t=6,8$ lattices for various different values of the quark masses. In the case of
$N_t=10$ we determined the equation of state exclusively with physical quark
masses. In order to satisfy the normalization condition of~(\ref{eq:nullpt}), we made heavier mass simulations at $T=100$ MeV up to the three
flavor point. For $N_t=12$ we made simulations at three temperature
values, this allows us to calculate the trace anomaly only.

The error bars on the figures are obtained by quadratically adding the statistical error and the systematic error of the spline method (see appendix~\ref{appendix1}). The temperature values have an error at the $2\%$ level arising from the scale setting procedure (see section~\ref{sec:lcp}). 

\subsection{The \texorpdfstring{$N_q=2+1$}{nq} flavor equation of state}

\begin{figure}[h!]
\centering
\vspace*{-0.2cm}
\epsfig{file=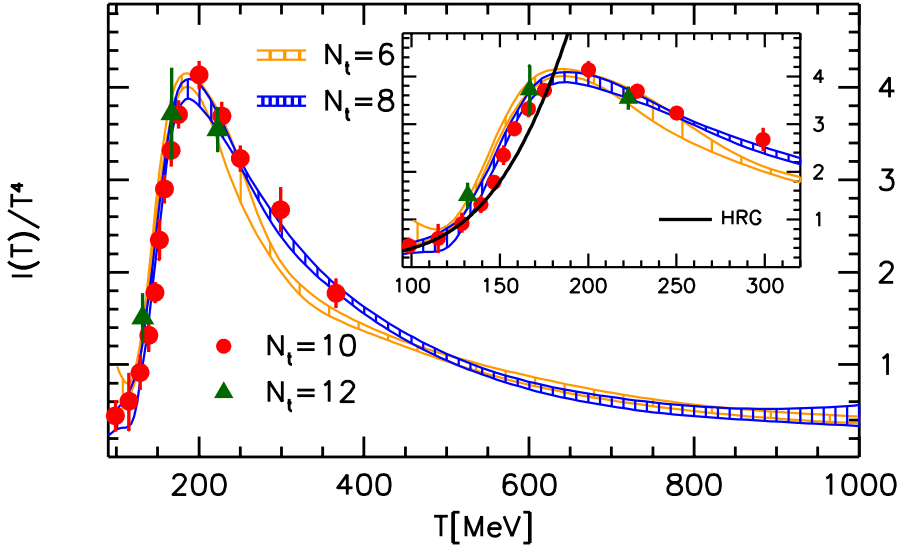,width=12.7cm,bb=18 360 592 718}
\vspace*{-0.1cm}
\caption[The normalized trace anomaly as a function of the temperature.]{
The trace anomaly $I=\epsilon-3p$ normalized by $T^4$ as a function of the temperature on $N_t=6,8,10$ and $12$ lattices.
}
\label{fig:eos_I}
\end{figure}
In figure~\ref{fig:eos_I} the normalized trace anomaly is shown as a function of the temperature. As it can be seen in the figure, results show essentially no dependence on the lattice spacing, as all four datasets lie on top of each other. Only the coarsest $N_t=6$ lattice shows some deviations around $\sim 300$ MeV. In the same figure, I also provide a zoom of the transition region; here the results from the HRG model are also plotted. A good agreement with the lattice results is found up to $T\sim 140$ MeV.

One characteristic temperature of the crossover transition
can be defined as the inflection point of the trace anomaly. This and other
characteristic features of the trace anomaly are summarized in the following table:
\begin{center}
\begin{tabular}{|l|r|}
\hline
Inflection point of $I(T)/T^4$&152(4) MeV\\
Maximum value of $I(T)/T^4$&4.1(1)\\
$T$ at the maximum of $I(T)/T^4$&191(5) MeV\\
\hline
\end{tabular}
\end{center}

\begin{figure}[h!]
\centering
\vspace*{-0.15cm}
\epsfig{file=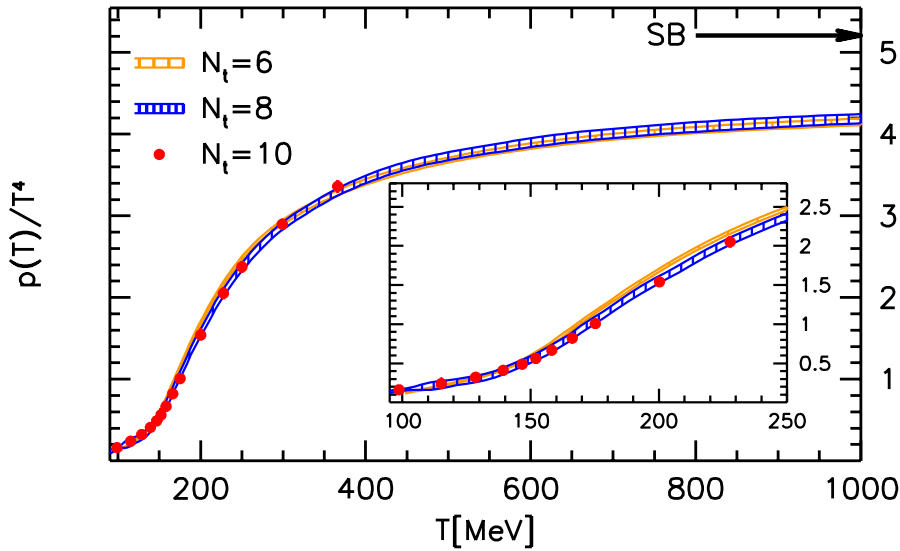,width=12.7cm,bb=18 360 592 718}
\caption[The normalized pressure as a function of the temperature.]{
The pressure normalized by $T^4$ as a function of the temperature on $N_t=6,8$ and $10$ lattices. The Stefan-Boltzmann limit $p_{SB}(T) \approx 5.209 \cdot T^4$ is indicated by an arrow.
}
\label{fig:eos_p}
\end{figure}

In figure~\ref{fig:eos_p} I show the main result of the present study: the pressure of
$N_q=2+1$ QCD as a function of the temperature.
Here results for three different lattice spacings are presented. The $N_t=6$ and $N_t=8$
are in the temperature range from 100 up to 1000 MeV. The results on $N_t=10$
are in the range from 100 up to 365 MeV. The zero point of the pressure is set by~(\ref{eq:nullpt}), as supported by the discussion in subsection~\ref{sec:setzero}. Note that from this condition we get a nonzero pressure for the $N_q=2+1$ system (i.e. $R=28.15$, cf.~(\ref{eq:nullpt})) even at the smallest temperature $T=100$ MeV.

The value of the pressure at 100 MeV is approximately two third of the value suggested by the HRG model. The origin of this difference cannot be clarified at the moment.
One expects that the lattice artefacts are considerably larger at low
temperatures, than what one estimates from the difference of $N_t=6,8$ and $10$
results. This is quite reasonable, since even our finest lattice at $T=100$ MeV
has $\sim 0.2$ fm lattice spacing, which is far from the regime, where lattice
results starts to scale. On the other hand, this discrepancy might also point to the
failure of the HRG model. In order to be on the safe side the size
of this unexplained difference is considered as an estimate of our systematic uncertainty in
the low temperature regime.

\begin{figure}[h!]
\centering
\vspace*{-0.15cm}
\epsfig{file=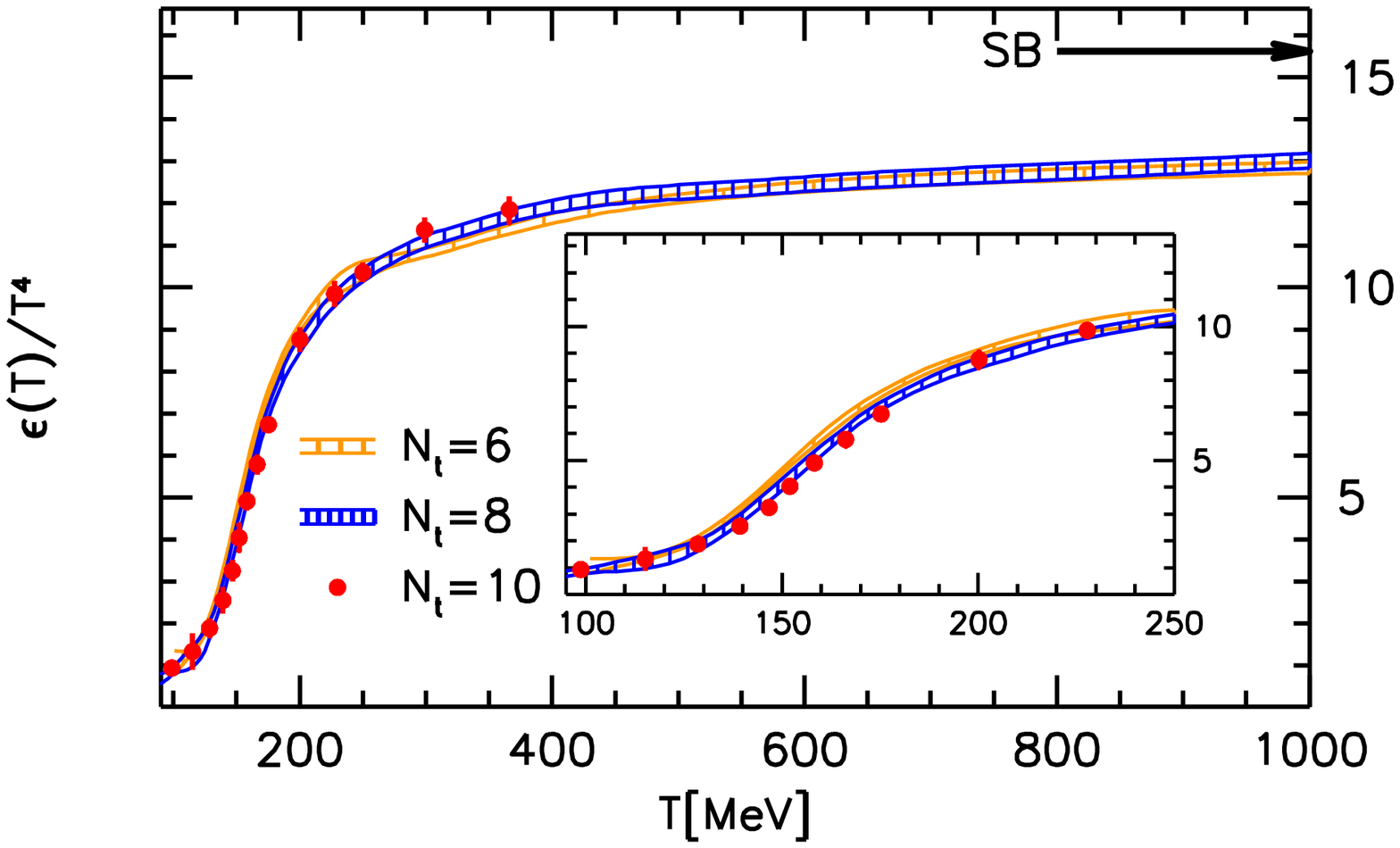,width=12.7cm,bb=18 360 592 718}
\epsfig{file=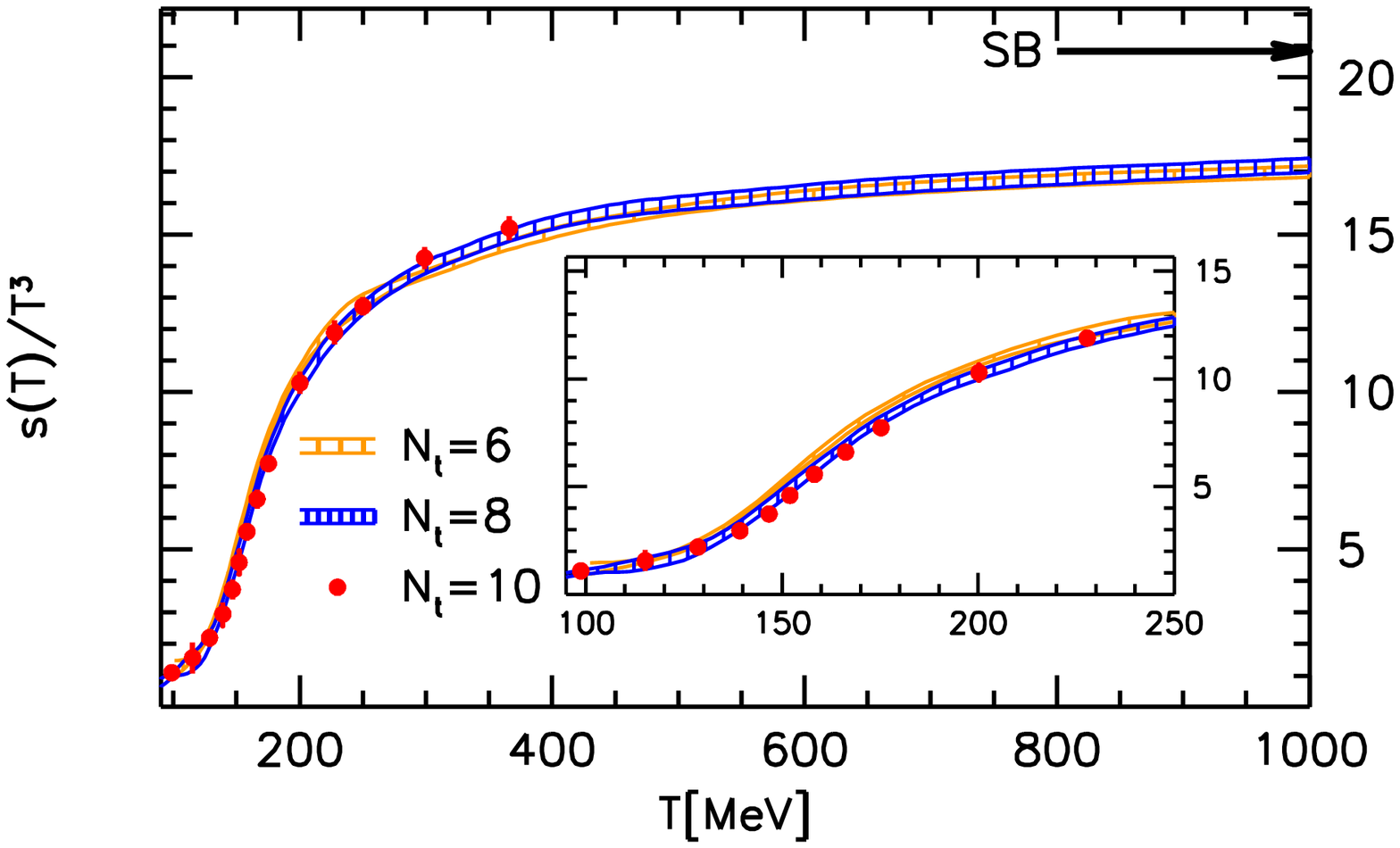,width=12.7cm,bb=18 360 592 718}
\caption[The normalized energy density and entropy density as functions of the temperature.]{
The normalized energy density and entropy density as functions of the temperature
on $N_t=6,8$ and $10$ lattices. The SB limits $\epsilon_{SB}= 3p_{SB}$ and $s_{SB}= 4p_{SB}/T$ are indicated by the arrows.
}
\label{fig:eos_es}
\end{figure}

\begin{figure}[h!]
\centering
\vspace*{-0.15cm}
\epsfig{file=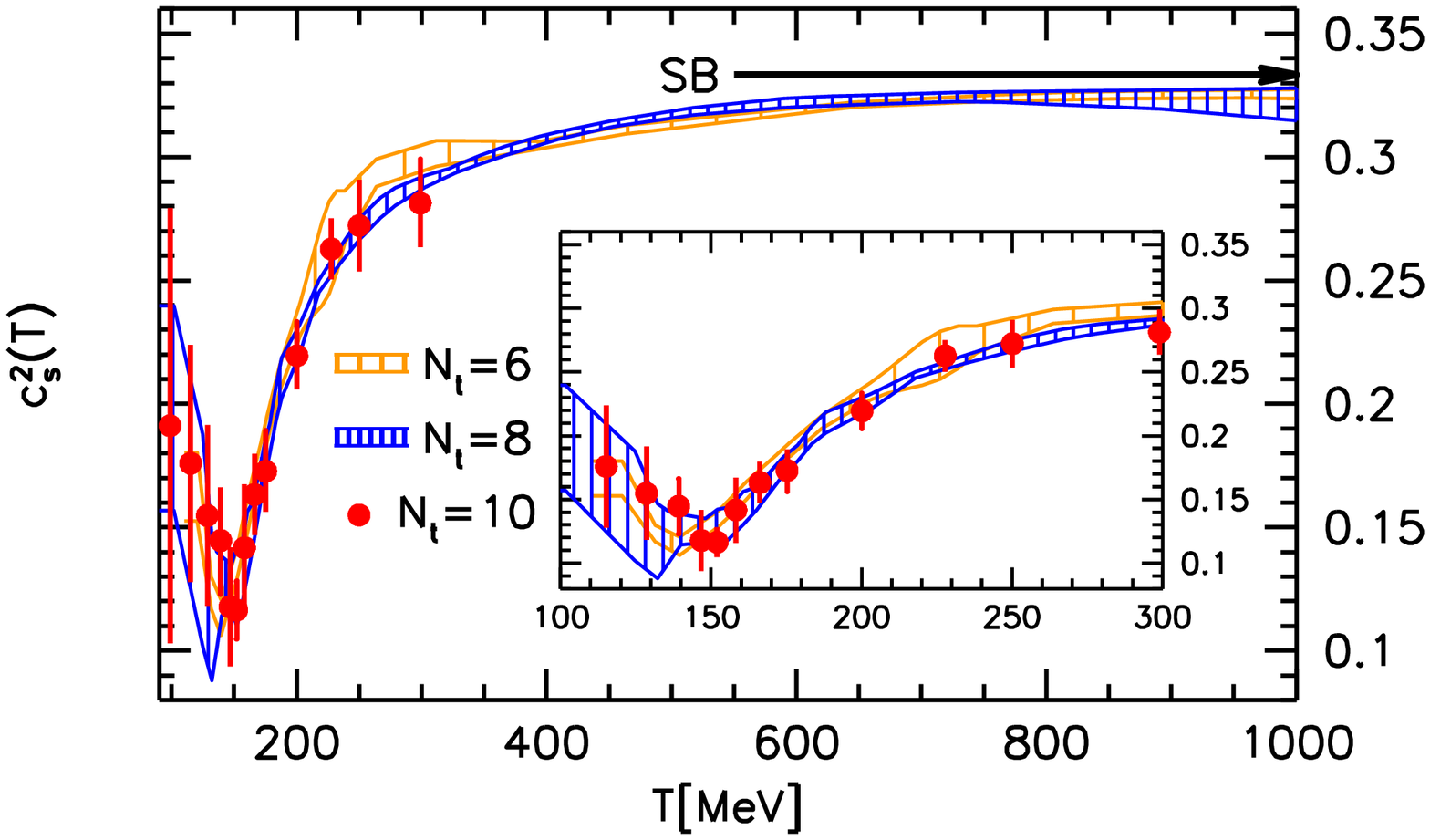,width=12.7cm,bb=18 360 592 718}
\epsfig{file=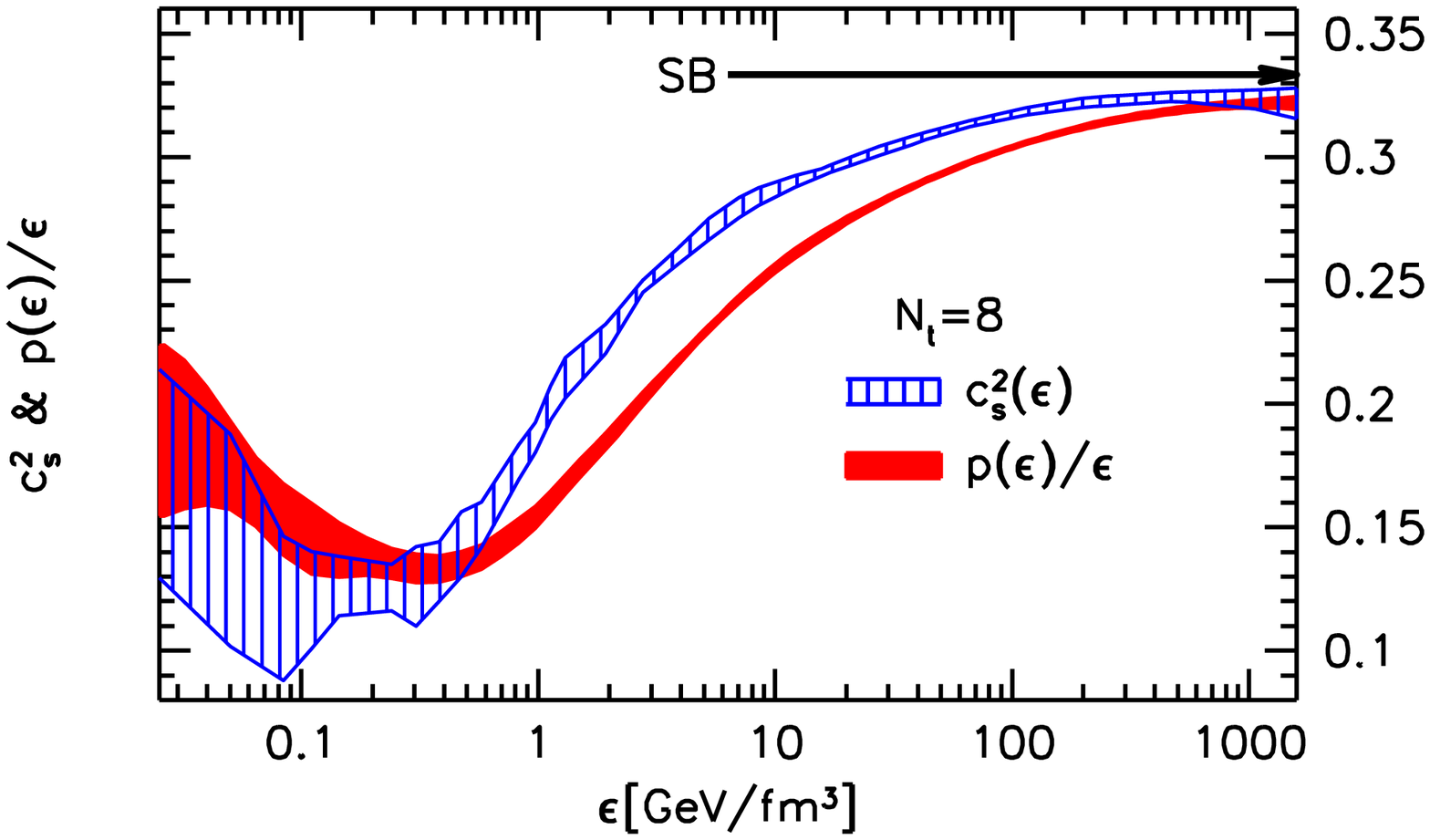,width=12.7cm,bb=18 360 592 718}
\caption[The square of the speed of sound and the ratio $p/\epsilon$ as functions of the temperature.]{
The square of the speed of sound as a function of the temperature on $N_t=6,8$ and $10$ lattices (upper panel) and $c_s^2$ and $p/\epsilon$ as a function of the energy density on $N_t=8$ lattices (lower panel). The SB limit of $1/3$ for both quantities is indicated by the arrows.
}
\label{fig:eos_cspe}
\end{figure}

In figure~\ref{fig:eos_es} I show the energy density (upper panel) and
the entropy density (lower panel), while in the upper panel of figure~\ref{fig:eos_cspe} the square of the speed of sound as a function of the temperature. Finally, in the lower panel of figure~\ref{fig:eos_cspe} the speed of sound and the ratio $p/\epsilon$ are presented as functions of the energy density.

Some characteristic properties of the speed of sound and the ratio $p/\epsilon$ are also of phenomenological interest. These are summarized in the following table:
\begin{center}
\begin{tabular}{|l|r|}
\hline
Minimum value of $c_s^2(T)$&0.133(5)\\
$T$ at the minimum of $c_s^2(T)$&145(5) MeV\\
${\epsilon}$ at the minimum of $c_s^2(T)$&0.20(4) GeV/fm$^3$\\
\hline
Minimum value of $p/\epsilon$&0.145(4)\\
$T$ at the minimum of $p/\epsilon$&159(5) MeV\\
${\epsilon}$ at the minimum of $p/\epsilon$&0.44(5) GeV/fm$^3$\\
\hline
\end{tabular}
\end{center}

\subsection{Continuum estimate and parametrization}
Based on the results for various lattice spacings we can also provide a continuum estimate
\footnote{For a rigorous continuum extrapolation one would need $N_t=12$ for the entire temperature region.}
for the quantities presented above. We take the average of the data at the smallest two lattice spacings and as an error we assign either the difference of the two or the statistical error depending on whichever is larger. As already mentioned, for low temperatures the lattice result for the pressure is significantly smaller than the prediction of the HRG model: at $T=100$ MeV the lattice result is $p(T)/T^4=0.16(4)$, whereas the model prediction is $p(T)/T^4=0.27$.  Therefore in the continuum estimate of the pressure we shift the central values of the lattice results up by the half of this difference ($0.06$) and this shift is then considered as a systematic error in the entire temperature range.

\begin{figure}[h!]
\centering
\vspace*{-0.15cm}
\epsfig{file=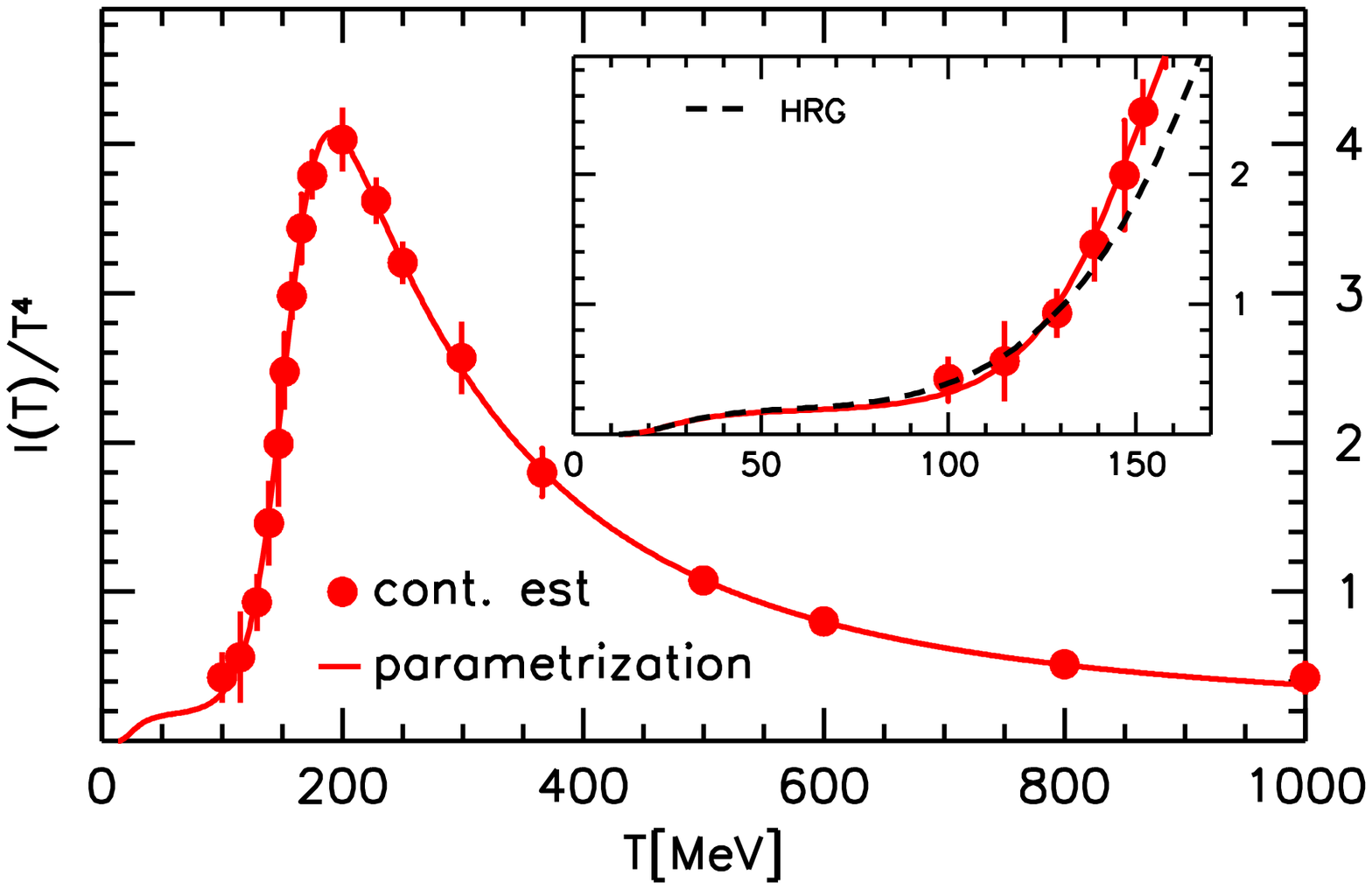,width=12.5cm,bb=18 360 592 718}
\caption[Continuum estimate and global parametrization of the trace anomaly.]{
Continuum estimate for the trace anomaly normalized by $T^4$ together with the parametrization of~(\ref{eq:par}) using the $N_q=2+1$ parameters from table~\ref{tab:par}.
}
\label{fig:par}
\end{figure}

The trace anomaly as a function of $T$ plays a central role in hydrodynamic models and thus we find it useful to give a global parametrization that describes $I/T^4$ in the whole temperature range considered. We considered the following fit function:
\be
\label{eq:par}
\frac{I(T)}{T^4}= 
\exp(-h_1/t - h_2/t^2)\cdot \left( h_0 + \frac{f_0\cdot [\tanh(f_1\cdot t+f_2)+1]}{1+g_1\cdot t+g_2\cdot t^2} \right),
\ee
where the dimensionless $t$ variable is defined as $t=T/(200 {\rm MeV})$. 
This function reproduces the continuum estimate for the normalized trace anomaly
in the entire temperature range $T=100\ldots 1000$ MeV.
The $\{f_0,f_1,f_2\}$ parameters describe the steep rise of the trace anomaly
in the transition region, whereas the $\{g_1,g_2\}$ parameterize the decrease
for higher temperatures. The parametrization also approximates the HRG model
prediction for $T<100$ MeV, this is described by the $\{h_0,h_1,h_2\}$
parameters. The parameters can be found in table~\ref{tab:par}.
\begin{table}[h!]
\centering
\begin{tabular}{|c||c|c|c|c|c|c|c|c|}
\hline
$N_q$ & $h_0$ & $h_1$ & $h_2$ & $f_0$ & $f_1$ & $f_2$ & $g_1$ & $g_2$ \\
\hline
$2+1$ & \multirow{2}{*}{0.1396} & \multirow{2}{*}{-0.1800} & \multirow{2}{*}{0.0350} & 2.76 & 6.79 & -5.29 & -0.47 & 1.04\\
\cline{1-1}\cline{5-9}
$2+1+1$ & & & & 5.59 & 7.34 & -5.60 & 1.42 & 0.50\\
\hline
\end{tabular}
\caption{
Parameters of the function in~(\ref{eq:par}) describing the trace anomaly in the $N_q=2+1$
and in the $N_q=2+1+1$ flavor cases.
}
\label{tab:par}
\end{table}

\noindent
For these temperatures the difference in the trace anomaly between
the parametrization and the HRG model is less than $\Delta(I(T)/T^4) \le 0.07$.
The parametrization together with our continuum estimate for the trace anomaly is shown in figure~\ref{fig:par}.
From this parametrization the normalized pressure can be obtained by
the definite integral of~(\ref{eq:pfromI}).
The so obtained function goes through the points of the continuum estimate of the pressure for temperatures $T=100\ldots 1000$ MeV and for $T<100$ MeV the deviation from the
HRG prediction is less than $\Delta(p(T)/T^4) \le 0.02$. 

\subsection{Light quark mass-dependence}

The multidimensional spline method gives the pressure (and the derived quantities) as a smooth function of both the temperature and the light quark mass. Thus it becomes possible to investigate the dependence of the EoS on $m_{ud}$.

\begin{figure}[h!]
\centering
\epsfig{file=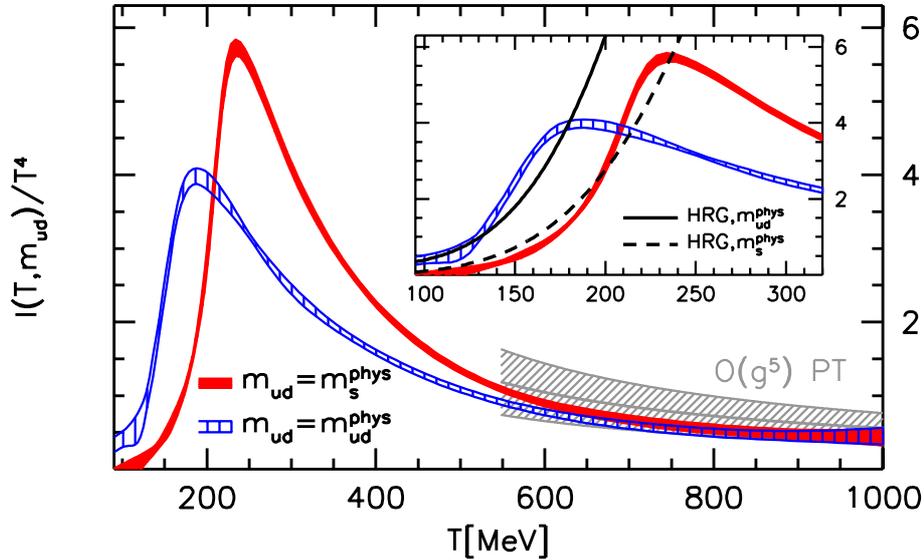,width=13cm,bb=18 360 592 718}
\caption[Light quark mass dependence of the trace anomaly.]{
The normalized trace anomaly for two different values of the light quark masses on $N_t=8$
lattices: the physical $m_{ud}= m_{ud}^{\rm phys}$ and the three degenerate
flavor $m_{ud}= m_s^{\rm phys}$ case.
}
\label{fig:mdep}
\end{figure}

In figure~\ref{fig:mdep}, the trace anomaly for two different values of the
light quark masses is plotted: for the physical case, where $m_{ud}= m_{ud}^{\rm phys}$
and for the three degenerate flavor case, where $m_{ud}= m_s^{\rm phys}$ (i.e. $R=28.15$ and $R=1$ in the notation of section~\ref{sec:spline_overview}, respectively). This latter case corresponds to a pion mass of approximately $m_\pi \sim 720$ MeV.
The results are from
our $N_t=8$ lattices, this is the smallest lattice spacing, where we have the
complete mass dependence of the equation of state.  As it is expected, the peak
position of the trace anomaly is shifted towards higher temperature values for
larger quark masses. The position in the three degenerate flavor case is $\sim
25$\% larger than at the physical point.  The height also increases by about
$\sim 40$\%. When zooming into the transition region, we also show the
comparison with the HRG model. For low temperatures one finds a reasonable
agreement also in the heavy quark mass case. As it is expected, the dependence
on the quark masses vanishes as one goes to higher temperatures. Therefore it
is plausible to compare the result with massless perturbation
theory. A good agreement can be observed with the highest order perturbative
calculation without non-perturbative input ($\mathcal{O}(g^5)$, see figure~\ref{fig:mdep}).

\subsection{Estimate for the \texorpdfstring{$N_q=2+1+1$}{nqq} flavor equation of state}

While at low temperatures the equation of state only contains terms that
originate from the light quarks, in high energy processes charm quarks can also
be created from the vacuum, and they can also be present in the initial or
final states.
An interesting issue regarding the EoS is whether the
charm quark can give an important contribution to the QCD EoS, in the range of
temperatures which are reached in heavy ion collisions. It is often assumed
that it can be neglected, the charm mass being too heavy to play any role at
$T\simeq 2T_c$. However, perturbative QCD predicts that its contribution to
thermodynamic observables is relevant at surprisingly low temperatures, down to
$T\simeq 350$ MeV \cite{Laine:2006cp}. Recent exploratory lattice studies have
confirmed these expectations \cite{Cheng:2007wu,Levkova:2009gq}, indicating a
non-negligible contribution of the charm quark to thermodynamics already at
$1.5 T_c$. These results have been obtained on rather coarse lattices
($N_t=4,~6$) and with the charm quark treated in the quenched approximation.

\begin{figure}[h!]
\centering
\epsfig{file=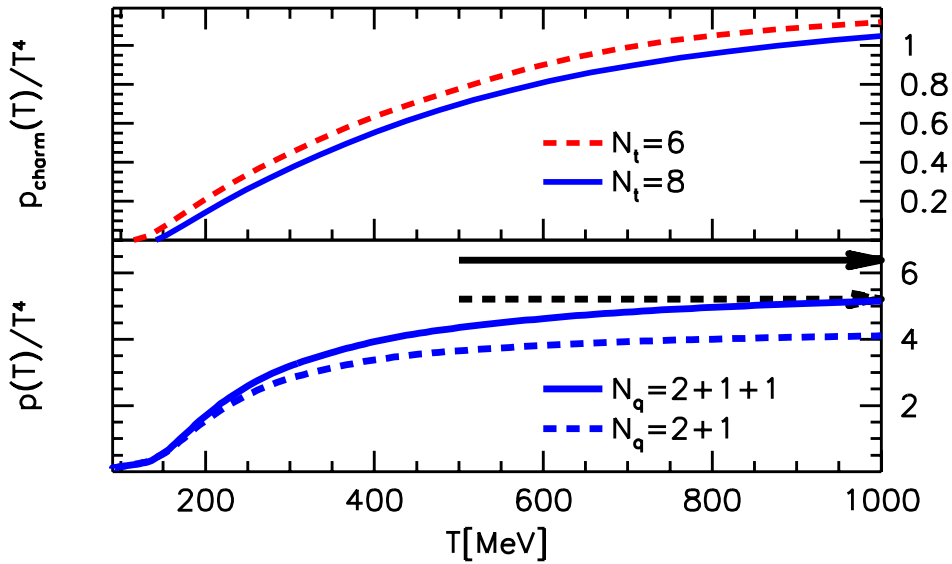,width=13cm,bb=18 360 592 718}
\caption[Contribution of the charm quark to the pressure for two different lattice spacings.]{Contribution of the charm quark to the pressure for two different lattice spacings (upper panel). The pressure normalized by $T^4$ for
$N_q=2+1+1$ and $N_q=2+1$ flavors on $N_t=8$ lattices (lower panel). The corresponding Stefan-Boltzmann limits are indicated by arrows. The charm to strange quark mass ratio is $Q=11.85$ on this plot.}
\label{fig:charm}
\end{figure}

We estimated the contribution of the
charm quark to the pressure on our $N_t=6$ and $8$ lattices at the partially quenched level (i.e. we used the same configurations as for the $N_q=2+1$ case).
The measurements were carried out for several values of the charm to
strange quark mass ratio $Q=m_c/m_s$. According to a recent high-precision lattice
calculation~\cite{Davies:2009ih} the physical value of this ratio is $Q^{\rm
phys}=11.85(16)$. For this central value we show the contribution $p_{\rm charm}$ as a
function of the temperature on the upper panel of figure~\ref{fig:charm}. We find that $p_{\rm charm}$ is non-zero already at temperatures $T\sim 200$ MeV. The total $N_q=2+1+1$
pressure is compared to the already presented $N_q=2+1$ pressure on the lower
panel of figure~\ref{fig:charm}. Using the parametrization in~(\ref{eq:par}) we also fit the $N_q=2+1+1$ data for the trace anomaly for the $N_t=8$ data. The resulting fit parameters can be found in table~\ref{tab:par}.

It is important to stress here that the estimate
for the charm contribution presented here suffers from two uncertainties. First, we
neglected the back-reaction of the charm quarks on the gauge field as the charm condensate was measured at the partially quenched level. Second, due to the large mass of the charm, large lattice artefacts can also be expected.
In order to provide a more precise calculation of the $N_q=2+1+1$ flavor
equation of state, one has to
relax the uncontrolled partial quenched approximation and introduce the charm
quark dynamically. Moreover, the lattice spacing has to be further reduced to
ensure a good resolution of the charmed excitations. The reduction of lattice artefacts can also be done by using improved fermion
actions, therefore these are also increasingly important when studying heavy quark degrees of freedom.\footnote{We remark that preliminary results regarding the dynamical contribution of the charm quark indicate that the above results may overestimate the $N_q=2+1+1$ pressure.
}

\subsection{Comparison with different fermion discretizations}

In the past years, the equation of state of QCD was studied with various different fermion discretizations. Among these are improved staggered actions like the p4, the asqtad and the hisq action (see section~\ref{sec:impract}). The disagreement in thermodynamic quantities between the previous two formulations and the stout smeared action used in the present thesis has a broad literature~\cite{Aoki:2009sc,Huovinen:2009yb,Borsanyi:2010bp}. The main difference can be described by a $\sim$ 20-30 MeV shift in the temperature. In particular, the transition temperatures obtained from various thermodynamic observables like the renormalized chiral condensate and the quark number susceptibilities differ significantly. 

\begin{figure}[h!]
\begin{center}
\epsfig{file=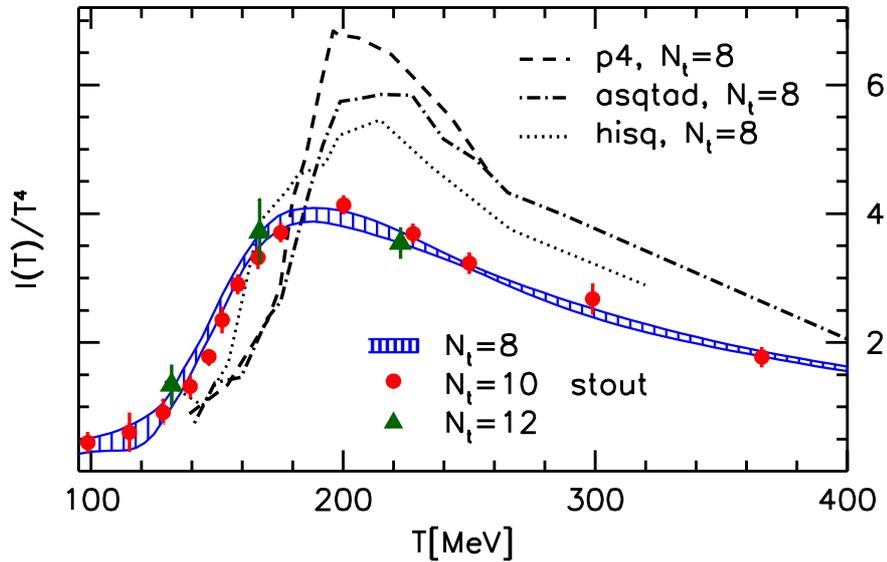,width=13cm,bb=18 360 592 718}
\caption[Comparison to recent results with different fermionic actions.]{
The normalized trace anomaly is compared to recent results obtained with different fermionic actions~\cite{Bazavov:2009zn,Cheng:2009zi,Bazavov:2010sb}.
}
\label{fig:cmp}
\end{center}
\end{figure}

It is interesting to look for this discrepancy in the equation of state as
well. In~\ref{fig:cmp} I show our results for the trace anomaly and a comparison with results obtained the p4, the asqtad and the hisq action. The data for the latter three are taken from~\cite{Bazavov:2009zn},~\cite{Cheng:2009zi} and~\cite{Bazavov:2010sb}, respectively. As it can be clearly seen, the upward going branch and the peak position are located at $\sim$ 20 MeV
higher temperatures for the p4 and the asqtad actions as compared to the stout results. Furthermore, the height of the peak is also about 50\%
larger for the former two. One can also note that a weaker disagreement is observed between the stout and the hisq results.
The origin of this discrepancy is most probably connected to lattice discretization errors and artefacts resulting from the taste splitting of the staggered tastes. Accordingly, results with both actions, as measured with smaller lattice spacings and smaller splitting seem to approach each other so that this discrepancy may be resolved in the near future.

\chapter{The equation of state at high temperatures}
\label{chap:PG}

Although results from RHIC support the idea that the high temperature QGP behaves like a strongly coupled liquid, there are observed phenomena (like jet quenching or elliptic flow) that can be described well also by perturbative methods (i.e. with weak couplings). The question, whether the deconfined phase of quarks and gluons is more like a weakly coupled plasma or a strongly coupled liquid is thus of high relevance. A possible way to test the validity of perturbative approaches is to make a comparison with non-perturbative results obtained from the lattice. However, while standard perturbation theory converges only at extremely high temperatures, available lattice results end at around $(5-10)\cdot T_c$. Including higher order terms in the expansion and applying resummation techniques can certainly push the lower limit of the validity of perturbation theory to smaller temperatures. Nevertheless, in order to carry out a proper comparison to perturbation theory, lattice measurements also have to be extended to higher temperatures. In the pure gluonic sector this has become recently possible.
In this chapter I show results obtained in the $SU(3)$ theory with the Symanzik improved gauge action up to previously unreached temperatures and also make a comparison to various perturbative approaches.

\section{Perturbative methods}

Perturbation theory has now been long used to calculate thermodynamic quantities in the weakly interacting (high temperature) limit. Most importantly, the pressure has been studied and recently determined up to order $\O(g^6 \log(g))$~\cite{Kajantie:2002wa}. Apparently, perturbative precision cannot be pushed further to higher loops due the appearance of serious infrared divergences. These divergences are in general characteristic to finite temperature field theories. In particular, these problems prevent one to calculate the pressure in the order $\O(g^6)$, namely there is an unknown coefficient in this contribution. Unfortunately, even computed to this high order, the expansion only converges at extremely high temperatures, since at low temperatures the various terms oscillate strongly and thus can produce large cancellations.

The disagreement between lattice data and standard perturbation theory has also been studied for other quantities like the expectation value $\expv{A_\mu^2}$ at large $\beta$ i.e. small lattice spacings~\cite{Lepage:1992xa}. The deviation from non-perturbative results apparently stems from the fact that the {\it bare} coupling constant is used as an expansion parameter in the perturbative series. One could very well define the perturbative expansion also in terms of a renormalized coupling constant $g_r$, which can be defined using some physical observable. Therefore, $g_r$ will be a running coupling $g_r(\mu)$, which depends on the relevant energy scale i.e. the renormalization scale $\mu$\footnote{The former identification of $\mu$ with the chemical potential is abandoned from now on.}.

From a different point of view, the reason for the poor convergence of perturbative series can also be due to various plasma effects that influence the high temperature description of the system. These effects can be included by means of a systematic resummation of the perturbative series, like in e.g. the Hard Thermal Loop (HTL) approach~\cite{Andersen:2010ct}. Whether such resummed expansions do reproduce lattice results at reasonably small values of the temperature is an open question and is one of the main aims of our study.

\begin{wrapfigure}{r}{8.4cm}
\centering
\vspace*{-0.1cm}
\includegraphics*[width=7.8cm]{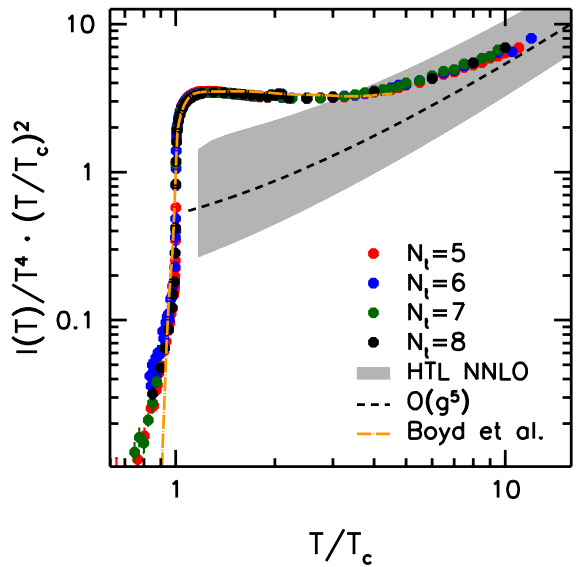}
\vspace*{-0.1cm}
\caption[Lattice results and perturbative predictions for the normalized trace anomaly multiplied by $T^2/T_c^2$.]{Our results for the normalized trace anomaly multiplied by $T^2/T_c^2$ for $N_t=5,6,7$ and $8$ (red, green and blue dots, respectively). Also plotted are lattice results of~\cite{Boyd:1996bx}, $g^5$ perturbation theory~\cite{Kajantie:2002wa} and the HTL approach~\cite{Andersen:2010ct}.}
\label{fig:tracea_NP}
\vspace*{-0.cm}
\end{wrapfigure}

Moreover, there are also indications that there is a non-perturbative contribution to the equation of state that can never show up in perturbation theory~\cite{Pisarski:2006yk}. For dimensional reasons, any finite order perturbative formula will only give logarithmic corrections to the $p(T)\sim T^4$ Stefan-Boltzmann law.
Instead of such logarithmic corrections, lattice data suggests that there is a two-dimensional ``condensate'', proportional to $T^2$, which gives a dominant contribution for temperatures up to $\sim 4 T_c$. This can be best observed in the trace anomaly $I=\epsilon-3p$ (see definition in section~\ref{sec:eosquant}). Specifically, it is instructive to study the combination $I/T^4\cdot(T/T_c)^2$, which is shown in figure~\ref{fig:tracea_NP}. Our results with the Symanzik improved gauge action for various lattice spacings are compared to results obtained with the Wilson gauge action~\cite{Boyd:1996bx}, the 5-loop perturbative expansion~\cite{Kajantie:2002wa} and the HTL NNLO scheme. While for the former the renormalization scale $\mu=2\pi T$ is used (black dashed line in the figure), for the latter a range of $\mu_{\rm HTL}=\pi T \ldots 4\pi T$ is considered (gray band).

Apparently, the combination $I/T^2$ as measured on the lattice is approximately constant in the temperature range $T_c<T<5T_c$ (there are however discrepancies between the Symanzik and Wilson results, see discussion later). While up to $5T_c$ lattice results seem completely incompatible with the perturbative predictions, at larger temperatures our results also account for the $T^4$-like steep rise in $I(T)$ indicating a qualitative agreement with perturbative methods. In view of the fact that the trace anomaly measures the deviation from the ideal gas (see discussion in section~\ref{sec:eosquant}), this suggests that besides the ideal (perturbative) contribution $\sim T^4$, $I$ also contains a non-ideal (non-perturbative) term $\sim T^2$. Thus we separate the trace anomaly into two parts:
\be
\frac{I(T)}{T^4} = i_{\rm pert}(T) + \frac{i_{\rm np}(T)}{T^2}
\ee
The pressure can be obtained from $I/T^4$ as a definite integral (see~(\ref{eq:pfromI})). At extremely high temperatures its value is given by the Stefan-Boltzmann limit $p_{\rm SB} = 8\pi^2/45T^4$. Integrating down from this point one obtains
\be
\frac{p(T)}{T^4} = p_{\rm SB} - \int\limits_{T}^\infty \left(\frac{i_{\rm pert}(T')}{T'} + \frac{i_{\rm np}(T')}{T'^3} \right) \d T' = \frac{p_{\rm pert}(T)}{T^4} - \int\limits_T^\infty \frac{i_{\rm np}(T')}{T'^3} \d T' 
\label{eq:pressureformula}
\ee 

The results for the trace anomaly in the high-temperature region allow for a fitting of the HTL renormalization scale $\mu_{\rm HTL}$ and the unknown coefficient ($q_c$ in the notation of~\cite{Kajantie:2002wa}) of the $\O(g^6)$ order contribution of perturbation theory. While $q_c$ has already been calculated by means of a fit to the lattice data of~\cite{Boyd:1996bx}, here we are able to repeat this fitting procedure at a much higher temperature, where the sixth order can be shown to be a minor correction to the fifth order.
Once the optimal coefficient of the $g^6$ term is known, the non-perturbative contribution can also be quantified through a fit to some specified parameters of the function $i_{\rm np}(T)$.

We remark that the observed non-perturbative pattern may be explained within a fuzzy bag model~\cite{Pisarski:2006yk}, in terms of a dimension-2 gluon condensate~\cite{Pisarski:2000eq,Kondo:2001nq}, in a system of transversely polarized quasi-particles~\cite{Castorina:2011ja} or within the gauge/string duality~\cite{Andreev:2007zv}. Here we do not go into the viability of such models and only identify it as the dominant non-perturbative contribution.

\section{Finite volume effects}

Existing lattice results for the pressure end at around $5T_c$. These include results in the pure gauge sector with the Wilson plaquette action~\cite{Boyd:1995zg,Boyd:1996bx} and also with various improved actions like the Symanzik action~\cite{Beinlich:1995ik,Beinlich:1997ia}, renormalization group-improved actions~\cite{Okamoto:1999hi} or fixed-point actions~\cite{Papa:1996an}.
The effect of changing the number of colors~\cite{Panero:2009tv} was also studied.
Results for the pressure of full QCD~\cite{Karsch:2000ps, Bernard:2006nj, Aoki:2005vt} are also present only up to $(5-10)\cdot T_c$, like for the study presented in the previous chapter~\cite{Borsanyi:2010cj}.

There are two main reasons for the absence of high temperature results: first, at increasingly high temperatures the signal/noise ratio in the trace anomaly decreases significantly. Consequently, it becomes more and more difficult to detect a nonvanishing value for $I/T^4$. Note that this small signal is just the information necessary to fit the unknown perturbative parameters, as discussed in the previous section. (This is the reason that the primary observable in our study is the trace anomaly and not the pressure, as in chapter~\ref{chap:EoS}). Second, since the lattice spacing varies with the temperature as $a=(N_t T)^{-1}$, in order to have a constant physical lattice size
, the number of lattice points in the spatial directions $N_s$ in principle has to increase like $T$. While the former problem can be avoided by accumulating larger statistics, the latter obstacle is more of a matter of principle.

There is, however, a {\it perturbative} argument that may help one to improve on the situation. In the high temperature region the relevant correlation length is 
the scale that corresponds to the electric and magnetic screening masses. These are given by $m_e\sim g T$ and $m_m\sim g^2 T$, respectively, and have been calculated on the lattice~\cite{Nakamura:2003pu}. Thus, according to the magnetic mass (which is the one that gives larger correlations), in the high temperature region
one has
\be
N_sa \cdot m_m > 1 \quad \to \quad N_s/N_t > T/m_m \sim \beta
\label{eq:constantasprat}
\ee
which implies that the aspect ratio has to be increased only proportionally to $\beta=6/g^2$. However it is important not to overlook the loophole in this argument. On the lattice at a given temperature, one can never know whether the perturbative region is already reached. Justifying the neglection of finite size effects based on a perturbative reasoning, therefore, is not really convincing. In order to establish the range of validity of the perturbative approach itself, it is indispensable to simulate the non-perturbative scale $\sim T_c$, too. In this sense the aspect ratio sets the maximum temperature where the system is still non-perturbative as $T<T_c\cdot N_s/N_t$.

Furthermore, it is well known that the finiteness of the spatial volume can also distort the $p=-f$ relation, as infrared corrections may appear. The impact of these corrections on the equation of state as a function of the aspect ratio has been analytically calculated recently~\cite{Gliozzi:2007jh}. According to this result, finite volume effects in the free energy density only drop below one percent if $N_s/N_t\ge8$ holds for the aspect ratio.

Keeping in mind the above considerations we perform three sets of simulations. First, we calculate the trace anomaly in the temperature range of $T/T_c= 0.7 \ldots 15$ (on $80^3 \times 5$, $96^3 \times 6$ and $112^3 \times 7$ lattices) and also extract a continuum limit from these results. These lattices with aspect ratio $N_s/N_t=16$ accommodate the non-perturbative scale $T_c^{-1}$ up to approximately $16T_c$. 
We also support the continuum extrapolation with an additional $N_t=8$ set of lattices ($64^3\times 8$) below $8 T_c$. This combined extrapolation is described in the beginning of section~\ref{sec:results_tracea}. From the trace anomaly various other thermodynamic functions can be determined according to the relations of section~\ref{sec:eosquant}.

In the next step we study the finite volume scaling of the trace anomaly on a non-continuum data set at $N_t=5$. We presents results using lattices of aspect ratio $N_s/N_t=4, 6, 8, 16$ and $24$. The latter $120^3 \times 5$ lattice accommodates the $T_c^{-1}$ scale up to $24 T_c$. Using these results we test finite size effects in the whole temperature region, see subsection~\ref{sec:voldep_tracea}. 

Since due to computational limitations we cannot increase the aspect ratio further, in our third set of simulations we calculate the continuum equation of state in a small box (on $40^3 \times 5$, $48^3 \times 6$ and $64^3 \times 8$ lattices) up to $1000 T_c$. In subsections~\ref{sec:fitg6} and~\ref{sec:fithtl} we use this data set to find the optimal free parameters of existing perturbative calculations, i.e. the already mentioned $q_c$ parameter of $\O(g^6)$ perturbation theory and the renormalization scale $\mu_{\rm HTL}$ of the HTL scheme. Using these small volume results we observe a good agreement with the newly fitted perturbative formulae, indicating that this approach successfully connects the low temperature non-perturbative region with the high temperature perturbative realm. The precision of our data points exceeds any previous calculation by about an order of magnitude.

In order for the large lattices to fit in the memory of our computer system, the renormalization of the trace anomaly was done via half-temperature subtraction, as explained in section~\ref{sec:renorm_pres}. Specifically, we calculate
\be
\frac{I(T)}{T^4} = \left(\frac{I(T)}{T^4} -\frac{1}{16}\frac{I(T/2)}{(T/2)^4}\right) + \frac{1}{16}\frac{I(T/2)-I(0)}{(T/2)^4}
\label{eq:traceameas}
\ee
Due to the fact that the trace anomaly very strongly depends on the volume around the critical temperature, in order to account for this dependence we measured the second term of the above expression on lattices with half the spatial extension. Thus both terms are measured with the same physical volume which ensures that the sum is smooth around $T=2n\cdot T_c$ for each $n\in \mathds{N}^+$ (otherwise the difference between the volumes shows up as small cusps at these temperatures). For the lattices with half the spatial size (i.e. the second term in~(\ref{eq:traceameas})) the subtraction is carried out in the standard way, i.e. at $T=0$. The continuum limit from this combined technique is equal to what one finds using the standard scheme.

\section{Scale setting}

Besides the proper treatment of finite volume effects another challenging issue was the accurate determination of non-perturbative beta function corresponding to the Symanzik improved action. 
Instead of setting the scale using $\sigma$ or $r_0$, it was advantageous to define the lattice spacing in terms of the transition temperature, as it was argued for in section~\ref{sec:scalesetting}. To this end we determined the critical couplings
$\beta_c$ up to $N_t = 20$ from the peak of the Polyakov loop susceptibility.
We define the improved coupling in the ``E'' scheme~\cite{Bali:1992ru}, generalized for the Symanzik gauge action as 
\be
\begin{split}
g_E^2(g^2) &= \frac{S_G^{\rm Symanzik}(g^2)}{d_1} = \frac{c_0 P_{1\times1}+c_1 P_{2\times1}}{d_1} \\
S_E^{\rm Symanzik}(g^2) &= d_1 g^2 + d_2 g^4 + \ldots
\end{split}
\ee
with coefficients $c_0$ and $c_1$ as introduced in~(\ref{eq:symanzikact}).

\begin{wrapfigure}{r}{8.4cm}
\centering
\vspace*{-0.5cm}
\includegraphics*[width=7.8cm]{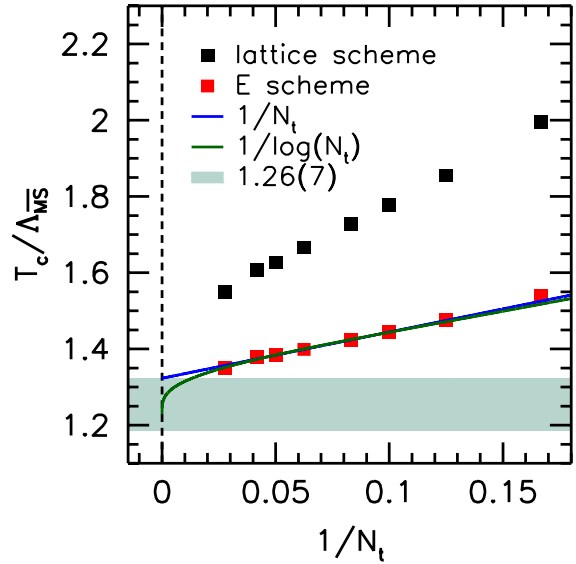}
\vspace*{-0.1cm}
\caption[Setting the scale by critical couplings.]{Setting the scale by critical couplings in the E scheme.}
\label{fig:scale_quenched}
\vspace*{-2cm}
\end{wrapfigure}

The expansion coefficients $d_i$ we calculated explicitly using measurements of $P_{1\times1}$ and $P_{2\times1}$ at high $\beta$, and checked against existing results in the literature~\cite{Weisz:1983bn}. We then matched the measurements of $\beta_c$ to the universal two-loop running and converted to the $\overline{MS}$ scheme~\cite{Skouroupathis:2007mq}. In figure~\ref{fig:scale_quenched} we plot as a function of $N_t^{-1}$ the lambda parameter in units of the transition temperature. We extrapolate to the continuum limit using a linear and a logarithmic fit, and define the systematic error of the result as half of the difference between the two. As a comparison we also show results for $T_c/\Lambda_{\overline{MS}}$ in the lattice scheme, where the convergence is expected to be worse.

Our final result we quote as
\be
\frac{T_c}{\Lambda_{\overline{MS}}} = 1.26(7)
\label{eq:latscaleset}
\ee

The error is overwhelmingly systematic and reflects the sensitivity to various continuum extrapolations. This is consistent with the combination of previous determinations: the Lambda parameter $\Lambda_{\overline{MS}} = 0.614(2)(5)r_0$ of~\cite{Gockeler:2005rv} can be translated to $\sigma$ units using $\sigma r_0 = 1.192(10)$ (based on~\cite{Guagnelli:1998ud}) and then used with $T_c / \sigma = 0.629(3)$ of~\cite{Boyd:1996bx}. Through our direct result one can easily translate the scale setting of the perturbative expressions to the lattice language. \vspace*{1.1cm}

\section{Results}
\label{sec:results_tracea}

\begin{figure}[h!]
\centering
\vspace*{-0.2cm}
\epsfig{file=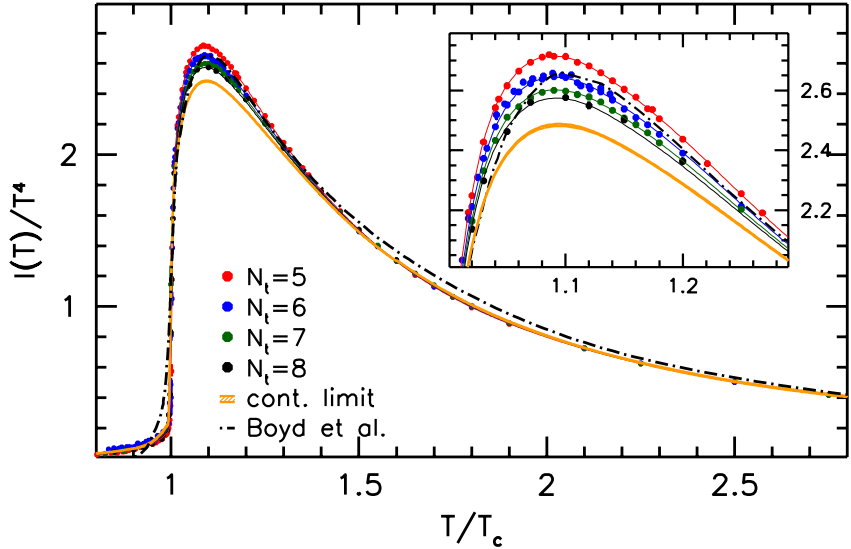,width=12.7cm,bb=18 360 592 718}
\vspace*{-0.1cm}
\caption[The trace anomaly for various lattice spacings in the transition region.]{
The trace anomaly for various lattice spacings in the transition region. The result of a combined spline fit for each lattice spacing, together with the continuum extrapolation is shown by the colored lines and the yellow band, respectively. For comparison results with the standard Wilson action~\cite{Boyd:1996bx} are also shown by the dashed-dotted line.}
\label{fig:i_trans}
\end{figure}

First we reproduce the results of~\cite{Boyd:1996bx} in the transition region. In figure~\ref{fig:i_trans} these results are compared to the trace anomaly measured on our first set of simulations, i.e. on large lattices ($N_s/N_t=16$) with $N_t=5,6,7$, supplemented by $N_t=8$, with $N_s/N_t=8$. From these four sets of results we perform a continuum extrapolation via a combined spline fitting method. The datasets for different lattice spacings are fitted together by an $N_t$-dependent spline function\footnote{The $N_t$-dependence is of the form $N_t^{-2}$, as motivated by the scaling properties of the Symanzik improved action.}. This ``multi-spline'' function -- defined upon a set of nodepoints $\beta_k$ with $k=1\ldots K$ -- is parameterized by two values at each nodepoint, written in the form $a_k+b_k N_t^{-2}$. We fit these $2K$ parameters to the measurements: the minimum condition for $\chi^2$ leads to a set of linear equations, which can be solved for the parameters.

As a result we have a smooth function interpolating our data for each $N_t$ (colored lines in the figure), together with a smooth, continuum extrapolated curve (yellow band in the figure). 
The statistical error of the fit is determined by a jackknife analysis, while the systematic error of the continuum result is estimated by the difference between the extrapolation from $N_t=5,6,7$ and $N_t=5,6,7,8$. 
Note that this method ensures an exact $N_t^{-2}$ scaling at each temperature. As visible in the figure, data points for various lattice spacings are on top of each other, with the exception of the transition region. This region is zoomed into in the inlay of the figure, showing that our data indeed exhibits the expected $N_t^{-2}$ scaling.

As figure~\ref{fig:i_trans} shows there is an apparent discrepancy between our continuum result and that of~\cite{Boyd:1996bx}, particularly around $T_c$. This might be attributed to the finite volume effects as well as to the difference in the scale setting procedure.

\subsection{Comparison to the glueball gas model}

\begin{figure}[h!]
\centering
\vspace*{-0.2cm}
\epsfig{file=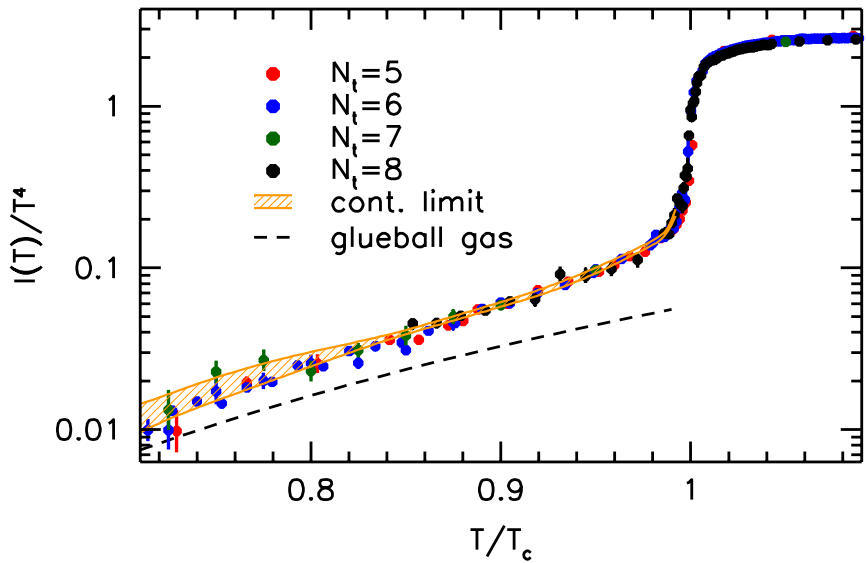,width=12.7cm,bb=18 360 592 718}
\vspace*{-0.1cm}
\caption[The trace anomaly in the confined phase and the glueball resonance model prediction.]{The trace anomaly in the confined phase measured on lattices with various lattice spacings and the continuum extrapolation (yellow band). The dashed line corresponds to the glueball resonance model, estimated from the twelve lightest glueballs.}
\label{fig:coldtra}
\vspace*{-0.0cm}
\end{figure}

In order to explore the thermodynamics of the confined phase, next we zoom into the low temperature region $T<T_c$. In this region one can also calculate the trace anomaly within the glueball resonance model. In figure~\ref{fig:coldtra} we plot our results together with the contribution of the first twelve glueballs of~\cite{Chen:2005mg}. There is an apparent deficit of the model prediction as compared to the lattice results. It has been suggested that this difference might be reduced if one allows for a temperature dependence of the glueballs~\cite{Buisseret:2009eb}, which has already been determined for the $0^{++}$ and $2^{++}$ states~\cite{Ishii:2002ww}.
This point was raised in connection to the lattice results of~\cite{Panero:2009tv} below $T_c$. 
Our data is in fairly good agreement with this scenario, however further studies are necessary to understand the question in more detail.

The glueball resonance model can also be used to set the integration constant in the pressure. 
We appoint as a reference temperature $T_{\rm ref}=0.75 T_c$, and shift the lattice results up to the resonance gas prediction at this temperature. 
This shift causes a $\sim 0.1\%$ change in our high temperature results for the pressure and is smaller than our statistical errors here.
This reference point is used also for the pressure as measured in our third set of simulations (i.e. on lattices with $N_s/N_t=8$).

\subsection{Volume dependence of the results}
\label{sec:voldep_tracea}

\begin{figure}[h!]
\centering
\vspace*{-0.2cm}
\epsfig{file=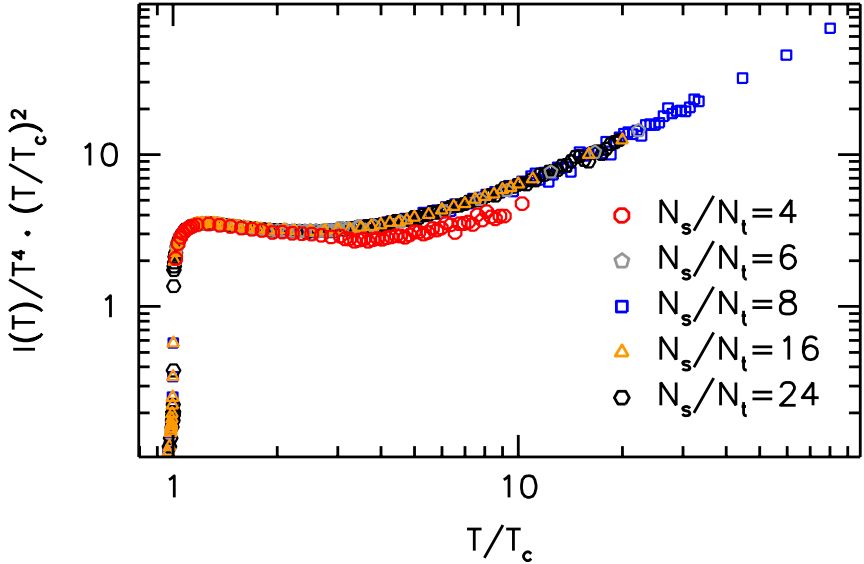,width=12.7cm,bb=18 360 592 718}
\vspace*{-0.1cm}
\caption[Volume dependence of the trace anomaly.]{Volume dependence of the trace anomaly on our $N_t=5$ lattices.}
\label{fig:voldep}
\vspace*{-0.0cm}
\end{figure}

The effect of the non-perturbative $\sim T^2$ contribution to the trace anomaly reduces at increasing temperatures. Moreover, the presence of this contribution becomes unnoticeable at sufficiently high $T$, regardless of whether or not the lattice size accommodates the inverse $T_c$ scale. One way to discuss the relevance of this non-perturbative scale is to compare the trace anomaly at various spatial volumes. This comparison is shown in figure~\ref{fig:voldep} for our $N_t=5$ lattices. 
The standard aspect ratio $N_s/N_t=4$ gives somewhat smaller values for $I/T^4$, but beyond $N_s/N_t=6$ we do not see any difference in the results above the transition region.

Based on our observations so far we conclude that:
\begin{enumerate}
\item
Our large volume results for the trace anomaly agree well with the perturbative formulas above $10T_c$.
\item
Within statistical accuracy we see no discrepancy between results from different volumes provided that $N_s/N_t\ge 6$.
\item
The significance of the non-perturbative contribution that is dominant in the transition region reduces as $\sim 1/T^2$.
\item
All of our volumes accommodate the perturbatively relevant scales, in particular the magnetic screening length $\sim (g^2T)^{-1}$.
\end{enumerate}
These considerations suggest that -- even if the lattice volumes are ever shrinking as the temperature is increased -- our results are able to describe the physical trace anomaly within the error bars shown. Thus it is reasonable to conjecture that our $N_s/N_t=8$ dataset reliably connects the transition region with the perturbative regime.

\subsection{Fitting improved perturbation theory}
\label{sec:fitg6}

Regardless of whether the conjecture of the last subsection is valid or not, we can make use of our small volume simulations at high temperature to compare to perturbative expansions, in particular, to extract some unknown coefficients of these formulas. We perform the continuum extrapolation in the same manner as for the large volume data, see section~\ref{sec:results_tracea}, using the $N_t=5,6$ and $8$ lattices. The systematic error we define in this case as the difference between the continuum extrapolated and the $N_t=8$ results.

\begin{figure}[h!]
\centering
\vspace*{-0.2cm}
\epsfig{file=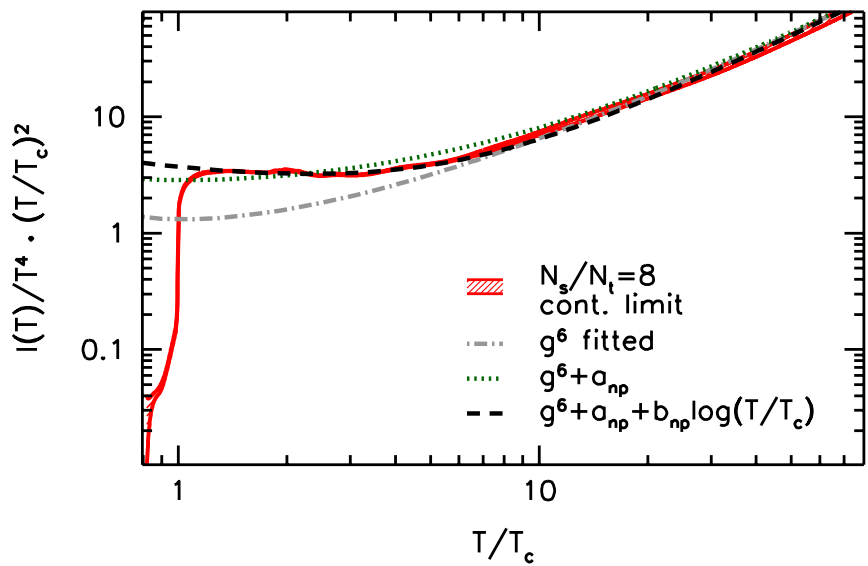,width=12.7cm,bb=18 360 592 718}
\caption[The trace anomaly compared to perturbation theory.]{The continuum limit obtained from the lattice results (red band), compared to fitted perturbation theory. We fit the $g^6$ coefficient (gray dashed-dotted line), and the non-perturbative contribution in two different forms (black dashed line and green dotted line). }
\label{fig:pertcomp}
\end{figure}

First we compare our results to $\O(g^6)$ improved perturbation theory~\cite{Kajantie:2002wa}. We perform a fit of the subtracted trace anomaly
\be
i_{\rm pert}(T,q_c,\mu)-i_{\rm pert}(T/2,q_c,\mu)
\label{eq:subtrtracea}
\ee
to the unknown coefficient $q_c$ of the $g^6$ term with a fixed renormalization scale of $\mu=2\pi T$. If we also allow for a variation of the scale, we find $\mu/2\pi T$ to be consistent with 1 within errors. These fits are carried out for our results between $10T_c < T < 1000 T_c$ and the systematic error is estimated by varying the endpoints of the fit interval. Beyond this we also consider as a source of systematic error the uncertainty of our lattice scale setting~(\ref{eq:latscaleset}). We quote as our final result for these parameters
\be
q_c =-3526(4)(55)(30)
\ee
with the numbers in the parentheses are from left to right the statistical error, the error coming from the lattice scale and that from the variation of the fit interval. A good fit quality is indicated as $\chi^2/{\rm dof}=0.7$. The fitted function is shown in figure~\ref{fig:pertcomp}.

While the $\sim T^{-2}$ behaviour of the trace anomaly in the low-temperature region has been seen in many studies, e.g.~\cite{Boyd:1996bx,Panero:2009tv}, its relative weight in the total observable has not yet been quantified.
Therefore we also consider it useful to estimate the non-perturbative contribution to the trace anomaly, which we assume to be of the form $i_{\rm np}(T) = a_{\rm np} + b_{\rm np}\log(T/T_c)$, i.e. we propose the following fit function:
\be
i_{\rm pert}(T) + \frac{a_{\rm np} + b_{\rm np}\log(T/T_c)}{T^2}
\label{eq:fitNP}
\ee
First we perform the fit for $a_{\rm np}$ with $b_{\rm np}=0$ kept fixed, then we carry out the fit for both non-perturbative coefficients. The fit interval is chosen to be $T_c < T < 60 T_c$. We find that the constant approximation is not able to resolve the trace anomaly in the low temperature region as $\chi^2/{\rm dof}\approx 150$. The logarithmic correction improves much on the situation and we get $\chi^2/{\rm dof}=4.7$. Moreover, the parameters are rather sensitive to the variation of the lower endpoint of the fit interval. Nevertheless, since there is no a priori constraint on the form of the fit function~(\ref{eq:fitNP}), we accept this as a first approximation to the non-perturbative contribution. We obtain the following coefficients:
\begin{align*}
\textmd{const.: }\quad a_{\rm np} &= 0.879(2)(40) & \hspace*{-1cm}b_{\rm np} &= 0 \\
\textmd{const.+log.: }\quad a_{\rm np} &= 1.371(1)(50) & \hspace*{-1cm}b_{\rm np} &= -0.618(2)(4) 
\end{align*}
with the errors coming from the statistics and the lattice scale, respectively.
We also plot both the constant and the logarithmic fit in figure~\ref{fig:pertcomp}.

\begin{figure}[h!]
\centering
\vspace*{-0.2cm}
\epsfig{file=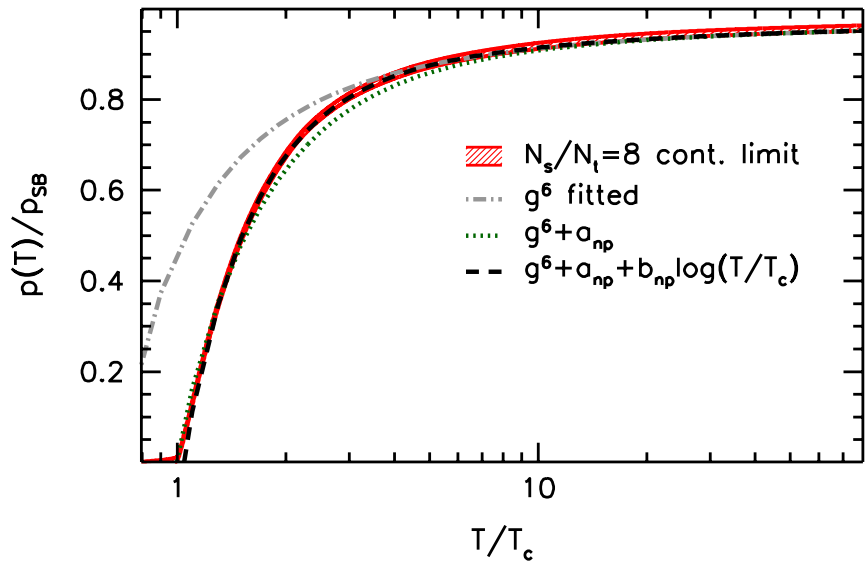,width=12.7cm,bb=18 360 592 718}
\caption[The normalized pressure compared to perturbation theory.]{The normalized pressure as measured on our small volume boxes. A comparison is shown to various fitted perturbative functions.}
\label{fig:pertcompp}
\end{figure}

\begin{figure}[h!]
\centering
\vspace*{-0.2cm}
\epsfig{file=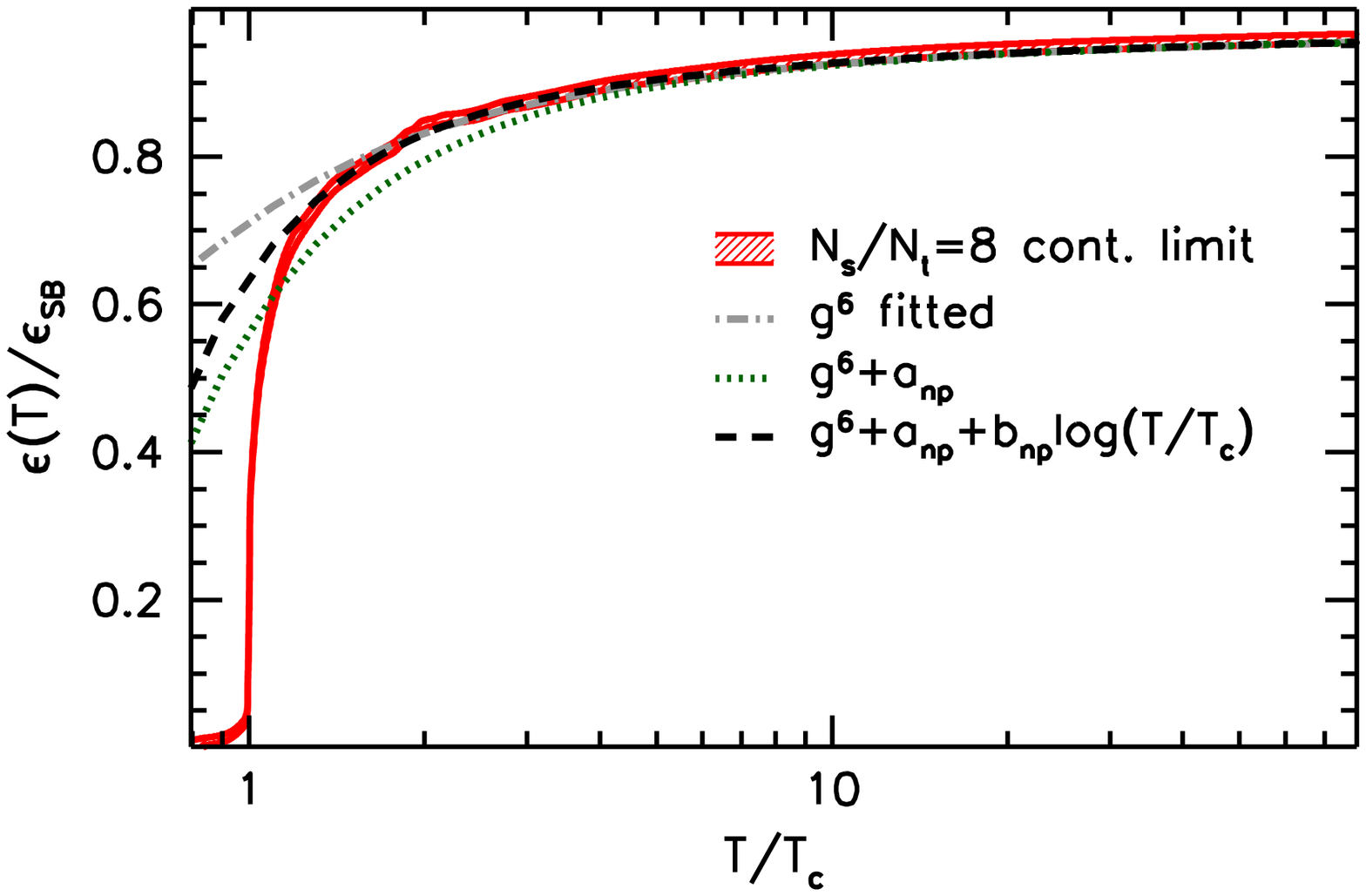,width=12.7cm,bb=18 360 592 718}
\caption[The normalized energy density compared to perturbation theory.]{The normalized energy density as measured on our small volume boxes.}
\label{fig:pertcompe}
\end{figure}

\begin{figure}[h!]
\centering
\vspace*{-0.2cm}
\epsfig{file=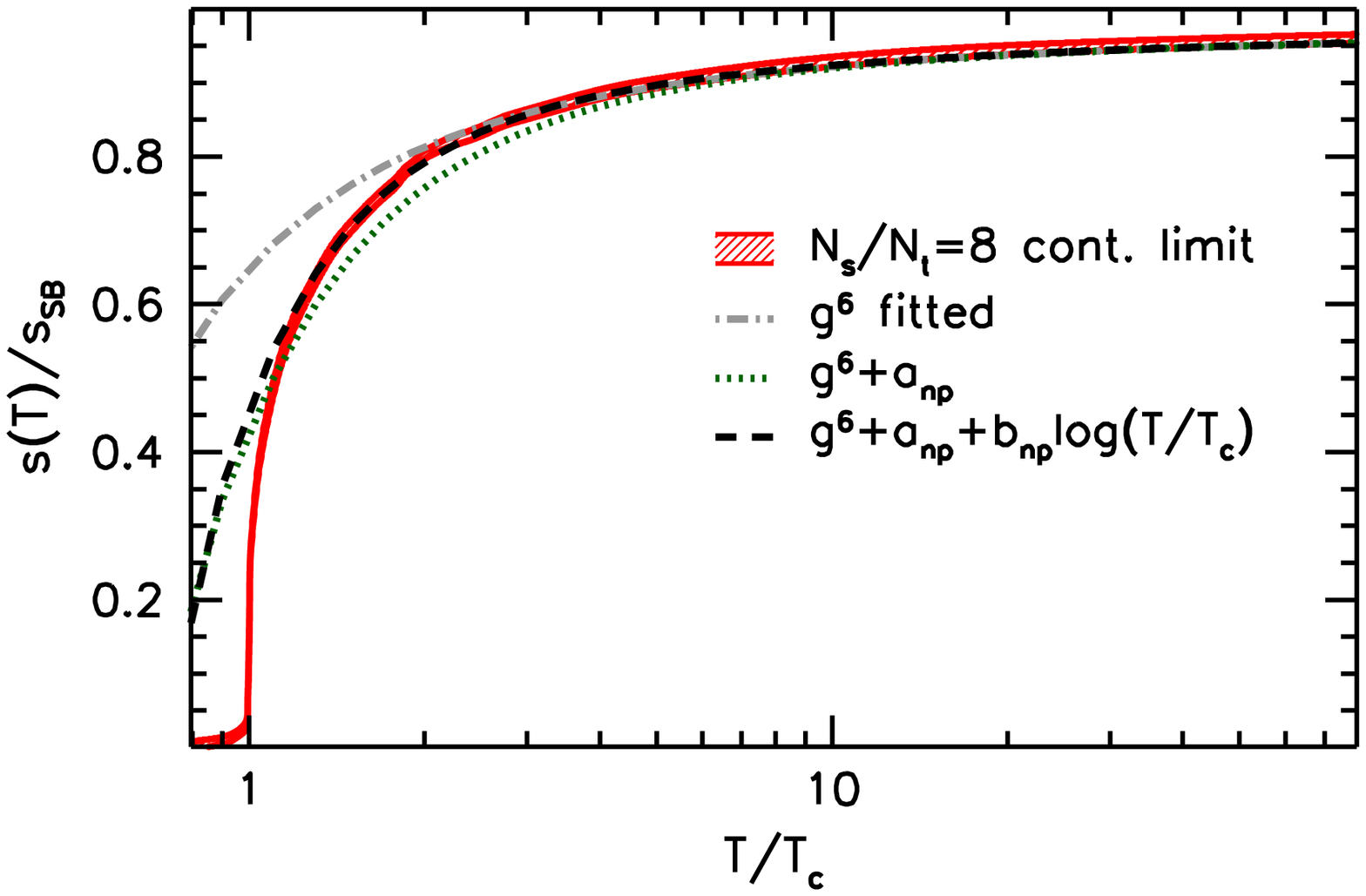,width=12.7cm,bb=18 360 592 718}
\caption[The normalized entropy density compared to perturbation theory.]{The normalized entropy density as measured on our small volume boxes.}
\label{fig:pertcomps}
\end{figure}

Using~(\ref{eq:pressureformula}) the fitted perturbative formulae for the pressure are also straightforward to write down. In figure~\ref{fig:pertcompp} we compare our small volume results to the so obtained predictions. Similar comparisons can be made for the case of the energy density and the entropy density also, where we find qualitatively the same behavior as for the pressure, see figures~\ref{fig:pertcompe} and~\ref{fig:pertcomps}.

\subsection{Fitting HTL perturbation theory}
\label{sec:fithtl}

\begin{figure}[ht!]
\centering
\vspace*{-0.2cm}
\epsfig{file=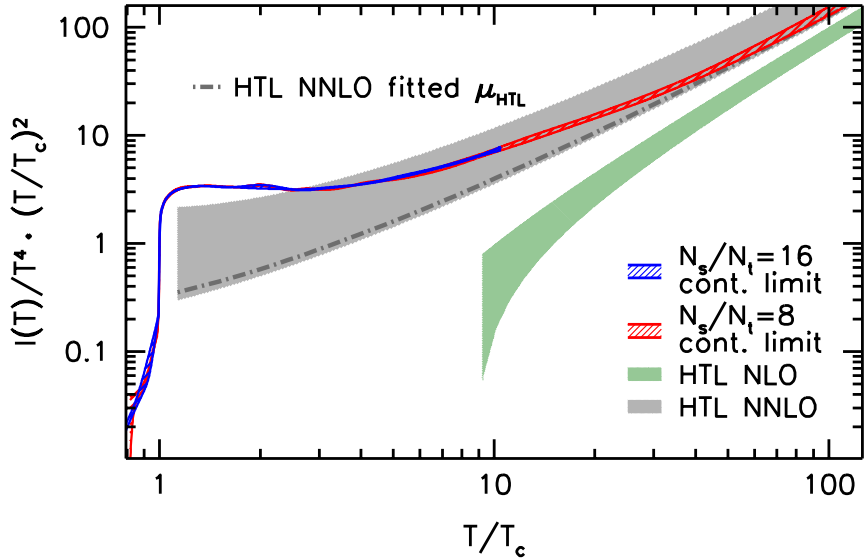,width=12.7cm,bb=18 360 592 718}
\caption[The trace anomaly compared to HTL perturbation theory.]{The trace anomaly measured on two different spatial volumes (red and blue bands), compared to the NLO and NNLO HTL expansion with varied renormalization scale $0.5<\mu_{\rm HTL}/2\pi T<2$ (green and gray shaded regions). The dashed-dotted line represents the expansion with the fitted scale of~(\ref{eq:fittedhtlmu}).}
\label{fig:htlcomp}
\end{figure}

Next we discuss the region of validity of the HTL resummed perturbation theory. In particular, we compare once again our $N_s/N_t=8$ lattice results to the NNLO expansion of the HTL scheme~\cite{Andersen:2010ct}. We consider the renormalization scale $\mu_{\rm HTL}$ as a free parameter of this expansion, and perform a fit to this parameter, i.e. our fit function to the subtracted trace anomaly is
\be
i_{\rm pert}(T,\mu_{\rm HTL})-i_{\rm pert}(T/2,\mu_{\rm HTL})
\ee
The fit is carried out for $T>100T_c$, and the endpoint is varied to obtain the systematic error coming from the fitting procedure. The sum of deviations for this fit is $\chi^2/{\rm dof}=0.6$, indicating a nice agreement between lattice results and the perturbative expansion. Our result for the renormalization scale is (in the same notation for the errors as before)
\be
\frac{\mu_{\rm HTL}}{2\pi T} = 1.75(2)(6)(50)
\label{eq:fittedhtlmu}
\ee
The fitted formula for the trace anomaly is shown in figure~\ref{fig:htlcomp}.

\chapter{Summary}

In this thesis I studied the finite temperature transition between confined and deconfined matter at zero and nonzero quark densities. The findings are relevant for the understanding of the evolution of the early Universe and contemporary and upcoming heavy ion experiments. Our results were obtained using large-scale simulations using lattices with various physical extents and lattice spacings, at physical values of the parameters of the theory. In all three projects we thoroughly checked finite size effects and discretization effects, thereby controlling the two most important possible sources of systematic error. In order to ensure that the extrapolations lead to the correct continuum limit, all the necessary additive and multiplicative renormalizations were applied to the studied observables.

In chapter~\ref{chap:phasediag} the study of the phase diagram of QCD at small chemical potentials was presented. We determined the curvature of the transition line which separates the chirally symmetric and asymmetric phases using two observables that are sensitive to this transition. Our results also contain information about the relative change in the width of the transition. This suggests that the strength of the transition decreases very slightly, therefore the weak crossover that is present at zero density persists up to quark chemical potentials $\mu_{u,d} \lesssim T_c$.

In chapter~\ref{chap:EoS} I showed results regarding the equation of state in full QCD. For this project I developed a new method to determine the pressure from the raw lattice data. This approach improves on the conventionally applied integral method such that it better estimates the systematic error of the result, and is also capable of quantifying the quark mass dependence of the pressure. From the pressure we derived various other thermodynamic observables, which can be directly used to set the equation of state in e.g. hydrodynamic models. We estimate the effect of a charm quark in the pressure and show that it is significant already at $T\sim 2T_c$. Furthermore, we observe that the pressure, the energy density and the entropy density are roughly 15-20$\%$ under their Stefan-Boltzmann limits at our highest temperature $\sim 6T_c$.

A these values of the temperature a reliable comparison with perturbation theory cannot be carried out since such perturbative expansions are only convergent at much larger temperatures. In chapter~\ref{chap:PG} I report results of the equation of state in pure gauge theory at extremely high temperatures $\sim 1000 T_c$. We perform simulations using lattices of size up to $N_s/N_t=24$ and gather statistics that exceed results in the literature by at least an order of magnitude to determine the high temperature signal in the interaction measure. At large $T$ we successfully match our results with perturbation theory and we are able to fit the unknown $\mathcal{O}(g^6)$ coefficient of the expansion. We also quantify the non-perturbative contribution to the interaction measure, which is proportional to $T^2$.

\phantomsection
\addcontentsline{toc}{chapter}{Acknowledgements}
% Thesis Acknowledgements ------------------------------------------------

%\begin{acknowledgementslong} %uncommenting this line, gives a different acknowledgements heading
\begin{acknowledgements}      %this creates the heading for the acknowlegments

First of all I would like to thank S\'andor Katz for his help and support. He always managed to find time to discuss my questions and to talk about interesting physics problems. The research projects suggested by him were not only exciting and very useful for my scientific development, but he also had a good sense of choosing them such that we would have interesting results. S\'andor is a colleague with whom I can collaborate very effectively. It would be a pleasure for me to maintain our scientific relationship.

I am indebted to Zolt\'an Fodor, who always had the experience and insight to recognize the potential in each research project, and also gave me the opportunity to develop my own ideas. During my semester in Wuppertal, his exactitude and discipline at work were exemplary for me. He provided an outstanding working environment, which was the soil for a team that is efficient in long-term collaboration.

I would like to say thanks to Yasumichi Aoki, Szabolcs Bors\'anyi, Antal Jakov\'ac, Stefan Krieg and K\'alm\'an Szab\'o, with whom we worked together in various projects presented in my thesis. I also owe many thanks to D\'aniel N\'ogr\'adi, Tam\'as Kov\'acs and B\'alint T\'oth for inspiring and useful discussions.

My research was carried out at the Department of Theoretical Physics at the E\"otv\"os University in Budapest and at the Institute for Theoretical Physics at the University of Wuppertal.
My work was partially supported by the grant (FP7/2007-2013)/ERC no. 208740. Computations were performed on the Blue Gene supercomputers at FZ Juelich, on
clusters at the University of Wuppertal and at the E\"otv\"os University, Budapest, furthermore on the QPACE
cluster in Regensburg. 

Finally I would like to thank my wife, Dorka for being always there for me, even when it comes to physics. Despite that she is working in such a different field, she is always interested to listen to me and think about some new idea that I have in mind.

\end{acknowledgements}
%\end{acknowledgmentslong}

% ------------------------------------------------------------------------

%%% Local Variables: 
%%% mode: latex
%%% TeX-master: "../thesis"
%%% End: 

% Thesis Acknowledgements ------------------------------------------------

%\begin{acknowledgementslong} %uncommenting this line, gives a different acknowledgements heading
\begin{acknowledgements_hu}      %this creates the heading for the acknowlegments

Mindenekel\H{o}tt t\'emavezet\H{o}mnek, Katz S\'andornak szeretn\'em megk\"{o}sz\"{o}nni f\'{a}radha-tatlan seg\'{i}ts\'eg\'et. Mindig tal\'alt id\H{o}t arra, hogy a k\'erd\'eseimen egy\"{u}tt gondolkoz-zunk, \'es hogy \'erdekes fizikai probl\'em\'akr\'{o}l besz\'elgess\"{u}nk. Az \'altala javasolt kutat\'asi t\'em\'ak nemcsak izgalmasak \'es a tudom\'anyos fejl\H{o}d\'esem szempontj\'ab\'{o}l nagyon hasznosak voltak, de azt is gyakran j\'{o}l meg\'erezte, hogy melyik t\'em\'ab\'{o}l fog sz\'elesebb k\"{o}rben \'erdekl\H{o}d\'esre sz\'amot tart\'{o} eredm\'eny kij\"{o}nni. Szem\'ely\'eben olyan munkat\'ar-sat ismertem meg, akivel nagyon j\'{o}l tudok egy\"{u}tt dolgozni, \'es akivel ezt a k\"oz\"os munk\'at a j\"{o}v\H{o}ben is folytatni szeretn\'em.

K\"{o}sz\"{o}netet szeretn\'ek mondani Fodor Zolt\'annak, aki tapasztalat\'aval \'es \'elesl\'at\'as\'aval mindig felismerte a kutat\'asi t\'em\'akban rejl\H{o} lehet\H{o}s\'egeket, \'es ugyanakkor szabad teret hagyott a saj\'at \"{o}tleteimnek is. A Wuppertalban elt\"{o}lt\"{o}tt szemeszter alatt munkafegyelme \'es precizit\'asa p\'eldamutat\'{o} volt sz\'amomra. Fodor Tan\'ar \'{U}r kiv\'al\'{o} munkafelt\'eteleket biztos\'{i}tott, amelyek egy hat\'ekonyan dolgoz\'{o} csapat l\'etrej\"{o}tt\'et \'es hossz\'{u} t\'av\'{u} egy\"{u}ttm\H{u}k\"{o}d\'es\'et teszik lehet\H{o}v\'e.

K\"{o}sz\"{o}nettel tartozom tov\'abb\'a Yasumichi Aokinak, Bors\'anyi Szabolcsnak, Jakov\'ac Antalnak, Stefan Kriegnek \'es Szab\'{o} K\'alm\'annak, akikkel a dolgozatomban bemutatott kutat\'asi t\'em\'akon k\"{o}z\"{o}sen dolgoztunk. K\"{o}sz\"{o}net illeti ezenk\'{i}v\"{u}l N\'{o}gr\'adi D\'anielt, Kov\'acs Tam\'ast \'es T\'oth B\'alintot, akikkel sokszor ny\'{i}lt lehet\H{o}s\'egem konzul-t\'alni. Megjegyz\'eseik \'es javaslataik gyakran hozz\'aseg\'itettek ahhoz, hogy egy-egy t\'emak\"ort jobban meg\'erthessek.

Munk\'amat az ELTE Elm\'eleti Fizikai Tansz\'ek\'en \'es a Wuppertali Egyetem Elm\'eleti Fizika Tansz\'ek\'en v\'egeztem. A kutat\'ast r\'eszben az (FP7/2007-2013)/ERC 208740-es sz\'am\'u EU p\'aly\'azat t\'amogatta. A sz\'am\'it\'asokat a Juelichi Kutat\'ocentrum BlueGene szupersz\'am\'it\'og\'epein, a Wuppertali Egyetem \'es az ELTE sz\'am\'it\'og\'epparkj\'an, valamint a Regensburgi Egyetem QPACE rendszer\'en v\'egezt\"uk.

V\'eg\"ul szeretn\'em megk\"osz\"onni feles\'egemnek, Dork\'anak, hogy b\'armikor besz\'elgethe-tek vele fizik\'ar\'ol is. Annak ellen\'ere, hogy annyira m\'as ter\"uleten dolgozik, mindig \'erdekl\H{o}d\'essel hallgat meg \'es \'erdemben reag\'al, ha egy \'uj \"otletemet mes\'elem el neki.

\end{acknowledgements_hu}
%\end{acknowledgmentslong}

% ------------------------------------------------------------------------

%%% Local Variables: 
%%% mode: latex
%%% TeX-master: "../thesis"
%%% End: 

%\include{Conclusions/conclusions}

\appendix

\chapter{Multidimensional spline integration}
\label{appendix1}

In this appendix I describe the method, which was used to obtain the pressure as a function of the parameters $\beta$ and $R$ using its measured derivatives~(\ref{eq:dbeta}) and~(\ref{eq:dR}). This method is not exclusively applicable to the case of the pressure in QCD; on the contrary, it can be used in a generalized context, e.g. for statistical physical problems, where using standard Monte-Carlo methods one is not able to measure the (logarithm of the) partition function $\log \Z$ itself, but only its derivatives with respect to the parameters of the theory. 

Therefore in the following discussion I introduce the algorithm for the general case, for reconstructing a multidimensional surface, using the gradient of the surface measured at some values of the coordinates. The algorithm includes an introduction of a set of nodepoints, upon which the multidimensional spline is determined. This determination is linear and thus straightforward to compute. By a systematic variation of the number and position of the nodepoints the method is then adapted to the particular problem. Unlike a multidimensional integration along some path, the present method results in a continuous, smooth surface, furthermore, it also applies to input data that are non-equidistant and not aligned on a rectangular grid. Function values, first and second derivatives and integrals of the surface are easy to calculate. Moreover, the proper estimation of the statistical and systematic errors is also incorporated in the method.

First I briefly summarize how a spline function can be introduced in an arbitrary number of dimensions $D$. Afterwards I show how the fit to the measurements is carried out. Since in practice $D>2$ is seldom necessary, in order not to complicate the notation, the method is discussed in detail for the case of two dimensions. Nevertheless, $D>2$ is also straightforward to implement. Finally I present the algorithm for the systematic placement of the nodepoints. The present method was recently published~\cite{Endrodi:2010ai}.

\section{Spline definition in arbitrary dimensions}

A detailed introduction to the theory of splines can be found in~\cite{Ahlberg,deBoor,Schumaker}.
For our purpose it suffices to know that in e.g. two dimensions a cubic spline is defined upon a grid $\{x_k, y_l\}$ with $0\le k <K$, $0\le l<L$. The spline surface is unambiguously determined by the values that it takes at the nodepoints $f_{k,l}=S(x_k,y_l)$ (and the boundary conditions, which we take to be ``natural''). A grid square $[x_k,x_{k+1}]\times[y_l,y_{l+1}]$ will be shortly referred to as $\{k,l\}$. The spline function itself is compactly written as\footnote{Note that in this formulation the cubic spline contains terms like $x^3y^3$, contrary to other definitions where a bicubic spline only has terms $x^{i}y^{j}$ with $0\le i+j\le3$.}
\be
S(x,y)=\sum\limits_{i=0}^{3}\sum\limits_{j=0}^{3} C_{i,j}^{k,l} t_{(k)}^i u_{(l)}^j,\quad \textmd{if } (x,y)\in\{k,l\}
\label{eq:sdef}
\ee

\begin{wrapfigure}{r}{6.7cm}
\centering
\vspace*{-1.1cm}
\includegraphics*[width=6.5cm]{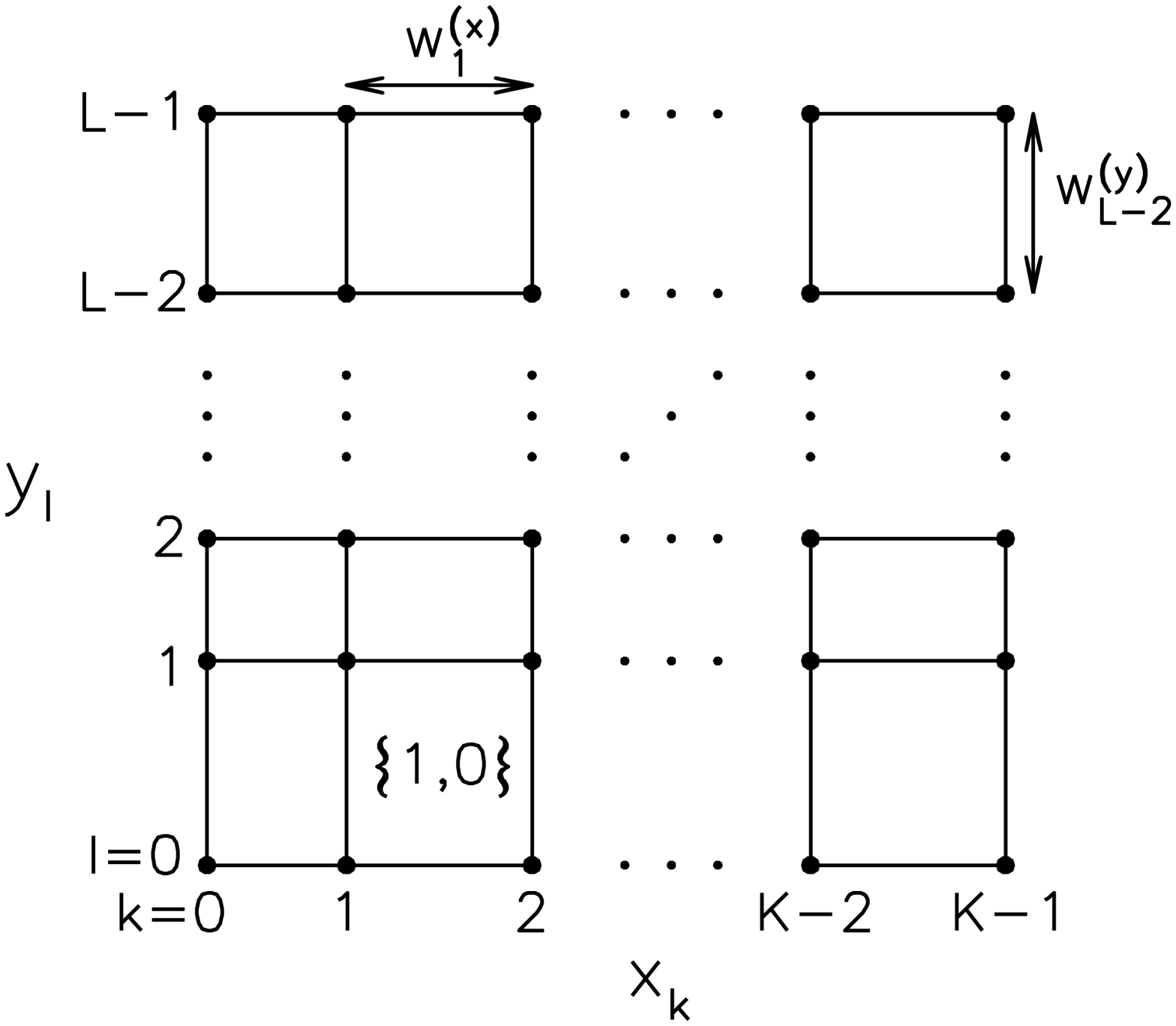}
\vspace*{-0.3cm}
\caption[Grid points in two dimensions.]{Grid points in two dimensions.}
\label{fig:2dspline}
\vspace*{-0.6cm}
\end{wrapfigure}

\noindent
where $C_{i,j}^{k,l}$ are the spline coefficients and $t_{(k)}$ and $u_{(l)}$ the dimensionless coordinates:
\be
t_{(k)} \equiv \frac{x-x_{k}}{w_{k}^{(x)}},\quad\quad u_{(l)} \equiv \frac{y-y_{l}}{w_{l}^{(y)}}
\label{eq:t2}
\ee
with the widths
\be
w_{k}^{(x)} \equiv x_{k+1}-x_{k},\quad\quad w_l^{(y)} \equiv y_{l+1}-y_{l}
\ee
The linear equations that determine the spline parameters $f_{k,l}$ from the coefficients for the $x$ and $y$ directions (summarized in matrix form as $X$ and $Y$, respectively) are independent from each other. Inverting these equations one obtains for the coefficients
\be
C_{i,j}^{k,l} = \sum\limits_{n_1=0}^{K-1}\sum\limits_{n_2=0}^{L-1} (Y^{-1})_{4l+j}^{n_2} (X^{-1})_{4k+i}^{n_1} f_{n_1,n_2}
\label{eq:c2}
\ee
Here the matrix $X$ ($Y$) only depends on the number $K$ ($L$) of grid points in the $x$ ($y$) direction and the widths $w_k^{(x)}$ ($w_l^{(y)}$).
Similarly, in arbitrary dimensions $D$ the different directions decouple and thus~(\ref{eq:c2}) is easily generalized to the case of more dimensions. In the following we introduce the fitting method in two dimensions; the generalization for higher dimensions is also straightforward to implement.

\section{Spline fitting}

For any value of the parameters $f_{k,l}$ the coefficients $C_{i,j}^{k,l}$ -- i.e. the whole spline surface $S(x,y)$ -- are known unambiguously. This way we consistently parameterized the spline surface (for the purpose of the QCD equation of state, $x$ plays the role of the temperature and $y$ the role of the quark mass ratio $R$). Now we want to set the spline parameters $f_{k,l}$ such, that derivatives of the spline surface are as close to some previously measured values as possible. Note that this way the function $S(x,y)$ will be undetermined up to an overall constant, since the translation $S(x,y)\to S(x,y)+A$ does not influence the derivatives. This symmetry will be taken into account in the following.

Let us consider $N$ number of points and say that at each point $(q^{(x)}_m,q^{(y)}_m)$ we measured the derivative in the $x$ and $y$ directions: $D^{(x)}_m$ and $D^{(y)}_m$ with errors $\Delta D^{(x)}_m$ and $\Delta D^{(y)}_m$ ($m=0\ldots N-1$). Being ``close'' can be quantified by minimizing
\be
\chi^2(f_{k,l})=\sum\limits_{m=0}^{N-1}\left[ \left( \frac{\frac{\partial S}{\partial x} - D^{(x)}_m}{\Delta D^{(x)}_m}\right)^2 + \left(\frac{\frac{\partial S}{\partial y} - D^{(y)}_m}{\Delta D^{(y)}_m}\right)^2 \right ]
\label{eq:chisqr}
\ee

Since $S$ and thus $\partial S/\partial x$ and $\partial S/\partial y$ are linear in $f_{k,l}$, this function has a quadratic dependence on the parameters $f_{k,l}$. This enables us to search for the minimum of $\chi^2(f_{k,l})$
\be
\frac{\partial \chi^2}{\partial f_{n_1,n_2}} = 0
\ee
by solving a system of linear equations
\be
M_{n_1,n_2}^{k,l} f_{k,l} = V_{n_1,n_2}, \quad n_1=0\ldots K-1,\;\; n_2=0\ldots L-1
\label{eq:sys}
\ee

Due to the above mentioned translational invariance of the solution, this system of equations is underdetermined and thus the inverse of $M$ does not exist. This can also be seen by checking that $M$ has a zero eigenvalue, corresponding to the eigenvector $(1,1,1,\ldots)$, or, in other words, each row of $M$ adds up to zero. Physically this means that one can set e.g. the first element of $f$ to zero, i.e. leave the first column of $M$. To obtain an invertible matrix one now has to drop one of its rows, for example the first\footnote{It can be checked that after eliminating the first column, any row of the matrix $M$ can be reproduced by a linear combination of the other rows, i.e. it is indeed correct to drop an arbitrary row.}. This way one arrives at a matrix $M'$ of size $(KL-1)\times (KL-1)$. In the same manner we define $V'$ to be the vector composed from the last $KL-1$ elements of $V$. One can then complement the solution $(M')^{-1}V'$ with a zero in the first element to obtain the final result which satisfies $f_{0,0}=0$.

In order to explicitly write the elements of the matrix $M$ and the vector $V$ let us analyze the dependence of $\chi^2$ on the parameters $f_{k,l}$. To this end we define $k(m)$ and $l(m)$ as the indices of the grid square that contains the $m$th measurement, i.e.
\be
(q^{(x)}_m,q^{(y)}_m) \in \{k(m),l(m)\}
\ee
and define $\xi_m$ and $\eta_m$ as the value of the dimensionless coordinate on that grid square corresponding to $q^{(x)}_m$ and $q^{(y)}_m$ (just as in~(\ref{eq:t2})):
\be
\xi_m \equiv \frac{q^{(x)}_m-x_{k(m)}}{w_{k(m)}^{(x)}},\quad\quad \eta_m \equiv \frac{q^{(y)}_m-y_{l(m)}}{w_{l(m)}^{(y)}}
\ee
Also, in order to be able to express $\partial S/\partial x$ and $\partial S/\partial y$ let us define the following matrices:
\be
\begin{split}
E_{m}^{n_1,n_2} &= \sum\limits_{i,j=0}^{3} (Y^{-1})_{4l(m)+j}^{n_2} (X^{-1})_{4k(m)+i}^{n_1} \cdot i\xi_m^{i-1} / w^{(x)}_{k(m)} \cdot \eta_m^j\\
F_{m}^{n_1,n_2} &= \sum\limits_{i,j=0}^{3} (Y^{-1})_{4l(m)+j}^{n_2} (X^{-1})_{4k(m)+i}^{n_1} \cdot \xi_m^i \cdot j\eta_m^{j-1} / w^{(y)}_{l(m)}
\end{split}
\ee
With the matrices $E$ and $F$ the expression for $\chi^2$ in~(\ref{eq:chisqr}) can be rewritten as:
\be
\chi^2(f_{k,l})= \sum\limits_{m=0}^{N-1}\left[ \left( \frac{E_m^{n_1,n_2}f_{n_1,n_2} - D^{(x)}_m}{\Delta D^{(x)}_m}\right)^2 + \left( \frac{F_m^{n_1,n_2}f_{n_1,n_2} - D^{(y)}_m}{\Delta D^{(y)}_m} \right)^2 \right]
\ee
which implies that in the system of linear equations~(\ref{eq:sys}) to be solved appear
\be
\begin{split}
M_{n_1,n_2}^{k,l} &= \sum\limits_{m=0}^{N-1} \left [ \left(\Delta D^{(x)}_m\right)^{-2} E_m^{k,l}E_m^{n_1,n_2} + \left(\Delta D^{(y)}_m\right)^{-2} F_m^{k,l}F_m^{n_1,n_2} \right]\\
V_{n_1,n_2} &= \sum\limits_{m=0}^{N-1} \left [ \left(\Delta D^{(x)}_m\right)^{-2} E_m^{n_1,n_2} D^{(x)}_m + \left(\Delta D^{(y)}_m\right)^{-2} F_m^{n_1,n_2} D^{(y)}_m \right]
\end{split}
\ee
Using the actual form of $M$ and $V$ the system of linear equations in~(\ref{eq:sys}) can be solved\footnote{The system of linear equations can be solved using e.g. the Lapack library.} for $f_{k,l}$. With the spline parameters the spline coefficients are also determined through~(\ref{eq:c2}). 

Since the number of measurements is $2N$ and we have $K\cdot L-1$ independent spline parameters, the degrees of freedom of this fit is given by the difference $\textmd{dof}=2N-K\cdot L+1$. This way the freedom corresponding to the translational symmetry is transformed out: since $f_{0,0}=0$, the result $S(x,y)$ is now set such that $S(x_0,y_0) = 0$ holds. Note that the spline function can also be transformed easily to satisfy some a priori known reference equation $S(x_r,y_r) = S_r$ by a simple translation $S\to S+(S_r-S(x_r,y_r))$.

Note that the $\chi^2$ in~(\ref{eq:chisqr}) represents a situation where the input data for the derivatives $D^{(x)}$ and $D^{(y)}$ are uncorrelated. However, this is usually not the case and one has to take into account the correlation between the measurements and therefore the $\chi^2$ function will contain additional terms. Usually measurements at different $m$ values are independent, but the correlation of $D_x(m)$ and $D_y(m)$ at the same $m$ is not negligible, since e.g. in a lattice calculation these are determined using the same configurations. This leads to the following $\chi^2_{corr}$:
\be
\begin{split}
\chi^2_{corr}=\sum\limits_{m=0}^{M-1}\Bigg[ &{Q_{(m)}^{-1}}_{00}\left(\frac{\partial S}{\partial x} - D_x(m)\right)^2 + \\
&2{Q_{(m)}^{-1}}_{01}\left(\frac{\partial S}{\partial x}-D_x(m)\right)\left(\frac{\partial S}{\partial y}-D_y(m)\right) +{Q_{(m)}^{-1}}_{11}\left(\frac{\partial S}{\partial y} - D_y(m)\right)^2 \Bigg]
\label{eq:chi2}
\end{split}
\ee
where $Q_{(m)}$ is the $2\times2$ correlation matrix consisting of the correlators of the two measured derivatives at the $m$th point. The corresponding system of linear equations has the same form as~(\ref{eq:sys}); only now on the left and right hand side enter
\be
\begin{split}
M_{n_1,n_2}^{k,l} = \sum\limits_{m=0}^{N-1} \big[ &{Q_{(m)}^{-1}}_{00} E_m^{k,l}E_m^{n_1,n_2} + 2{Q_{(m)}^{-1}}_{01} (E_m^{k,l}F_m^{n_1,n_2}+E_m^{n_1,n_2}F_m^{k,l}) \\
&+ {Q_{(m)}^{-1}}_{11} F_m^{k,l}F_m^{n_1,n_2} \big]\\
V_{n_1,n_2} = \sum\limits_{m=0}^{N-1} \big [ &{Q_{(m)}^{-1}}_{00} D_x(m)E_m^{n_1,n_2} + 2{Q_{(m)}^{-1}}_{01} (D_x(m)F_m^{n_1,n_2} \\
&+E_m^{n_1,n_2}D_y(m))+ {Q_{(m)}^{-1}}_{11} D_y(m)F_m^{n_1,n_2} \big]
\end{split}
\ee

\section{Stable solutions}

The above prescribed method is bound to give the function $S(x,y)$ for which the sum of deviations $\chi^2$ is the smallest. The solution on the other hand also depends on the number and the position of the nodepoints, and these have to be tuned appropriately in order to determine the surface that fits best.

If the number of nodepoints $K\cdot L$ is small compared to the number of measurements $N$, then the reduced sum of deviations will be large ($\chi^2/\textmd{dof} \gg 1$) and the spline function $S(x,y)$ may not be a good approximation to the surface sought for. On the other hand if $K\cdot L$ is large\footnote{Obviously the inequality $K\cdot L-1<2N$ should hold otherwise the problem is underdetermined.}, the best fit will become an oscillatory function which has the correct derivatives everywhere (i.e. $\chi^2/\textmd{dof}\approx 1$), but is probably not the ``right'' solution, especially when one knows that $S$ should be monotonic\footnote{For example this is the case for the pressure $\log\Z$ as a function of the temperature.} in e.g. $x$. Note that this unwanted feature is a characteristic of spline functions even in one dimension.

\begin{figure}[h!]
\centering
\includegraphics*[width=7.0cm]{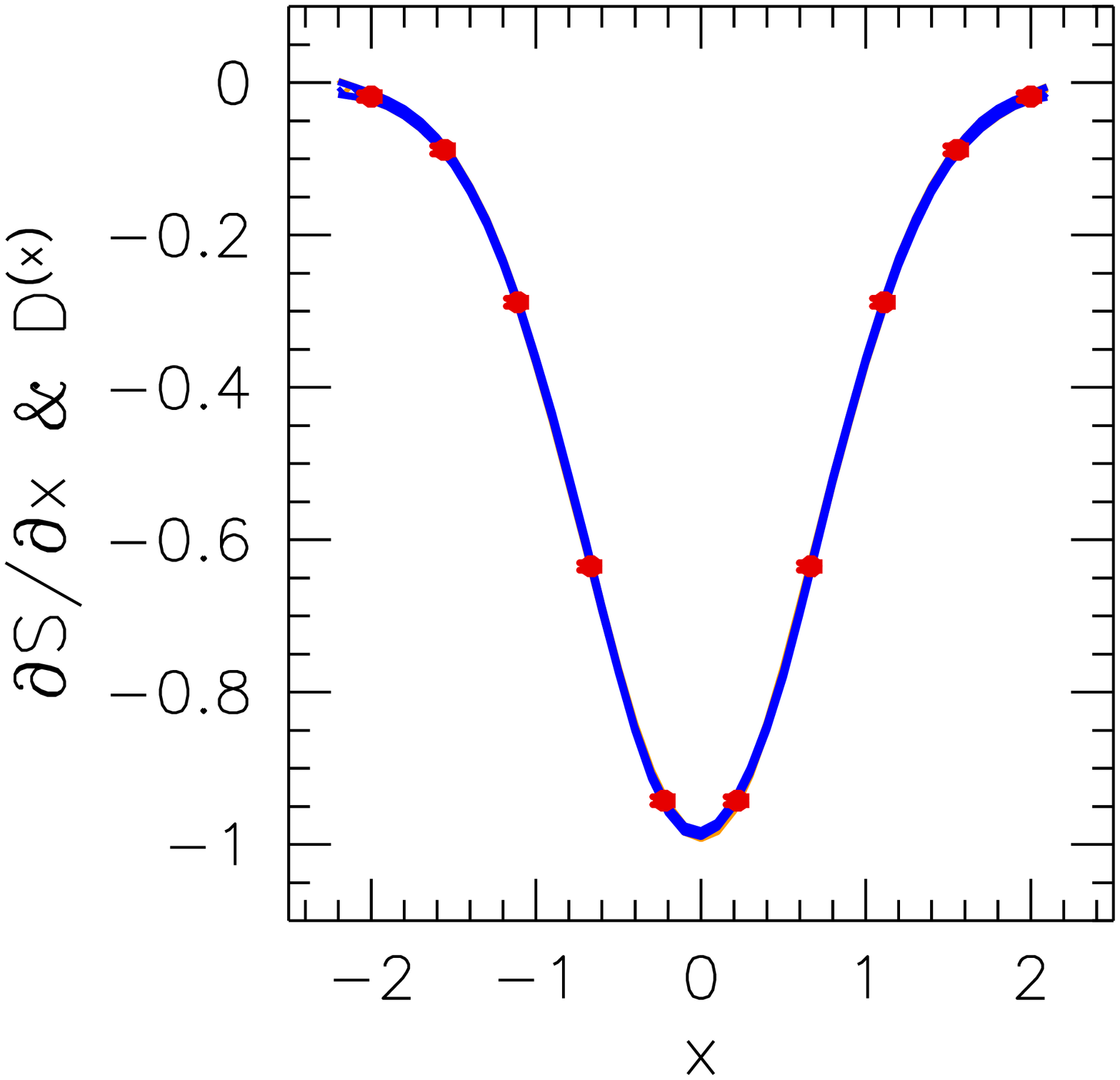} \quad
\includegraphics*[width=7.0cm]{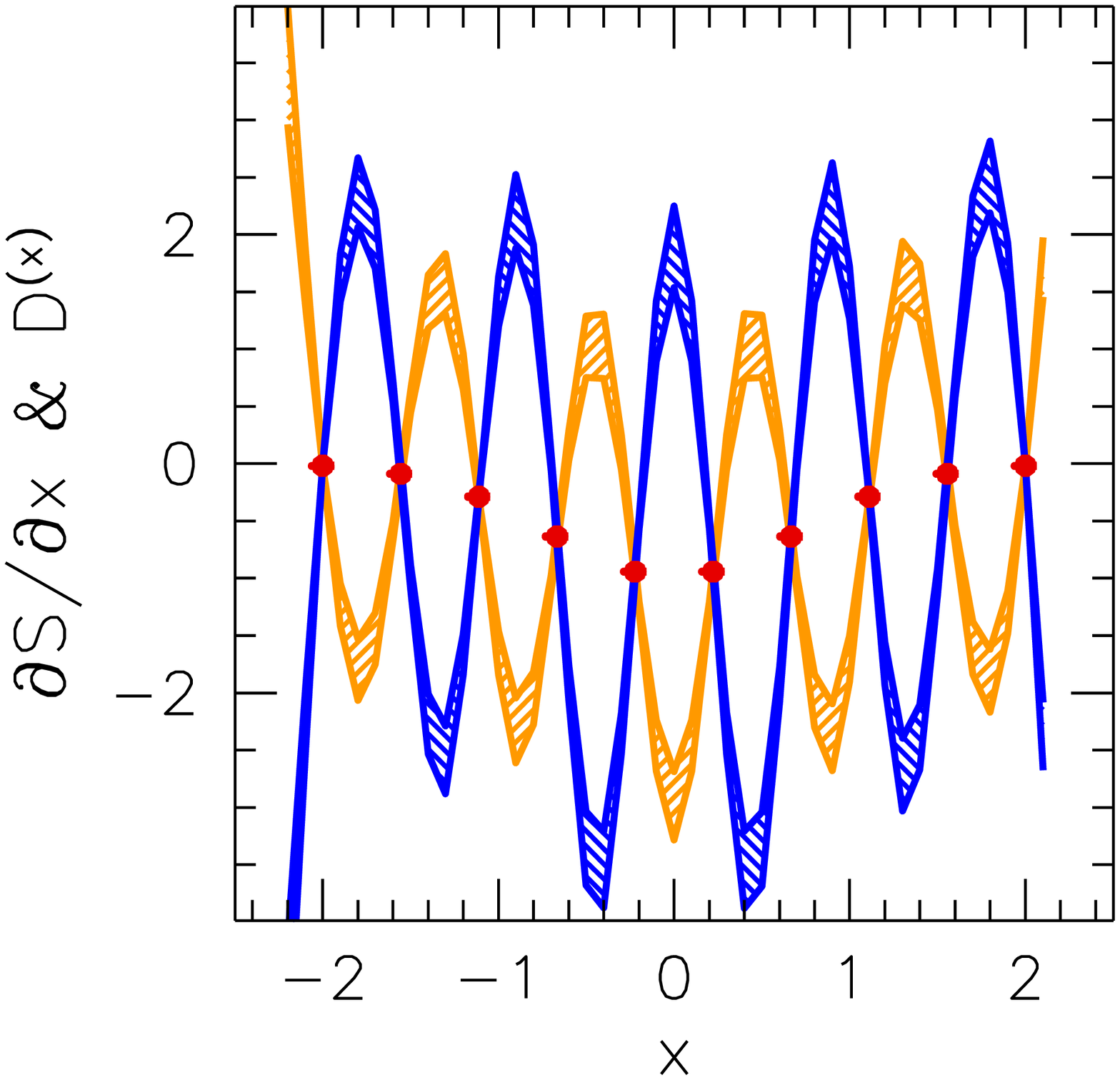}
\caption[A method to filter out unstable solutions.]{One-dimensional slice of a two-dimensional fit. Shown are the input data $D^{(x)}$ and the derivative $\partial S/\partial x$ of the spline surface obtained using two nodepoint sets. The two sets differ only in one gridpoint, see~(\ref{eq:nodep}). On the left side the number of nodepoints in the $x$ direction is $K=10$; a slight change in the nodepoints results in no visible change in the solution (orange and blue curves are on top of each other). On the other hand, on the right side $K=14$, and a similar change dramatically distorts the solution. The corresponding values for $\mathcal{D}$ are $0.004$ (left) and $\sim10^7$ (right).}
\label{fig:stable}
\end{figure}

Thus we need a measure of how ``right'' the solution is. A useful way to define this property is to investigate how much $f$ changes as the nodepoints are modified, since the oscillatory solutions are unstable even under a small change in the nodepoints. This way we can filter out the stable, realistic solutions (see illustration on figure~\ref{fig:stable}).

We define modified nodepoint sets $x^{(\alpha)}$ as
\be
\begin{split}
x^{(\alpha)}_k &= x_k+\varepsilon, \quad\quad \textmd{if } k=\alpha \\
x^{(\alpha)}_k &= x_k, \quad\quad\quad\;\,\, \textmd{otherwise}
\end{split}
\label{eq:nodep}
\ee
with $\varepsilon$ some small number, e.g. $\varepsilon=(x_{K-1}-x_0)/K/10$, and in the same manner for $y^{(\beta)}$. Now we carry out the fitting procedure for each of the modified nodepoint sets, resulting in the modified solutions $f^{(\alpha)}$ and $f^{(\beta)}$. The sum of the relative differences between the original solution $f$ and the modified solutions
\be
\mathcal{D} \equiv \frac{1}{K} \sum_{\alpha = 0}^{K-1} \frac{1}{K\cdot L} \sum\limits_{k,l} \left|\frac{f^{(\alpha)}_{k,l} - f_{k,l}}{f_{k,l}}\right| + \frac{1}{L} \sum_{\beta = 0}^{L-1} \frac{1}{K\cdot L} \sum\limits_{k,l} \left|\frac{f^{(\beta)}_{k,l} - f_{k,l}}{f_{k,l}}\right|
\label{eq:D}
\ee
will indeed serve as an indicator of the stability of the fit. If this relative change $\mathcal D$ is under a few percents then the fit can be considered stable. 

\section{Systematics of the method}

The statistical error $\sigma_{stat}$ of the result from the spline fitting method described above can be determined using the standard jackknife algorithm, i.e. the system of linear equations~(\ref{eq:sys}) needs to be solved for each jackknife sample\footnote{Note that this can be performed in a single Lapack call that solves the equations for different vectors on the right hand side.}. The systematic error on the other hand can be determined by varying the nodepoints $\{x_k,y_l\}$. Based on experience the number of nodepoints may range from $M/2$ to $M$; usually an equidistant nodepoint-set can already produce a small value for $\chi^2/\textmd{dof}$, but increasing the density of nodepoints in areas where the function changes rapidly (i.e. where the measured derivatives are large) can further help to improve the fit quality.

Accordingly, a straightforward way to determine the systematic error is to generate various nodepoint sets with different number (and position) of gridpoints.
Then, for each set $\tau$ of the nodepoints the fit is carried out resulting in a spline function $S_\tau$ and an indicator $G_\tau = \left(\chi^2_{corr}/\textmd{dof}\right)^{-1}$ of the fit quality. Results which are in the above detailed sense not stable should be filtered out at this point.
Then at each point the systematic error $\sigma_{sys}$ is determined by
\be
\sigma_{sys} (x,y) = \sqrt{ \left\langle S_{\tau}(x,y)^2 \right \rangle_G - \left\langle S_{\tau}(x,y) \right \rangle_G^2 }
\ee
with
\be
\left\langle \mathcal{O}_{\tau} \right \rangle_G = \sum\limits_\tau \mathcal{O}_\tau G_\tau \Big/ \sum\limits_\tau G_\tau
\ee
Thus the total error can be estimated to be
\be
\sigma_{tot}=\sqrt{\sigma_{sys}^2+\sigma_{stat}^2}
\ee

The above discussed method can be compared to conventional integration in two dimensions~\cite{Endrodi:2010ai}. Based on this comparison we conclude that the multidimensional spline algorithm effectively processes the input data and results in much smaller statistical errors than the usual integral method. Furthermore, the main advantage of the spline fitting procedure is that it better estimates the systematics of the result. While for ordinary integration only a single path in the space of the parameters is considered, the spline method is in some sense equivalent to taking into account all possible integration paths at the same time. Our results indicate that the contribution coming from these generalized paths cannot be neglected in order to estimate the systematics.

%\cleardoublepage
\bibliographystyle{mybib} %{Classes/CUEDbiblio}
\renewcommand{\bibname}{References} % changes default name Bibliography to References
\bibliography{thesis} % References file
%\phantomsection
%\addcontentsline{toc}{chapter}{References} %adds References to contents page

\newpage
\thispagestyle{empty}
\begin{center}
\vspace*{-2.6cm}
{\large \bf QCD thermodynamics with dynamical fermions}\\
\vspace*{0.3cm}
{\bf Gergely Endr\H{o}di }\\
\vspace*{0.2cm}
-- Summary --
\vspace*{-0.1cm}
\end{center}
In this thesis I studied the finite temperature transition between confined and deconfined matter at zero and nonzero quark densities. The findings are relevant for the understanding of the evolution of the early Universe and contemporary and upcoming heavy ion experiments. Our results were obtained using large-scale simulations using lattices with various physical extents and lattice spacings, at physical values of the parameters of the theory. In all three projects we thoroughly checked finite size effects and discretization effects, thereby controlling the two most important possible sources of systematic error. In order to ensure that the extrapolations lead to the correct continuum limit, all the necessary additive and multiplicative renormalizations were applied to the studied observables.

In chapter~\ref{chap:phasediag} the study of the phase diagram of QCD at small chemical potentials was presented. We determined the curvature of the transition line which separates the chirally symmetric and asymmetric phases using two observables that are sensitive to this transition. Our results also contain information about the relative change in the width of the transition. This suggests that the strength of the transition decreases very slightly, therefore the weak crossover that is present at zero density persists up to quark chemical potentials $\mu_{u,d} \lesssim T_c$.

In chapter~\ref{chap:EoS} I showed results regarding the equation of state in full QCD. For this project I developed a new method to determine the pressure from the raw lattice data. This approach improves on the conventionally applied integral method such that it better estimates the systematic error of the result, and is also capable of quantifying the quark mass dependence of the pressure. From the pressure we derived various other thermodynamic observables, which can be directly used to set the equation of state in e.g. hydrodynamic models. We estimate the effect of a charm quark in the pressure and show that it might be important already at $T\sim 2T_c$. Furthermore, we observe that the pressure, the energy density and the entropy density are roughly 15-20$\%$ under their Stefan-Boltzmann limits at our highest temperature $\sim 6T_c$.

A these values of the temperature a reliable comparison with perturbation theory cannot be carried out since such perturbative expansions are only convergent at much larger temperatures. In chapter~\ref{chap:PG} I report results of the equation of state in pure gauge theory at extremely high temperatures $\sim 1000 T_c$. We perform simulations using lattices of size up to $N_s/N_t=24$ and gather statistics that exceed results in the literature by at least an order of magnitude to determine the high temperature signal in the interaction measure. At large $T$ we successfully match our results with perturbation theory and we are able to fit the unknown $\mathcal{O}(g^6)$ coefficient of the expansion. We also quantify the non-perturbative contribution to the interaction measure, which is proportional to $T^2$.

\newpage
\thispagestyle{empty}
\begin{center}
\vspace*{-2.6cm}
{\large \bf QCD termodinamika dinamikus fermionokkal}\\
\vspace*{0.3cm}
{\bf Endr\H{o}di Gergely }\\
\vspace*{0.2cm}
-- \"Osszefoglal\'o --
\vspace*{-0.1cm}
\end{center}

Disszert\'aci\'omban az er\H{o}sen k\"olcs\"onhat\'o anyag bez\'art, illetve plazma \'allapota k\"oz\"otti v\'eges h\H{o}m\'ers\'ekleti \'atmenetet vizsg\'altam nulla \'es v\'eges kvark s\H{u}r\H{u}s\'egekn\'el. Az eredm\'enyek mind a korai Univerzum fejl\H{o}d\'es\'enek, mind pedig napjaink neh\'ezion-\"utk\"oztet\'esi k\'is\'erleteinek szempontj\'ab\'ol relev\'ansak. Az eredm\'enyeinket nagy sz\'am\'it\'asig\'eny\H{u} r\'acs szimul\'aci\'ok haszn\'alat\'aval kaptuk; a szimul\'aci\'okat sz\'amos k\"ul\"onb\"oz\H{o} m\'eret\H{u} \'es r\'acs\'alland\'oj\'u r\'acson v\'egezt\"uk el, az elm\'elet param\'etereinek fizikai \'ert\'eke mellett. Mindh\'arom bemutatott projektben megm\'ert\"uk a v\'eges m\'eret effektusokat \'es a diszkretiz\'aci\'os hib\'akat, ez\'altal \"ugyelve a k\'et legfontosabb lehets\'eges szisztematikus hibaforr\'asra. A vizsg\'alt mennyis\'egek minden sz\"uks\'eges addit\'iv \'es multiplikat\'iv renorm\'al\'as\'at elv\'egezt\"uk.

A~\ref{chap:phasediag}. fejezetben a QCD f\'azisdiagramj\'anak vizsg\'alat\'at mutattam be. Meghat\'aroztuk az 
\'atmeneti vonal g\"orb\"ulet\'et k\'et olyan mennyis\'eg seg\'its\'eg\'evel, amelyek \'erz\'ekenyek erre az \'atmenetre. Az eredm\'enyek az \'atmenet er\H{o}ss\'eg\'enek relat\'iv v\'altoz\'as\'ar\'ol is sz\'amot adnak. Meg\'allap\'itottuk, hogy az \'atmenet csek\'ely m\'ert\'ekben gyeng\"ul, \'igy a $\mu=0$-n\'al megval\'osul\'o crossover alacsony k\'emiai potenci\'aln\'al $\mu_{u,d} \lesssim T_c$ is fennmarad.

Az~\ref{chap:EoS}. fejezetben a QCD \'allapotegyenlet\'et vizsg\'altam dinamikus fermionok jelenl\'et\'eben. Ehhez a projekthez kifejlesztettem egy \'uj m\'odszert a nyom\'as nyers r\'acs adatokb\'ol t\"ort\'en\H{o} meghat\'aroz\'as\'ara. Ez a technika a hagyom\'anyosan alkalmazott integr\'al m\'odszern\'el effekt\'iveb-ben becsli meg a szisztematikus hib\'at, \'es a nyom\'as kvarkt\"omeg-f\"ugg\'es\'et is k\"ozvetlen\"ul megadja. A nyom\'asb\'ol sz\'amos m\'as termodinamikai mennyis\'eget is meghat\'aroztunk, amelyek k\"ozvetlen\"ul felhaszn\'alhat\'oak az \'allapotegyenlet be\'all\'it\'as\'ara p\'eld\'aul hidrodinamikai modellekben. Megm\'er-t\"uk a charm kvark j\'arul\'ek\'at a nyom\'ashoz, \'es azt tal\'altuk, hogy m\'ar $\sim2 T_c$-n jelent\H{o}s j\'arul\'ekr\'ol van sz\'o. Meg\'allap\'itottuk tov\'abb\'a, hogy a nyom\'as, az energias\H{u}r\H{u}s\'eg \'es az entr\'opias\H{u}r\H{u}s\'eg k\"or\"ulbel\"ul 15-20$\%$-kal alulm\'ulj\'ak Stefan-Boltzmann limeszeiket a legnagyobb h\H{o}m\'ers\'eklet\"unk-n\'el ($T\sim 6T_c$) is.

Ezeken a h\H{o}m\'ers\'ekleteken nem lehets\'eges \"osszehasonl\'itani a r\'acs eredm\'enyeket a perturb\'aci\'osz\'am\'it\'assal, ugyanis a perturbat\'iv sorfejt\'esek csak rendk\'iv\"ul magas h\H{o}m\'ers\'ekleten konverg\'alnak. A~\ref{chap:PG}. fejezetben ez\'ert extr\'em magas h\H{o}m\'ers\'ekletekig ($T\sim 1000 T_c$) megm\'ert\"uk a nyom\'ast tiszta m\'ert\'ekelm\'eletben; ehhez az alkalmazott r\'acsok t\'erir\'any\'u kiterjed\'es\'et rendk\'iv\"ul nagynak (ak\'ar $N_s/N_t=24$-nek) kellett megv\'alasztani. Igen nagy sz\'amban gener\'altunk konfigur\'aci\'okat; a kapott statisztikus hiba az irodalomban fellelhet\H{o} eredm\'enyeket legal\'abb egy nagys\'agrenddel fel\"ulm\'ulja. 
Magas h\H{o}m\'ers\'ekleten \"osszehasonl\'itottuk az eredm\'enyeket a jav\'itott, illetve fel\"osszegzett perturb\'aci\'osz\'am\'it\'as k\'epleteivel, \'es illesztett\"uk az $\mathcal{O}(g^6)$ rend szabad egy\"utt-hat\'oj\'at a r\'acs adatok seg\'its\'eg\'evel. V\'eg\"ul k\'et egyszer\H{u} f\"uggv\'eny form\'aj\'aban meghat\'aroztuk a k\"olcs\"onhat\'asi m\'ert\'ekhez j\'arul\'o nemperturbat\'iv tagot, amely $T^2$-tel ar\'anyos.

\end{document}